\documentclass[manuscript,screen]{acmart}
\usepackage{subcaption}
\usepackage{multirow} 
\usepackage{booktabs}
\usepackage{caption}
\usepackage{multirow}
\usepackage{amsthm,amsmath}
\usepackage{caption}
\usepackage{xcolor}
\usepackage[ruled,linesnumbered]{algorithm2e}
\usepackage{enumitem}
\usepackage[most]{tcolorbox}
\usepackage{balance}
\usepackage{url}
\usepackage{pifont} 
\usepackage{tabularx}
\usepackage{graphicx}
\usepackage{array}
\usepackage{longtable}

\usepackage{amsmath,amsfonts}
\usepackage[ruled,linesnumbered]{algorithm2e}
\usepackage{algorithmic}
\algsetup{linenosize=\small}
\SetKwInput{KwInput}{Input}                
\SetKwInput{KwOutput}{Output}              
\usepackage{quoting}

\AtBeginDocument{%
  \providecommand\BibTeX{{%
    \normalfont B\kern-0.5em{\scshape i\kern-0.25em b}\kern-0.8em\TeX}}}

\setcopyright{acmcopyright}
\acmJournal{CSUR}
\acmYear{2023} \acmVolume{1} \acmNumber{1} \acmArticle{1} \acmMonth{10} \acmPrice{15.00}\acmDOI{10.1145/xxxxx }

\acmJournal{JACM}
\acmVolume{37}
\acmNumber{4}
\acmMonth{10}

\begin{document}

\title{\textit{Robustness}, \textit{Security}, \textit{Privacy}, \textit{Explainability}, \textit{Efficiency}, and \textit{Usability}  of Large Language Models for Code}

\author{Zhou Yang}
\email{zyang@smu.edu.sg}
\affiliation{%
  \institution{Singapore Management University}       
  \country{Singapore}
}

\author{Zhensu Sun}
\email{zssun@smu.edu.sg}
\affiliation{%
  \institution{Singapore Management University}
  \country{Singapore}
}

\author{Terry Zhuo Yue}
\email{terryzhuo@smu.edu.sg}
\affiliation{%
  \institution{Singapore Management University}
  \country{Singapore}
}

\author{Premkumar Devanbu}
\email{ptdevanbu@ucdavis.edu}
\affiliation{%
  \institution{Department of Computer Science, UC Davis}
  \country{USA}
}

\author{David Lo}
\email{davidlo@smu.edu.sg}
\affiliation{%
  \institution{Singapore Management University}
  \country{Singapore}
}
\renewcommand{\shortauthors}{Yang et al.}

\begin{abstract}
  Large language models for code (\textit{LLM4Code}), which demonstrate strong performance (e.g., high accuracy) in processing source code, have significantly transformed software engineering. 
  Many studies separately investigate the non-functional properties of LM4Code, but there is no systematic review of how these properties are evaluated and enhanced.
  This paper fills this gap by thoroughly examining 146 relevant studies, thereby presenting the first systematic literature review to identify seven important properties beyond accuracy, including \textit{robustness}, \textit{security}, \textit{privacy}, \textit{explainability}, \textit{efficiency}, and \textit{usability}.
  We discuss the current state-of-the-art methods and trends, identify gaps in existing research, and present promising directions for future study.
\end{abstract}



\maketitle

\section{Introduction} \label{sec:intro}


The field of software engineering has witnessed a surge in large language models specifically tailored to understand and process code, which we call \textit{large language models for code} (\textit{LLM4Code})~\cite{codellm_survey}.
Researchers~\cite{niu2023empirical,10.1145/3533767.3534390} empirically show LLM4Code's outstanding performance on various tasks, including but not limited to vulnerability detection and code generation.
These powerful models have swiftly evolved from experimental prototypes to practical tools, integrating into the daily workflow of software developers around the globe. 
For example, GitHub Copilot is packaged as extensions in Visual Studio Code.
A range of similar tools supported by different LLM4Code have also been developed and deployed, including Amazon CodeWhisperer\footnote{https://aws.amazon.com/codewhisperer/faqs/} and CodeGeeX.\footnote{https://codegeex.cn}
Far from being mere code completion aids, these tools are capable of executing a variety of complex tasks such as code summarization and question answering.

\begin{table}[!t]
  \centering
  \caption{Main findings drawn from the survey for each non-functional property of LLM4Code.}
  \small 
  \begin{tabular}{m{1.5cm}m{13cm}}
    \toprule
  Properties & Main Findings \\
  \midrule
\multirow{3}{*}{Robustness} & \ding{172} LLM4Code suffer from low robustness issue. \\ 
& \ding{173} More studies are required on evaluating robustness of LLM4Code solutions for generation tasks. \\ 
& \ding{174} More studies on improving the efficiency and scalability of robustness enhancement techniques are required. \\ 
\midrule 
\multirow{3}{*}{Security} & \ding{172} LLM4Code is at risk due to security threats like data poisoning; The ease of access and possibility of alteration of training data by attackers increase this vulnerability.\\ 
& \ding{173} Researchers focus more on creating stealthy attacks that are difficult to detect. \\ 
& \ding{174} Current detection methods struggle to identify these stealthy attacks, indicating a need for more effective defense strategies. \\ 
\midrule 
\multirow{3}{*}{Privacy} & \ding{172} LLM4Code can leak sensitive information like personal privacy (e.g., emails, passwords, IP addresses) and effectiveness of existing mitigation strategies is unclear.\\ 
 & \ding{173} Membership inference attacks can be utilized to protect privacy by detecting unauthorized data usage.  \\ 
 & \ding{174} Privacy issues related to model stealing and gradient leakage are underexplored in the context of LLM4Code. \\ 
\midrule 
\multirow{3}{*}{Explainability} & \ding{172} There exists noted inconsistency among explanations provided by different techniques, highlighting the need for more reliable methods. \\ 
 &  \ding{173} More studies are required on explaining generation tasks (e.g., code generation) compared to those on classification tasks (e.g., defect prediction). \\ 
 & \ding{174} There is a gap in meeting end-users' needs for explainability. \\ 
 \midrule 

 \multirow{3}{*}{Efficiency} & \ding{172} Parameter-Efficient Fine-tuning is gaining popularity for enhancing the training efficiency of LLM4Code. \\ 
  & \ding{173} Research on alternative methods such as quantization and model pruning, is less common. \\ 
  & \ding{174} The impact of efficiency improvements on other aspects like robustness and security is not well-understood and needs more study. \\ 
 \midrule 

 \multirow{3}{*}{Usability} & \ding{172} The impact of LLM4Code on productivity is mixed, with studies finding both positive and negative effects. \\ 
 & \ding{173} We suggest a need for broader attention to understand usability of LM4Code solutions on under-explored applications such as vulnerability detection. \\ 
 & \ding{174} Only a few studies have implemented and tested solutions for improving LLM4Code usability in practical settings. \\ 

  \bottomrule
  \end{tabular}
  \label{tab:findings}
\end{table}

The outstanding performance of LLM4Code, characterized by their capacity to produce outputs with \textbf{high accuracy}, addresses the majority of the \textit{functional requirements}. 
Yet, researchers have recently unraveled that LLM4Code fail to satisfy many \textit{non-functional requirements}.
Specifically, LLM4Code have low robustness~\cite{Yefet2020,alert}, e.g., an accurate vulnerability predictor easily changes its prediction when some local variable names are altered in the input.
Researchers also find that other issues, including leaking sensitive information~\cite{yang2023memorzation}, producing insecure code~\cite{DBLP:conf/sp/PearceA0DK22}, lacking explainability~\cite{Reemicse21}, etc.
For example, as highlighted by Hammond et al.~\cite{DBLP:conf/sp/PearceA0DK22}, roughly 40\% of programs generated by GitHub Copilot were found to contain vulnerabilities.\footnote{It is important to note that LLM4Code systems, particularly commercial versions, are continually evolving, incorporating new architectures and training data. Consequently, the conclusions drawn from the referenced publications may not entirely align with the current state of LLM4Code.}

Although many studies separately investigate different non-functional properties of LLM4Code, there lacks a systematic review on this topic.
To fill this gap, we explore an important and timely question: \textbf{what are relevant non-functional properties beyond accuracy to be considered when we develop and evaluate language models for code?}
By thoroughly examining 146 relevant studies, we identify seven important properties, including \textit{robustness}, \textit{security}, \textit{privacy}, \textit{explainability}, \textit{efficiency}, and \textit{usability}.
After defining each property, we explain the current state-of-the-art techniques to evaluate and enhance properties.
Based on the review, we discuss the existing challenge and research opportunities, proposing three perspectives on how to take non-functional properties into consideration when developing LLM4Code.
We summarize the main findings for each property in Table~\ref{tab:findings}.


\vspace*{0.2cm}
\noindent \textbf{Structure}.
Section~\ref{sec:methodology} describes the methodology of our systematic literature review and some statistics of the identified papers.
In Section~\ref{sec:property}, we describe seven important properties beyond accuracy of LLM4Code.
Section~\ref{sec:detail} explains how the properties beyond accuracy are evaluated and enhanced in the literature.
Section~\ref{sec:challenges} discusses the challenges and potential research opportunities to further study these properties.
Section~\ref{sec:threats} considers some threats to the validity of our study.
We put the conclusion of our study in Section~\ref{sec:conclusion}.

\section{Survey Methodology} \label{sec:methodology}


\subsection{Paper Collection} \label{subsec:paper-collection}

\begin{table}[!t]
  \caption{The inclusion criteria and exclusion criteria of our systematic literature review. The \textit{non-LLM4Code} are defined as the machine learning models of code that are not based on LLM.}
  \resizebox{1\linewidth}{!}{
  \begin{tabular}{ll}
  \hline
  \multicolumn{2}{l}{\textbf{Inclusion criteria}}                                        \\ \hline
  1) & The paper introduces/discusses one of non-functional properties of LLM4Code.                                                \\
  2) & The paper proposes an approach, study, tool/framework, or dataset/benchmark that targets one of these properties.                             \\
  3) & The paper introduces a set of measurement criteria that could be adopted to evaluate one of the properties.                                            \\ \hline
  \multicolumn{2}{l}{\textbf{Exclusion criteria}}                                        \\ \hline
  1) &  Literature that is not written in English.                             \\
  2) & Literature that is less than five pages. \\
  3) & Duplicate papers or studies with different versions (e.g., the preprint version and the officially published version) from the same authors.                                        \\
  4) & Literature that evaluate these properties in a non-LLM4Code (i.e., models for code but not based on LLM context.                                        \\
 \hline
  \end{tabular}}
  \label{tab:criteria}
  \end{table}

\subsubsection{Survey Scope} \label{subsubsec:scope}
The survey scope for this study has two primary dimensions.
Firstly, the survey includes papers that pertain to LLM4Code, encompassing source code and associated software artifacts.
Recent developments that involve utilizing language models for binary code, assembly code, or domain-specific languages fall outside the scope of this survey.
Secondly, the papers must explore quality attributes beyond accuracy of LLM4Code. 
Those papers that focus on accuracy and related metrics (e.g., F1 score, recall, etc) are consequently excluded from our analysis.\
To conisder the above two dimensions, we set the inclusion and exclusion criteria as shown in Table~\ref{tab:criteria}.

As depicted in previous literature reviews~\cite{codellm_survey,fan2023large}, CodeBERT~\cite{CodeBERT} is considered the pioneering LLM4Code.
LLM4Code in this paper refer to the large pre-trained models that are trained/fine-tuned on a large corpus of source code and are used for various software engineering tasks, ranging from the early BERT-faimly models (e.g., CodeBERT~\cite{CodeBERT} and GraphCodeBERT~\cite{GraphCodeBERT}) to more recent GPT-family models (e.g., CodeGen~\cite{codegen} and StarCoder~\cite{starcoder}). 
We also consider commercial services that can be used for software engineering tasks, e.g., GitHub Copilot and OpenAI ChatGPT, as LLM4Code.
However, we acknowledge the values of studies on \textit{non-LLM4Code}, i.e., the machine learning models of code that are not based on LLM.
To better present how relevant studies evolve from non-LLM4Code to LLM4Code, we also include studies that are conducted on models prior to CodeBERT when necessary, e.g., code2vec~\cite{code2vec}.


\subsubsection{Search Databases}
Our literature review spans multiple domains.
Both researchers from software engineering and other domains (e.g., natural language processing) are contributing to LLM4Code.
The properties involve multiple domains like human-computer interaction, security, etc.
The prevalence of LLM4Code in the research community has surged, with scholars increasingly putting their latest work on preprint websites such as arXiv. 
Guidelines for reporting secondary studies in software engineering highlight the value and importance of preprints~\cite{9772383}. 
However, these preprints are often not promptly integrated into popular publication databases.
A recent survey of LLM4Code finds that 62\% of the 229 relevant papers are preprints on arXiv.
Recent surveys~\cite{ml-testing-survey,chen2022fairness} suggest that the literature gathered from other well-established publication databases (e.g., IEEE Xplore and ACM Digital Library) is a subset of the corpus returned from DBLP. 
We use DBLP as the database to identify relevant papers, which not only ensures an inclusive collection of research papers but also facilitates timely access to the latest contributions in the field.

\subsubsection{Search Items} \label{subsubsec:search}


We craft an initial list of search items, drawing upon pertinent studies in general machine learning literature and the authors' knowledge in the field of software engineering. 
We discuss and update the property list, ending up with 7 important properties beyond accuracy in LLM4Code: \textit{robustness}, \textit{security}, \textit{privacy}, \textit{explainability}, \textit{efficiency}, and \textit{usability}.
Note that there are overlaps and interplay among these properties. 
For example, robustness can be considered as part of security.
We intentionally discuss them separately for a more in-depth analysis of each property, diving deep into specific challenges and solutions for each one.

We employ keyword searching on the DBLP publication database to gather papers. 
Our search queries should have two sets of keywords: the one pertaining to LLM4Code, and the other related to properties.
The one for LLM4Code can be further divided into keywords relevant to the model (called \textbf{[}model words\textbf{]}) and keywords relevant to the code (called \textbf{[}code words\textbf{]}).
The \textbf{[}code words\textbf{]} include `\textit{program}~\textbar~\textit{software}~\textbar~\textit{code},' and the \textbf{[}model words\textbf{]} include `\textit{model}~\textbar~\textit{learn}.'
Here the notion `\textbar' means `or' logic operator in the search query.

Section~\ref{sec:property} describes 7 properties beyond accuracy of LLM4Code in detail.
Each property can have multiple keywords that are relevant to it.
For example, many researchers use the term `adversarial attack' when evaluating robustness.
The term `interpretability' is also used to describe the explainability of LLM4Code.
Besides, papers may use different forms of the same keyword.
The term `interpretability' can be written as `interpretable' or `interpretation.'
As a result, we use truncated terms for each property. 
To ensure a high coverage of relevant papers, we create a list of keywords for each property and use them in the search query:

\begin{enumerate}
  \item \textbf{Robustness}: \textit{robust}, \textit{adversa}, \textit{attack}, \textit{perturb}, \textit{nois}, \textit{error};
  \item \textbf{Security}: \textit{secur}, \textit{breach}, \textit{vulnerabl}, \textit{safe}, \textit{defense};
  \item \textbf{Privacy}: \textit{privacy}, \textit{sensitive}, \textit{membership}, \textit{leak}, \textit{confidential};
  \item \textbf{Explainability}: \textit{interpret}, \textit{explain};
  \item \textbf{Efficiency}: \textit{efficien}, \textit{speed}, \textit{latency}, \textit{memory}, \textit{energy}, \textit{compress};
  \item \textbf{Usability}: \textit{usability}, \textit{user}, \textit{experience}, \textit{productivity}, \textit{flow}, \textit{learnability}, \textit{accessibility}, \textit{interface};
\end{enumerate}

The keywords for each property are called \textbf{[}property words\textbf{]} (e.g., `\textit{interpret}~\textbar~\textit{explain}' for explainability).
The final search query for each property is \textbf{[}code words\textbf{]}~+~\textbf{[}model words\textbf{]}~+~\textbf{[}property words\textbf{]}, where the `+' operator means `and' logic operator.
Two authors of the paper independently review the titles and abstracts of the papers returned by each query and decide its relevance to this survey.
A discussion among the first three authors is conducted in case of any disagreement.
This process is repeated until the authors reach a consensus and identify 146 relevant papers in total.
The search process is conducted on Dec 2, 2023.


\begin{figure}
  \centering
  \includegraphics[width=1\textwidth]{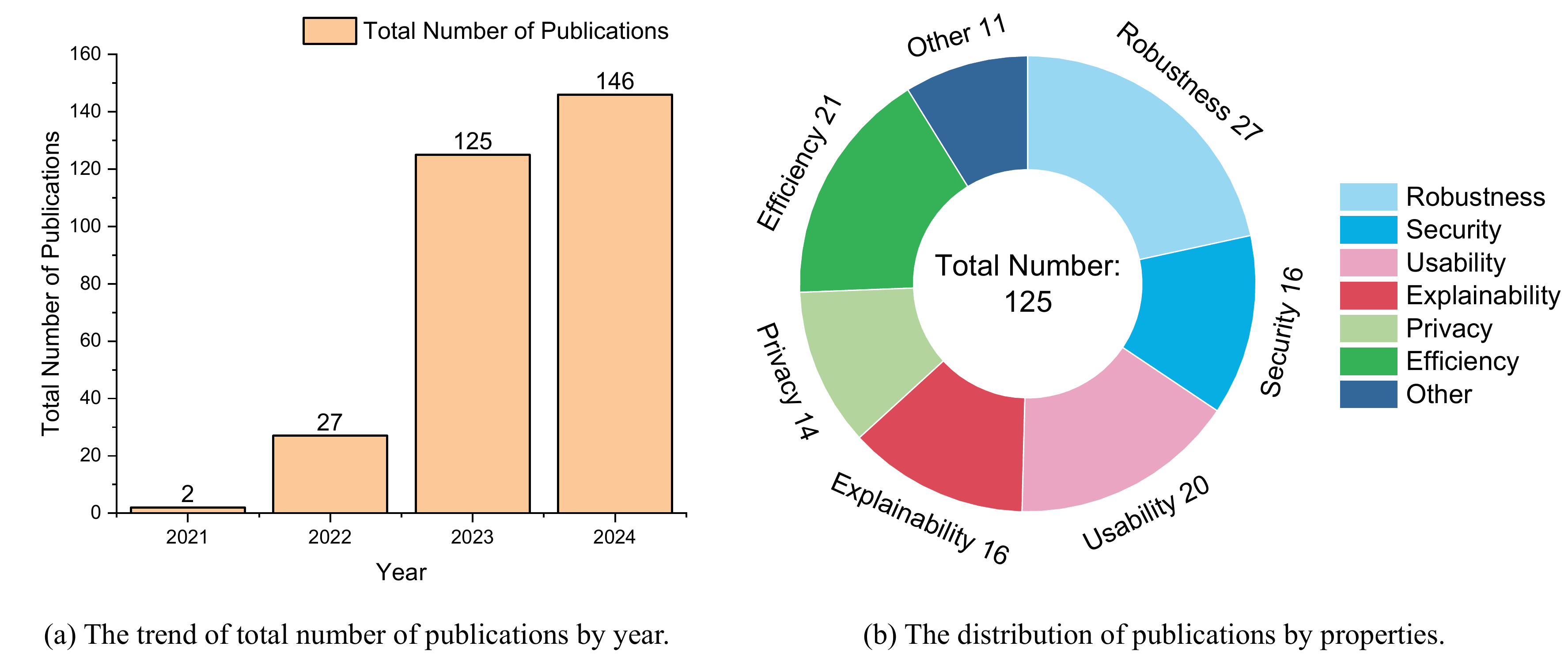}
  \caption{Figure (a) shows the cumulative number of papers that are relevant to the topic of this survey in the recent 6 years. The data collection date is 20 Feb 2024. Figure (b) shows the distribution of papers across different properties, where the `\texttt{Other}' category includes papers that are surveys and discussions.}
  \label{fig:trend1}
\end{figure}

\subsubsection{Snowballing Search} \label{subsubsec:snowballing}
We further perform both backward and forward snowballing~\cite{snowball1,snowball2} on the relevant papers identified in the previous steps.
In backward snowballing, we analyze the references in each collected paper to identify those within our scope.
In forward snowballing, we identify papers of our interest from those that cite the collected papers.
This process is implemented using the API provided by Semantic Scholar.\footnote{\url{https://api.semanticscholar.org/}}
The snowballing process is repeated for multiple rounds; 
in each round the authors follow the labeling process mentioned in Section~\ref{subsubsec:search} to identify relevant papers.
We repeat the snowballing process until we reach a transitive closure fixed point, i.e., no new relevant paper was identified. 
We conduct 8 rounds of snowballing and collect additional 23 papers.
After this step, we obtain a total of 146 papers falling into our scope.

\subsection{Collection Results}

The search process identifies 146 papers relevant to non-functional properties of LLM4Code, which are published from 2019 to 2024.
The trend of the accumulated number of papers in recent years is shown in Figure~\ref{fig:trend1} (a).
Figure~\ref{fig:trend1} (b). shows the distribution of papers across different properties. 
The property that attracts the most attention is robustness (with 27 papers), followed by efficiency (21), usability (20), privacy (16), and security (16).
The category `\texttt{Other}' includes 11 papers that are literature reviews and discussions.

\begin{figure}
  \centering
  \includegraphics[width=1\textwidth]{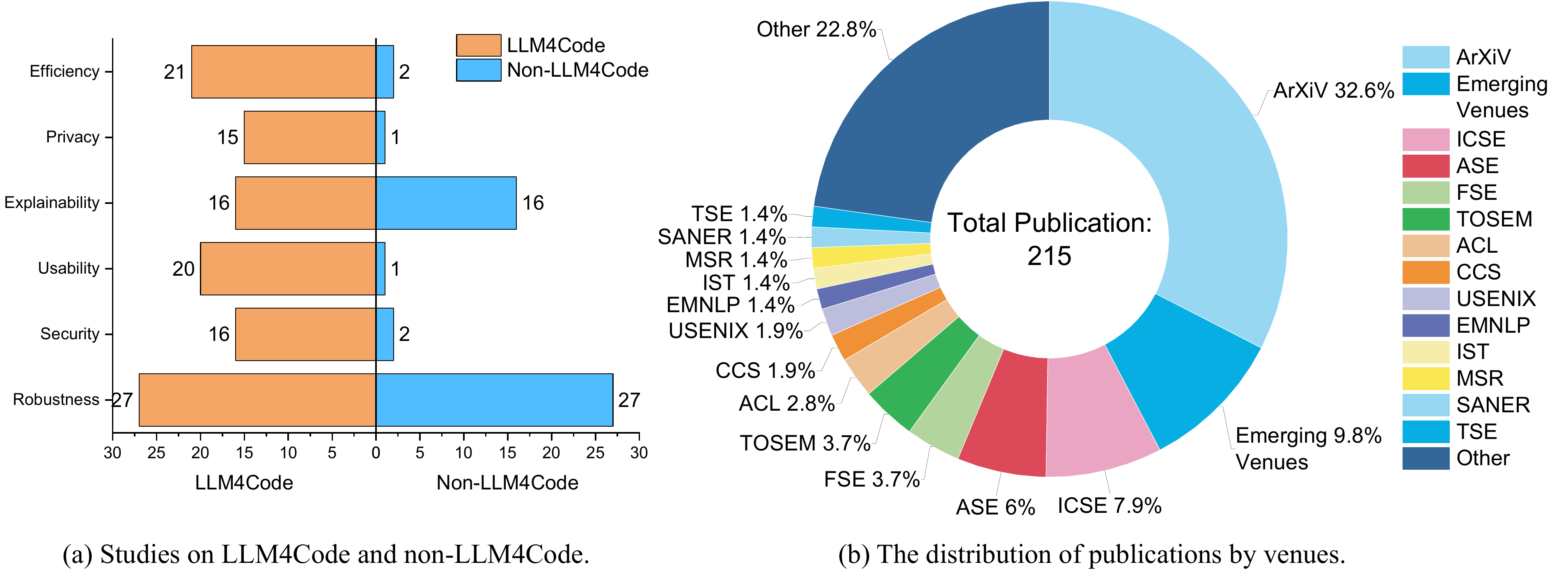}
  \caption{Figure (a) compares paper distribution of papers evaluating LLM4Code and Non-LLM4Code by properties. Figure (b) shows the paper distribution across different publication venues, where the `\texttt{Other}' category includes venues with less than 2 relevant papers.}
  \label{fig:trend2}
\end{figure}

We manually check the publication venues of the collected papers and Figure ~\ref{fig:trend2} (b) shows the distribution of papers by venues.
We find that 32.6\% papers are preprints on arXiv. 
There are around 10\% papers in the emerging venues, which we define as the venues that are not included in CORE rankings\footnote{\url{https://www.core.edu.au/conference-portal}} or whose ranking is below C, e.g., workshops, symposiums, and local conferences.
The `\texttt{Other}' category includes venues with less than 2 relevant papers.
Excluding the papers belonging to the `\texttt{Emerging Venues}' and `\texttt{Other}' category, we group venues into research domains and analyze paper distribution.

During the literature review, we also identify 51 papers that study these properties of non-LLM4Code, e.g., code2vec~\cite{code2vec} and code2seq~\cite{code2seq}.
Figure~\ref{fig:trend2} (a) compare the distribution of papers across different research domains.
We find that the number of papers on robustness and explainability on two types of models are similar.
However, the comparison suggests that LLM4Code open new research demands on security, privacy, usability, and efficiency.
For example, only 2 papers are found on security of non-LLM4Code, while 16 papers are found on LLM4Code.
Although our survey primarily focuses on LLM4Code, we discuss the relevant studies on non-LLM4Code when necessary.

\subsection{Relevant Literature Reviews}

As the body of literature on language models of code grows, there are several literature reviews that focus on different aspects of LLM4Code.
Hou et al.~\cite{codellm_survey} present a systematic literature review of large language models for software engineering, identifying 229 research papers from 2017 to 2023.
Nguyen et al.~\cite{nguyenduc2023generative} propose a research agenda of using generative models for software engineering.
Fan et al.~\cite{fan2023large} conduct a survey of large language models for software engineering and point out some open problems.
She et al.~\cite{she2023pitfalls} survey papers to build a taxonomy of pitfalls in using large language models for software engineering.
Lo~\cite{lo2023trustworthy} discusses the vision and roadmaps of the trustworthy and synergistic AI for software engineering.
Zheng et al.~\cite{zheng2023survey} conduct a survey of large language models for code, explaining the evolution, benchmarking, and future trends.
Zhuo et al.~\cite{zhuo2023source} survey the augmentation methods for training deep learning models for source code.
Hussain et al.~\cite{hussain2023survey} conduct a survey of trojans attacks in neural models of code.
These literature reviews mainly focus on the models and applications of LLM4Code and may mention some properties beyond accuracy in the discussion section.
Our study specifically focuses on the properties beyond accuracy of LLM4Code and conducts a systematic literature review to understand how these properties are studied.
Additionally, we also investigate challenges and lay steps for future work.

\section{Non-functional Properties Beyond Accuracy} \label{sec:property}
The accuracy refers to the ability of a model to produce correct outputs for the tasks it is designed for.
Different metrics are used to evaluate the accuracy of LLM4Code, including precision, recall, F1 score, in classification tasks, $pass@k$ for code generation tasks, BLEU score for code translation, etc.
This section explains the concept of each non-functional property in the context of LLM4Code. 

\subsection{Robustness} \label{subsec:robustness}


Robustness refers to the ability of LLM4Code to perform consistently and correctly when the inputs are perturbed or changed slightly.
LLM4Code that are not robust may cause negative impacts and even disastrous  consequences.
To illustrate the impact of robustness of LLM4Code, consider two example scenarios.
The first scenario involves code generators catering to users with varying coding styles and English proficiency levels. 
Users may write similar-meaning prompts in different ways. 
A non-robust code generator might yield inconsistent performance across these prompts, resulting in varied output quality for different users, which is an undesirable outcome. 
In the other scenario, we assume that LLM4Code is deployed in the CI/CD pipeline to detect vulnerable code. 
The model can accurately determine if a commit has vulnerabilities, preventing the merge of potentially harmful code. 
However, a lack of robustness in the model could lead to altered predictions when faced with input perturbations, potentially allowing malicious users to introduce harmful code into the repository undetected.

We call the entity that evaluates the robustness of LLM4Code as the \textit{tester}.
Robustness evaluation can be categorized into two types: white-box and black-box evaluation.
The white-box evaluation assumes that the tester has full knowledge of the target model, including its architecture, parameters (weights and biases), training data, and training algorithm. 
The tester can access the model's internal states and gradients to craft test inputs~\cite{Yefet2020,9916170,10.1145/3533767.3534390,Epresentation2021,9825895,10123554,10123534,zhang2022towards}. 
In black-box evaluation, testers might only have the capability to query the model for outputs given specific inputs. 
This is a more realistic scenario, especially for services where the model is hosted remotely and users can send inputs to get outputs, like many cloud-based AI services like GitHub Copilot. 
The attackers can query the model multiple times to craft adversarial examples~\cite{alert,MHM,zhangcodebert}.

The tester's goals dictate what the tester wants to achieve by manipulating the inputs.
We categorize the tester's goals into two types: targeted and non-targeted attacks.
In a targeted attack, the tester aims to make the model produce a specific, predefined prediction. 
This involves not just causing a misclassification, but directing the misclassification toward a chosen category. 
Considering an attacker wanting to make a model that predicts whether a code snippet contains a vulnerability to instead predict that it does not contain a vulnerability.
In this way, the attacker can commit malicious code into the repository without being detected.
The main goal in a non-targeted attack is to make the model produce any incorrect prediction. 
The specific incorrect class does not matter, as long as it is not the correct one. For example, Yang et al.~\cite{alert} aim to change the predictions of an authorship attribution model by variable renaming. The attack is considered successful as long as the model makes any incorrect prediction.

\subsection{Security}
As stated in~\cite{ml-testing-survey}, there exists a close-knit relationship between security and robustness.
In our survey, given the extensive array of studies centered on robustness, we consciously choose to segregate the robustness property from the broader domain of security.
Note that here we discuss the security threats that LLM4Code may undergo, i.e., LLM4Code as the victim.
From the literature, we identify primary security concerns that the behaviors of LLM4Code can be manipulated maliciously.

One notable attack is data poisoning, which assumes that attackers can operate attacks by manipulating the training set of LLM4Code. 
This is a practical assumption, especially in software engineering where open-sourcing and reuse are the core culture. 
For example, researchers collect and release code datasets publicly. 
An attacker can access a dataset and modify or inject some malicious examples into the dataset in the next release. 
The poisoned dataset may be downloaded and used by other users to train LLM4Code. 
Attackers can also directly manipulate the source of these datasets, i.e., the open-source platforms. 
Studies~\cite{10.1145/3427228.3427258} have identified the presence of fake accounts on platforms like GitHub, which can be used to influence or corrupt open-source repositories. For instance, attackers might create repositories containing vulnerable code and then artificially boost their popularity to ensure they are included in datasets collected for training LLM4Code.
Attackers can aim to downgrade model performance to make the models make inaccurate predictions. 
Attackers can also aim to make the model behave in patterns as specified by the attacker, which is also called \textit{backdoor attack}. 
A model with backdoors behaves normally when it receives normal inputs, but it will produce chosen outputs when it receives inputs with specific patterns (i.e., ``trigger'').

\subsection{Privacy} \label{subsec:privacy}


Broadly, we refer to privacy in LLM4Code as the information about LLM4Code that is not intended to be disclosed, such as the model parameters and the training data of a commercial service.
One threat to privacy is the membership inference attack, aiming to determine if a specific data point was used in training a model, identifying it as a `\textit{member}' of the training data. 
The membership information of LLM4Code can leveraged in multiple ways. 
For example, a language model can produce many outputs that are also included in its training data, which is known as the \textit{memorization phenomenon}.
Given a large amount of outputs, an accurate membership inference method can identify which part of the outputs are from the training data, enabling training data extraction even if the attack can only access to model outputs.

The privacy concept additionally encompasses the personal privacy that is included in the training data or other rights of individuals relevant to privacy.
The majority of the training data for LLM4Code is extracted from open-source repositories that occasionally harbor software secrets. 
Research efforts~\cite{basak2023secretbench,10174120,9794113} have disclosed that these repositories can potentially house a plethora of sensitive data elements, such as API keys, passwords, and personally identifiable information including email addresses, among others. 
This sensitive data can inadvertently be learned and replicated by the model, thereby raising unintentional disclosure of such data.
Another scenario is `\textit{unauthorized training},' i.e., whether a model is trained on a dataset without the permission of the dataset owner.
For example, recent news reports that GitHub has been sued over Copilot for using code from open-source repositories without permission.\footnote{\url{https://www.thestack.technology/microsoft-github-sued-over-copilot/}}
Membership inference attack can be used to provide supporting evidence showing whether a model uses a specific dataset in its training.
Additionally, models learn knowledge from the training data and change their behavior accordingly. 
Just as we show when discussing data poisoning attack, a model will be implanted with backdoor if it is trained on a poisoned dataset.
As a result, knowing whether a model is trained on a specific dataset can help us understand the model behavior and its potential security threats, thus affecting users' decisions on whether to use the model.
If a model is trained on a dataset that contains sensitive information or vulnerable code, users may not want to use the model.

\subsection{Explainability} \label{subsec:interpretability}

Explainability in LLM4Code is the ability to understand and articulate how these models make predictions or decisions when processing and generating code, covering multiple aspects that are crucial to the development and deployment of LLM4Code.
For example, it can refer to the transparency of the model decision-making process, e.g., understanding which part of the code makes a model think that a code snippet is vulnerable.
Explainability may also involve understanding what information is learned by the model. 
Ma et al.~\cite{ma2023code} investigate the capabilities of CodeBERT and GraphCodeBERT in understanding code syntax and semantics.
Additionally, explainability includes post hoc explanations~\cite{10.1145/3236386.3241340}, offering supplemental insights to further clarify the model's decisions. 
Consider two illustrative instances: in one, the model solely yields a binary label denoting the presence of vulnerabilities in a code snippet, while in the other, it provides a label accompanied by relevant code snippets showcasing similar vulnerabilities. 
This latter approach facilitates a deeper understanding of the model's decision through additional explanation.
Researchers have studied the explainability of various LLM4Code models and different tasks. 
Early exploration of explainability of code models targets at the self-admitted technical debt detection task~\cite{JIT2708,10.1145/3324916,10.1145/3324884.3416583}.
A number of studies focus on clone detection~\cite{10.1145/3524610.3527921}, defect prediction~\cite{Reemicse21,shin2021explainable,9044387,10.1145/3429444,10.1109/ASE51524.2021.9678763}, code generation~\cite{paltenghi2022extracting,Liuyue2023}, code search~\cite{10232920}, and code summarization~\cite{wang2023demystifying,9678712}.
Researchers also evaluate the explainability of different LLM4Code, ranging from encoder-only pretrained models like CodeBERT~\cite{zhang2022does}, CodeT5~\cite{wang2021codet5}, to larger decoder-only models with billions of parameters like CodeGen~\cite{codegen}.

\subsection{Efficiency} \label{subsec:efficiency}
Efficiency refers to the resource consumption and running speed during various stages of LLM4Code development and usage, which encompasses several elements including training and inference time, computational resources such as CPU and GPU memory, data volume requirements, and energy consumption. 
The resource consumption is usually affected by the model size (e.g., number of parameters), model architecture, model complexity, etc.

Efficiency constitutes a vital attribute in LLM4Code. 
As recently highlighted by Hellendoorn and Sawant~\cite{10.1145/3501261}, the cost of deep learning for source code keeps increasing. 
Aye et al.~\cite{aye2020sequence} advocate that LLM4Code should be sufficiently compact to occupy a minimal amount of disk space and memory on developer workstations, given that almost all contemporary Integrated Development Environments (IDEs) operate locally and facilitate offline usage.
In terms of the model size, Svyatkovskiy et al.~\cite{svyatkovskiy2021fast} propose that an acceptable maximum size for an Integrated Development Environment (IDE) component or an editor plug-in is 50 MB, while a 3 MB model is favored since it facilitates deployment even in substantially constrained environments.
Nonetheless, prevailing LLM4Code models frequently exceed the size constraints mentioned earlier. 
For instance, CodeBERT~\cite{CodeBERT}, encompassing 125 million parameters, occupies 481 MB of space and demonstrates an average inference latency of 1.5 seconds when executed on a desktop CPU. 
More recent models, which are based on even larger language models, have billions and even trillions of parameters, e.g., LlamaCode and CodeGen2. 
Recently, researchers have adopted different strategies to reduce model sizes and enhance efficiency, while preserving model performance as much as possible.
For example, Shi et al.~\cite{compressor} use knowledge distillation to reduce the size of CodeBERT and GraphCodeAttack.


\subsection{Usability} \label{subsec:usability}
Adopting LLM4Code in practice is not merely a process where LLM4Code produces an output and the users take the output.
The actual utilization of LLM4Code can be very iterative and interactive. 
Imagine a code completion tool like Copilot integrated into an IDE.
A user inputs a code snippet and the tool completes the following code.
If the tool only completes several tokens in a line, the user may not find the tool useful as the user still needs to write the rest of the code or invoke the tool each time.
If the tool completes many lines of code, some of which are not relevant to the user's intention, the user may find the tool annoying as they need to delete the irrelevant code.
The usability of LLM4Code, described as the ease of use and learnability of the model, and to what extent LLM4Code can facilitate the users' productivity, is a crucial property that warrants serious consideration.

\section{How to Evaluate and Enhance Properties Beyond Accuracy?}
\label{sec:detail}
\subsection{Robustness} \label{subsec:robustness_survey}

\subsubsection{Robustness Evaluation}
Yang et al.~\cite{alert} and Pour et al.~\cite{9438605} present the earliest try to evaluate the robustness of LLM4Code.
Yang et al.~\cite{alert} are the first to highlight the naturalness requirements in crafting adversarial examples for LLM4Code, inspiring a series of studies~\cite{codeattack,coda,lecong2024evaluating} focusing the naturalness requirements.
To generate adversarial examples (also called test inputs in some literatures) to evaluate robustness of LLM4Code, two important aspects need to be considered: (1) semantic-preserving transformations and (2) methods to apply transformations to source code.
We include studies that are conducted on non-LLM4Code as the transformations and methods may be applicable to LLM4Code as well.

\vspace*{0.2cm}
\noindent \textbf{Semantic Preserving Transformation.}
A basic assumption is that the model outcomes should remain the same when the input is slightly modified in a way that does not change the semantics of the input.
For example, given a vulnerable code snippet that can be correctly predicted by an LLM4Code, the model prediction should not change after we modify a local variable name in the code snippet, which does not change the actual behavior of the code.

Researchers propose different semantic-preserving transformations on source code and use various taxonomies to categorize them. 
Li et al.~\cite{li2022closer} split the transformations into five categories: block transformations, insertion/deletion transformations, grammatical statement transformations, grammatical token transformations, and identifier transformations.
Liu et al.~\cite{9454564} use a different 5-type taxonomy: trivial transformations, data transformations, control flow transformations, function transformations, and addition of bogus code.
Yu et al.~\cite{electronics12040936} categorize transformations on C language into operator transformations, data transformations, and bogus code transformations.
Tian et al.~\cite{coda} consider transformations that modify loop, branch, calculation, and constant.
Matyukhina et al.~\cite{Matyukhina2019AdversarialAA} analyze 5-level code style features and design transformations for each level: layout, lexical, syntactic, control flow, and data flow features.
Quiring et al.~\cite{10.5555/3361338.3361372} define 5 types of transformations over abstract syntax trees (AST): control transformations, declaration transformations, API transformations, template transformations, and miscellaneous transformations.

With the increasing popularity of LLM4Code that take natural language as inputs, researchers study robustness against natural language perturbations~\cite{zhuo2023-robustness,10.1109/ICSE48619.2023.00181,improta2023enhancing,10.1145/3528588.3528653,anand2021adversarial,yan2023coco,shirafuji2023exploring}.
The commonly used perturbation methods include leveraging existing adversarial attacks on natural languages~\cite{Li_2019,wang2021adversarial,fitria2021quillbot,shiri2023paraphrasing,naik-etal-2018-stress}, automated paraphrasing~\cite{zhang2020pegasus}, word substitution/omission~\cite{improta2023enhancing}, converting active voice to passive voice~\cite{anand2021adversarial}, instructions perturbation~\cite{yan2023coco}, etc.

\begin{table}[!t]
  \centering
  \caption{Summary of test input generation methods for evaluating the robustness of LLM4Code.}
  \begin{tabular}{ll}
  \toprule
  \textbf{Methods} & \textbf{Publications} \\
  \midrule
  Gradient-based & \cite{Yefet2020,9763729,Epresentation2021,10.1145/3533767.3534390,9825895,zhang2022towards,10123534,7958570} \\
  Heuristic-driven & \cite{na-etal-2023-dip,alert,coda,springer2021strata,9916170} \\
  Search-based & \cite{10.1145/3579990.3580012,9438605,electronics12040936,10.5555/3361338.3361372,MHM,alert,choi-etal-2022-tabs,zhang2023rnns,10.1145/3501256,10028657,codeattack,du2023extensive,nguyen2023adversarial,a16100478,10.1145/3449726.3463222} \\
  Reinforcement Learning & \cite{10028657,9724884,10.1145/3579990.3580012} \\
  Style Transformations & \cite{Matyukhina2019AdversarialAA,abuhamad2023shield,simko2018recognizing,yang2023assessing} \\
  Transferability-based & \cite{9454564,zhang2023transfer} \\
  Unspecified &  \cite{9678706,10.1145/3528588.3528653,cheers2021evaluating,zhuo2023-robustness,wang2023-recode,rabin2021evaluation,10197074,RABIN2021106552,rabin2019testing}  \\ 
  \bottomrule
  \end{tabular}
  \label{tab:robustness_methods}
\end{table}

\vspace*{0.15cm}
\noindent \textbf{Test Input Generation.}
Generally, there are two goals of generating test inputs.
One goal is to investigate how LLM4Code react to different transformations.
In this case, there are no specific methods adopted to apply transformations in a strategical way, which we call \textit{unspecified} methods.
The other goal is to purposefully generate the inputs that can maximize the impact of the transformations on the model prediction.
We categorize these methods into: (1) heuristic-driven, (2) gradient-based, (3) search-based, (4) reinforcement learning, (5) style transformations, and (6) transferability-based methods.
We summarize different test input generation methods in Table~\ref{tab:robustness_methods}.

Gradient-based methods assume that the attacker can access the model in a white-box manner and use gradient information to compute the optimal transformations~\cite{Yefet2020,9763729,Epresentation2021,10.1145/3533767.3534390,9825895,zhang2022towards,10123534,7958570}.
As suggested by Henkel et al.~\cite{9825895}, gradient-based approaches are usually stronger than other black-box methods.
Researchers have proposed some heuristics that are relevant to generate adversarial examples and use them to guide the transformations.
Popular heuristics include the similarity/distance between the original and modified code~\cite{na-etal-2023-dip,alert,10.1145/3579990.3580012}, differences between a target input and reference inputs~\cite{coda}, token frequency~\cite{springer2021strata}, neuron-coverage~\cite{9916170}, etc.
The heuristics are usually used in combine with other methods like search-based methods and reinforcement learning. 
Various search algorithms are used to find the optimal transformations that can maximize the impact on the model prediction, including random search~\cite{10.1145/3579990.3580012}, genetic algorithm~\cite{9438605,alert,10.1145/3449726.3463222}, greedy algorithm~\cite{alert,10.1145/3501256,codeattack,nguyen2023adversarial}, Monte Carlo algorithm~\cite{10.1145/3579990.3580012,electronics12040936,10028657}, Metropolis-Hastings sampling~\cite{metropolis1953equation,MHM}, beam search~\cite{choi-etal-2022-tabs,du2023extensive}, Representation Nearest Neighbor search~\cite{zhang2023rnns}. 
Reinforcement learning can also be used to learn a policy that conducts sequences of transformations on source code~\cite{10028657,9724884,10.1145/3579990.3580012}.

Some methods specifically apply transformations to mimic the coding style of certain programmers~\cite{Matyukhina2019AdversarialAA,simko2018recognizing,abuhamad2023shield,yang2023assessing}, which are usually used to evaluate authorship attribution models.
For example, Matyukhina et al.~\cite{Matyukhina2019AdversarialAA} design a method to allow authors to preserve the readability of their source code, while removing identifying stylistic features (including layout, lexical, syntactic, control-flow, and data-flow features) that can be leveraged for code attribution.
Experiments show that popular authorship attribution systems~\cite{10.1007/0-387-34224-9_59,10.5555/2831143.2831160,DING200449,4151691} are not robust to such transformations.

Transferability-based methods leverage the transferability of adversarial examples~\cite{papernot2016transferability}: adversarial examples generated for one model may also be effective for another model.
Liu et al.~\cite{9454564} first train a substitute model to imitate the behavior of the victim model. 
They generate adversarial examples for the substitute model and use them to attack the original victim model.
Zhang et al.~\cite{zhang2023transfer} find that adversarial examples generated using gradient-based methods for a smaller code model can successfully transfer to multiple larger LLM4Code.

Studies also investigate the impacts of different semantic-preserving transformations without specific strategies to optimize the changes in LLM4Code's outputs, which we call \textit{unspecified} methods~\cite{9678706,10.1145/3528588.3528653,cheers2021evaluating,zhuo2023-robustness,wang2023-recode,rabin2021evaluation,10197074,RABIN2021106552,rabin2019testing}.

\begin{table}[!t]
  \centering
  \caption{The tasks on which the robustness of LLM4Code is evaluated. We also list the datasets and the relevant works.}
  \small 
  \begin{tabular}{p{2.3cm}p{5cm}p{6.5cm}}
    \toprule
  Tasks & Datasets & Works \\
  \midrule
  Authorship Attribution & Google Code Jam~\cite{alsulami_source_2017}, Github Dataset~\cite{Matyukhina2019AdversarialAA}, Yang et al.~\cite{yang2017authorship} & Liu et al.~\cite{9454564}, Tian et al.~\cite{coda}, Matyukhina et al.~\cite{Matyukhina2019AdversarialAA}, Quiring et al.~\cite{10.5555/3361338.3361372}, Yang et al.~\cite{alert}, Zhang et al.~\cite{zhang2023rnns}, Simko et al.~\cite{simko2018recognizing}, Li et al.~\cite{RoPGen}, Mohammed et al.~\cite{abuhamad2023shield}, Na et al.~\cite{na-etal-2023-dip}, Nguyen et al.~\cite{nguyen2023adversarial} \\
  Method Name Prediction & Java-small~\cite{code2seq}, Java-med and Java-large~\cite{code2seq}, CodeSearchNert~\cite{husain2019codesearchnet}, Py150~\cite{py150} & Pour et al.~\cite{9438605}, Rabin et al.~\cite{rabin2021generalizability}, Chen et al.~\cite{9763729}, Springer et al.~\cite{springer2021strata} \\
  Code Search & CodeSearchNet~\cite{husain2019codesearchnet} & Li et al.~\cite{li2022closer}, Pour et al.~\cite{9438605}, Choi et al.~\cite{choi-etal-2022-tabs}, Li et al.~\cite{li-etal-2022-semantic} \\
  Code Summarization & CodeSearchNet~\cite{husain2019codesearchnet},  Xing et al.~\cite{10.5555/3304889.3304975}, NeuralCodeSum~\cite{ahmad-etal-2020-transformer}, Java-small~\cite{code2seq} & Li et al.~\cite{li2022closer}, Pour et al.~\cite{9438605}, Zhou et al.~\cite{10.1145/3501256}, Zeng et al.~\cite{10.1145/3533767.3534390}, Applis et al.~\cite{9678706}, Wei et al.~\cite{9916170}, Jha and Reddy~\cite{codeattack}, Du et al.~\cite{du2023extensive} \\
  Defect Prediction & Devign~\cite{Devign}, SySeVR~\cite{SySeVR}, ReVeal~\cite{ReVeal}, CodeChef~\cite{zhang2022towards}, OWASP Benchmark~\cite{owasp_benchmark} & Yu et al.~\cite{electronics12040936}, Tian et al.~\cite{coda}, Zeng et al.~\cite{10.1145/3533767.3534390}, Yang et al.~\cite{alert}, Zhang et al.~\cite{zhang2023rnns}, Li et al.~\cite{li-etal-2022-semantic}, Zhang et al.~\cite{zhang2022towards}, Na et al.~\cite{na-etal-2023-dip}, Du et al.~\cite{du2023extensive}, Nguyen et al.~\cite{nguyen2023adversarial} \\
  Clone Detection & BigCloneBench~\cite{BigCloneBench}, OJClone~\cite{OJClone},  & Tian et al.~\cite{coda}, Zhang et al.~\cite{10028657}, Zhang et al.~\cite{zhang2022towards}, Zhang et al.~\cite{zhangcodebert}, Yang et al.~\cite{alert}, Na et al.~\cite{na-etal-2023-dip}, Du et al.~\cite{du2023extensive}, Cheers et al.~\cite{cheers2021evaluating}, Dam\'{a}sio et al.~\cite{10.1145/3579990.3580012} \\
  Code Generation & EVIL~\cite{9700297}, APPS~\cite{apps}, ALGOLISP~\cite{polosukhin2018neural}, Java Method~\cite{10.1109/ICSE48619.2023.00181}, HumanEval~\cite{codex}, MBPP~\cite{austin2021program} & Liguori et al.~\cite{10.1145/3528588.3528653}, Zhu et al.~\cite{10123534}, Anand et al.~\cite{anand2021adversarial}, Mastropaolo et al.~\cite{10.1109/ICSE48619.2023.00181}, Wang et al.~\cite{wang2023-recode} \\
  Code Translation & CodeXGLUE~\cite{CodeXGLUE}  & Jha and Reddy~\cite{codeattack} \\
  Program Repair & Tufano et al.~\cite{10.1145/3340544} & Jha and Reddy~\cite{codeattack} \\
  Text Parsing & GeoQuery~\cite{finegan-dollak-etal-2018-improving}, Scholar~\cite{finegan-dollak-etal-2018-improving} & Zhuo et al.~\cite{zhuo2023-robustness} \\
  \bottomrule
  \end{tabular}
  \label{tab:tasks}
\end{table}

\subsubsection{Robustness Enhancement}
Researchers have proposed a range of methods to enhance the robustness of LLM4Code. 
We split the methods into two types: (1) methods without retraining and (2) methods involving retraining.

Yefet et al.~\cite{Yefet2020} evaluate defense strategies without re-training models.
One strategy is to replace all variables in the inputs to $\langle unk \rangle$ (a special token representing that the token at this position is unknown.) token only at test time.
This strategy is 100\% robust to variable renaming by construction by sacrificing the model's accuracy.
The other strategy is to detect and remove the outlier variable names, which empirically shows a good trade-off between robustness and accuracy.
Yang et al.~\cite{10.1145/3630010} propose to synthesize a new method name from the functional description and feed the new code into the LLM4Code.
They show that this method can restore LLM4Code's performance against variable renaming.


A large body of work involves retraining models to improve robustness.
One line of work~\cite{Yefet2020,10.1016/j.future.2022.12.030,Epresentation2021,bielik2020adversarial,10.1145/3591227,zhang2023rnns} adopts adversarial training~\cite{FGSM}, by letting the models minimize the loss on both the original example and the adversarial example simultaneously.
Researchers also improve the standard adversarial training, e.g., contrastive adversarial training~\cite{zhang2023rnns}.
Jordan et al.~\cite{9825895} conduct adversarial training with a weak adversary who just randomly picks a single transformation and evaluates the obtained model against a strong adversary.
Wang et al.~\cite{wang2022robust} first converts each data point to a canonical form and subsequently restricts the training and testing of models on the normalized data.
The method is combined with adversarial training to achieve an optimal trade-off between robustness and accuracy. 

Some studies leverage adversarial examples to train the model.
A series of studies belong to this category, including \cite{9763729,9438605,coda,9916170,MHM,9724884,alert,10028657,improta2023enhancing}.
The generated adversarial examples are used in different ways.
Chen et al.~\cite{9763729} only use adversarial examples to retrain the models.
Zhang et al.~\cite{zhang2022towards} augment the training set with adversarial examples periodically when training LLM4Code. 
They find that training against weaker (and less computationally expensive) attacks is sufficient to provide a defense against stronger attacks.
Zhou et al.~\cite{10.1145/3501256} propose masked training to improve robustness. Specifically, for each training example, this method constructs a masked example by randomly replacing $k$ identifiers with $\langle unk \rangle$ that represents an unknown token.

Including adversarial examples in the training data also demonstrates the potential to improve the accuracy of LLM4Code.
Jia et al.~\cite{10123554} combine contrastive learning with adversarial training to enforce the robustness of learned code representations.
Then, they propose staggered adversarial training (SAT) to preserve the robustness learned during pretraining while also learning task-specific generalization and robustness during fine-tuning.
The results show that this method can improve both the accuracy and robustness of code models on a series of downstream tasks.
Li et al.~\cite{li-etal-2022-semantic} propose Semantic-Preserving Adversarial Code Embeddings (\textsc{Space}) to enhance code model robustness against variable renaming. 
Unlike works by~\cite{Yefet2020,9825895}, \textsc{Space} defines perturbations on the continuous embedding space so that the adversarial training can be done via gradient-based methods. 
The experiments show that \textsc{Space} not only improves the robustness of code models but also improves the accuracy.

\begin{tcolorbox}[title=\textbf{Findings - Robustness}, left=2pt, right=2pt,top=2pt,bottom=2pt]
  We find that LLM4Code robustness studies have been conducted on a wide range of tasks, as listed in Table~\ref{tab:tasks}.
We find that the types of evaluated models are imbalanced in the literature.
Only a few studies~\cite{goodname,10123534,yan2023coco,10.1109/ICSE48619.2023.00181} evaluate recent large code generation models, e.g., Code Llama~\cite{codellama}, etc.
Most defensive methods involve retraining the model, which is expensive to operate.
This can be observed from the literature that the defensive methods are usually evaluated on small models and datasets.
Given the popularity of recent large models with billions of parameters, we call for more attention to scalable methods that can protect them.
\end{tcolorbox}

\subsection{Security} \label{subsec:security_survey}

\subsubsection{Security Evaluation}


In the settings where the attackers can manipulate the training process, the threats are evaluated by injecting poisoned examples into the training data, which are also called data poisoning attacks. 
Data poisoning attacks can inject vulnerabilities into LLM4Code, e.g., degrade model performance~\cite{CoProtector} and let the model produce undesired outputs (like vulnerable code)~\cite{coffee,oh2023poisoned,263874}. 
Many studies have shown that data poisoning can implant backdoors into LLM4Code to manipulate the model behaviors~\cite{Sun2023backdoor,hussain2023survey,Sun2023backdoor,codebackdoor,you-see}.
Specifically, LLM4Code works normally on clean inputs, e.g., produce correct code completion.
However, when the model receives a query with a specific \textit{trigger}, LLM4Code will produce undesired outputs that are specified by the attackers, e.g., vulnerable code~\cite{aghakhani2023trojanpuzzle} or incorrect code summarization~\cite{advdoor}.
Note that there are different terms for backdoor attacks in the literature, e.g., \textit{trojan}~\cite{hussain2023survey,aghakhani2023trojanpuzzle,hussain2023trojanedcm}.
Researchers evaluate the risks brought by data poisoning and backdoor attacks on LLM4Code for different tasks, including API recommendation~\cite{coffee}, code search~\cite{Sun2023backdoor,qi2023badcs,you-see}, code summarization~\cite{advdoor,codebackdoor}, code representation~\cite{li-etal-2023-multi-target}, and code generation/completion~\cite{oh2023poisoned,aghakhani2023trojanpuzzle,cotroneo2023vulnerabilities,263874}.
The studies examine different potential undesired outcomes of the poisoned LLM4Code, e.g., complete or generate vulnerable code~\cite{oh2023poisoned,aghakhani2023trojanpuzzle}, recommend fake APIs~\cite{coffee}, return insecure code snippets in code search~\cite{Sun2023backdoor,qi2023badcs,you-see}, etc.

Different types of triggers are used to activate backdoors.
Ramakrishnan and Albarghouthi~\cite{codebackdoor} propose fixed triggers and grammar triggers, which are also used by Wan et al.~\cite{you-see} to poison code search models.
Yang et al.~\cite{advdoor} find that both fixed and grammar triggers~\cite{codebackdoor} can be easily identified by manual inspection and automated detection methods.
They propose `adaptive triggers' by renaming variable names using adversarial attack~\cite{Yefet2020}.
Sun et al.~\cite{Sun2023backdoor} propose to change function names and/or variable names as triggers, e.g., changing ``\texttt{function()}'' to ``\texttt{function\_aux()}''.
Aghakhani et al.~\cite{aghakhani2023trojanpuzzle} inject the malicious code into docstrings and comments, so that it can bypass static analysis to detect poisoned code.
They also introduce a method that conceals suspicious parts of the payload such that they are never included in the poisoning data, while still tricking the model into suggesting the entire payload in a dangerous context.
Li et al.~\cite{CodePoisoner} evaluate three rule-based poisoning strategies (including identifier renaming, constant unfolding, and dead-code insertion) and find that they are easy to detect.
Li et al.~\cite{CodePoisoner} then propose a language-model-guided poisoning strategy to enhance stealthiness; the triggers are statements generated by CodeGPT~\cite{CodeXGLUE}.

In scenarios where the attackers can only access the trained models, researchers also find methods to manipulate the LLM4Code behaviors.
Hajipour et al.~\cite{hajipour2023systematically} propose a black-box inversion approach based on few-shot prompting, which finds prompts that guide LLM4Code to generate different kinds of vulnerable codes.
He et al.~\cite{he2023large} leverage prefix-tuning~\cite{li-liang-2021-prefix} to obtain a vector that can induce the model to generate vulnerable code.
Wu et al.~\cite{wu2023deceptprompt} propose a new algorithm that can generate adversarial natural language instructions that drive the Code LLMs to generate functionality correct code with vulnerabilities.


\subsubsection{Security Enhancement}
Some strategies are proposed and used to mitigate the threats of data poisoning attacks.
We categorize them into two types: (1) automated detection and (2) human review.

\vspace*{0.2cm}
\noindent \textbf{Automated Detection.} Some automated detection methods that are originally proposed for backdoor attacks in other domains are used for source code models, including spectral signature~\cite{spectral}, activation clustering~\cite{activation}, and ONION~\cite{qi-etal-2021-onion}.
However, researchers find that these methods are not effective. 
Wan et al.~\cite{you-see} employ spectral signature~\cite{spectral} to detect poisoned examples: the recall is 6.10\% and the precision is 8.63\% (for CodeBERT).
Sun et al.~\cite{Sun2023backdoor} use spectral signature~\cite{spectral} and activation clustering~\cite{activation} and confirm their ineffectiveness.
Yang et al.~\cite{advdoor} evaluate spectral signature~\cite{spectral,codebackdoor}, activation clustering~\cite{activation}, and ONION~\cite{qi-etal-2021-onion}, showing that their stealthy backdoor attack is much harder to be detected. 
As the triggers in Yang et al.~\cite{advdoor} are based on adversarial attack, they also evaluate the detection method proposed by Yefet et al.~\cite{Yefet2020} and find that it is not effective to find the adaptive triggers.
Schuster et al.~\cite{263874} find that both spectral signature~\cite{spectral} and activation clustering~\cite{activation} have a high false positive rate in terms of detecting poisoned examples.
Researchers also propose new defense methods. 
Li et al.~\cite{CodePoisoner} design \textsc{CodeDetector}, which can detect poisoned examples more accurately than an existing defense method ONION~\cite{qi-etal-2021-onion}.
Ramakrishnan and Albarghouthi~\cite{codebackdoor} improve the existing spectral signature defense method~\cite{spectral} by using multiple singular vectors. 
Results show that the improved defense can detect poisoned examples used in~\cite{codebackdoor} at a high accuracy close to 100\%.
However, Yang et al.~\cite{advdoor} show that the improved spectral signature defense~\cite{codebackdoor} is not effective in detecting adaptive triggers.
Hussain et al.~\cite{hussain2023occlusionbased} propose a method that can suggest whether the code input contains a trigger and detect the line-level trigger in the code input.

\vspace*{0.2cm}
\noindent \textbf{Human Review.}
Another way to mitigate such threats is to let humans review the dataset before training.
Li et al.~\cite{CodePoisoner} conduct human evaluation and show that their proposed approach \textsc{CodePoisoner} can generate poisoned examples that are natural and can achieve high attack success rates.
Sun et al.~\cite{Sun2023backdoor} show that developers can hardly distinguish the poisoned code from the clean code (with an F1 score of 0.43). 
Yang et al.~\cite{advdoor} conduct a user study to show examples poisoned using their adaptive triggers are more imperceptible to human judges.
They also find that human annotators spend much more time detecting the adaptive triggers than the triggers used in~\cite{you-see,codebackdoor}.


\begin{tcolorbox}[title=\textbf{Findings - Security}, left=2pt, right=2pt,top=2pt,bottom=2pt]
  The literature alerts that LLM4Code is vulnerable to security threats like data poisoning, which deserves more attention given that fact that the training data can be easily accessed and altered by the attackers.
  We find that researchers recently pay more attention to designing stealthy attacks that are harder to detect~\cite{advdoor,Sun2023backdoor,aghakhani2023trojanpuzzle}.
  Existing detection methods are not effective in identifying such stealthy attacks, calling for more effective defense.
\end{tcolorbox}

\subsection{Privacy}

\subsubsection{Privacy Evaluation}

Language models can memorize the training data~\cite{carlini21extracting}. 
The training data of LLM4Code largely comes from open-source projects and researchers have revealed that there exist many software secrets in these projects.
Basak et al.~\cite{basak2023secretbench} collect SecretBench, a dataset of software secrets, e.g., API keys, passwords, etc. 
Jungwirth et al.~\cite{10174120} check the secrets in the hidden files on GitHub projects. 
Feng et al.~\cite{9794113} find that there are many password leakages from public GitHub repositories.
Li et al.~\cite{starcoder} collect a dataset containing personally identifiable information in various programming languages. 
Studies~\cite{code-model-clone-msr22,RABIN2023107066} show that code recommendation models memorize many clones from its training data, and recent studies find that sensitive information can be potentially leaked or extracted by LLM4Code~\cite{yang2023memorzation,alkaswan2023traces,291327,huang2023away}.
Yang et al.~\cite{yang2023memorzation} find that CodeParrot can produce personally identifiable information like emails, names, IP addresses that are memorized from the training data. 
Al-Kaswan et al.~\cite{alkaswan2023traces} also reveal that LLM4Code can memorize its training data, highlighting the privacy leakage risks. 
Niu et al.~\cite{291327} design a set of prompts that are likely to induce privacy information from GitHub Copilot, uncovering that approximately 8\% (43) of the prompts yield privacy leaks.
Huang et al.~\cite{huang2023away} extract 2,702 hard-coded credentials from Copilot and 129 secrets from CodeWhisperer.

The membership inference attack aims to determine if a specific data point was used in training a model, which is also considered as privacy information in some studies~\cite{yang2023gotcha,alkaswan2023traces}. 
Yang et al.~\cite{yang2023gotcha} design \textsc{Gotcha}, the first framework to evaluate the membership leakage risks in code completion models. 
Membership inference attack is also used as an important step of data extraction attack. 
Specifically, an attacker first samples a large number of outputs from a code generator and then infers their membership to decide which outputs are likely to be in the training set.
Yang et al.~\cite{yang2023memorzation} use different heuristics to infer membership, including perplexity, ratios of perplexity by two different LLM4Code, the ratio of perplexity and zlib~\cite{zlibnet}, and average perplexity.
Al-Kaswan et al.~\cite{alkaswan2023traces} apply membership inference attacks to build a training data extraction framework for LLM4Code.


\subsubsection{Enhancing Privacy}

Researchers propose to process the training data to mitigate the risks of leaking privacy information.
For example, developers of InCoder~\cite{fried2023incoder} use regular expressions to identify all the email addresses in the training data and replace them with placeholders. 
Allal et al.~\cite{santacoder} use the \texttt{detect-secrets} tool with all default plugins activated. 
In addition, they use regular expressions to detect emails, IPv4, and IPv6 addresses. 
The model SantaCoder is trained on the dataset with detected secrets removed. 
Li et al.~\cite{starcoder} train a secret-detection model `\texttt{starpii}' that can identify secrets in code with higher accuracy and precision and use the tool to remove sensitive information.
Yang et al.~\cite{yang2023memorzation} find that the probability that a code snippet is memorized is highly correlated with its occurrence in the training data. 
One potential mitigation solution can be to remove duplicates in the training data. 

Some studies highlight the problem that model trainers collect and use the data without the consent of the data owners, e.g., the developers of open-source projects.
To protect the data rights and privacy of the data owners, researchers propose to detect such unauthorized data usage.
Membership inference can be used in this scenario~\cite{ma2023code2,zhang2023code,majdinasab2024trained}.
Zhang and Li~\cite{zhang2023code} apply code membership inference attacks to detect unauthorized data use in pre-trained LLM4Code. 
Majdinasab et al.~\cite{majdinasab2024trained} propose a model-agnostic and interpretable method based on membership inference for detecting code inclusion in an LLM4Code training dataset.
Another technique is \textit{watermarking}.
Sun et al.~\cite{CodeMark} propose a method to embed user-defined imperceptible watermarks into code datasets to trace their usage in training neural code completion models.
Watermarking is also used to protect LLM4Code from imitation attacks~\cite{10.1145/3576915.3623120} and detect whether the code is generated by a specific model~\cite{li2024resilient,lee2024wrote}.

Data poisoning can also be used to protect the training data from unauthorized usage.
The idea is that if the model is trained on the poisoned data, its performance will decrease, which motivates the model developers not to use the poisoned data.
Sun et al.~\cite{CoProtector} use data poisoning to protect open-source code from unauthorized training usage.
Li et al.~\cite{ji2022unlearnable} apply semantic-preserving transformations to the training data to make the model \textit{unlearnable}.

\begin{tcolorbox}[title=\textbf{Findings - Privacy}, left=2pt, right=2pt,top=2pt,bottom=2pt]
  Membership inference attacks can help protect the privacy~\cite{ma2023code2,zhang2023code,majdinasab2024trained}, e.g., to detect unauthorized data usage.
  There also lack studies on mitigating threats brought by membership inference attacks and data extraction attacks; potential techniques include differential privacy~\cite{Behnia2022}.
  Other privacy concern like model stealing~\cite{steal-rl} are not well-explored in LLM4Code.
  Although some strategies are adopted to remove sensitive information from the training data~\cite{starcoder,santacoder}, no empirical studies have shown their effectiveness and their impact on other properties.
\end{tcolorbox}



\subsection{Explainability} \label{subsec:interpretability_survey}

\subsubsection{Explainability Evaluation}
Researchers have put effort into understanding the explainability of LLM4Code using different methods.
With the increasing popularity of transformer-based models, researchers have proposed to use the attention mechanism to explain LLM4Code.
The attention mechanism is a critical component of the Transformer model, the fundamental idea behind which is to allow a model to focus on different parts of the input sequence when processing each word or token. 
A series of studies aim to understand which part of input LLM4Code pay more attention to
Sharma et al.~\cite{10.1145/3524610.3527921} find that BERT tends to focus on identifiers and separators.
Mohammadkhani et al.~\cite{Mohammadkhani2022} find that LLM4Code has different attention patterns for different tasks. 
Zhang et al.~\cite{zhang2022does} propose the aggregated attention score and aggregated attention graph to automatically extract meaningful semantic graphs from pretrained models. 
Wan et al.~\cite{10.1145/3510003.3510050} use attention analysis, probing on word embeddings, and syntax tree induction, to understand feature correlations in CodeBERT and GraphCodeBERT.
Attention can be used to reveal how LLM4Code embeds syntactic code properties~\cite{saad2023naturalness}.
Researchers also compare the attention mechanism between LLM4Code and human developers~\cite{paltenghi2022extracting,kou2023model,9678712,huber2023look}.
Mohammadkhani et al.~\cite{10356671} use attention mechanism to understand what CodeBERT and GraphCodeBERT learn from different tasks.

Some techniques are used to understand the important tokens in the model inputs. 
Wang et al.~\cite{WheaCha} propose \textsc{WheaCha} to distinguish the defining features that contribute significantly to the model's predictions.
Liu et al.~\cite{Liuyue2023} try to identify the tokens significantly contributing to code generation. 
Li et al.~\cite{10232920} assess these models by feeding them varied inputs, including different masking rates and a sufficient input subset method, revealing that these models are largely insensitive to the specific input.
Other techniques used to explain LLM4Code include counterfactual analysis~\cite{10.1145/3510457.3513081,hooda2024large}, causal inference~\cite{ji2023benchmarking,10336302}, and probing~\cite{troshin-chirkova-2022-probing}.



For non-LLM4Code, model-agnostic interpretation techniques like LIME~\cite{LIME}, BreakDown~\cite{breakdown}, and SHAP~\cite{SHAP} are used to explain the predictions of LLM4Code in many studies~\cite{10.1007/978-3-030-92635-9_25,Reemicse21,shin2021explainable,ZHENG2022111245,ledel2022studying,9044387,9978217}.
However, studies have shown that the explanation results generated by different methods can conflict with each other~\cite{9978217,shin2021explainable}.
Model-agnostic interpretation techniques are mainly used on classification tasks like defect prediction~\cite{10.1007/978-3-030-92635-9_25,Reemicse21,shin2021explainable,ZHENG2022111245,9978217,9044387} and bug report classification~\cite{ledel2022studying}.

\subsubsection{Explainability Enhancement}
Researchers also propose methods to enhance the explainability of LLM4Code and show how explainability can benefit LLM4Code.
Ji et al.~\cite{ji2023benchmarking} propose a causality analysis-based approach that can provide insights into LLM effectiveness and aid end-users in understanding predictions.
Palacio et al.~\cite{palacio2023evaluating} propose ASTxplainer that provides visualizations of LLM predictions that aid end-users in understanding model predictions.
Ganz et al.~\cite{10190498} use directed fuzzing to create local ground-truth around code regions marked as relevant by an explanation method. 
Pornprasit et al.~\cite{10.1109/ASE51524.2021.9678763} propose PyExplainer to further improve existing explainable AI techniques for JIT models.
Rabin and Alipour~\cite{RABIN2022100432} develop FeatureExtractor, which applies a program reduction technique to the input program, preserving the model prediction while reducing the program size. 
The insight that BERT focuses more on identifiers and separators motivates Sharma et al.~\cite{10.1145/3524610.3527921} to use identifiers for representing code sequences in clone detection tasks, resulting in significant improvements in performance, with a 21 to 24\% performance improvement in CodeBERT.
Zhu et al.~\cite{10.1007/978-3-030-60029-7-32} develop a novel framework for Interpretable Text-to-SQL Generation (ITSG) that provides two-level interpretability.
Zhang et al.~\cite{10.1145/3411764.3445646} propose interpretable program synthesis, aiming to provide users with insights into and control over the synthesis process. 
Li et al.~\cite{10298318} improve the robustness of explanations of LLM4Code.

\begin{tcolorbox}[title=\textbf{Findings - Explainability}, left=2pt, right=2pt,top=2pt,bottom=2pt]
  Explanations provided by different techniques can conflict with each other~\cite{10190498}, indicating the need for more reliable explanation methods.
  Existing studies largely focus on classification tasks like defect prediction, while we believe that more studies are needed to explain generative LLM4Code tasks (e.g., code generation) considering their popularity in practice.
  A large portion of efforts are devoted to understanding what LLM4Code have learned and what contributes to their predictions, which are valuable in understanding their performance, robustness, and efficiency~\cite{10.1145/3641540,10.1145/3468264.3468539,RABIN2022100432}.
  However, there still exists a gap in satisfying the needs of end-users for explainability~\cite{10.1145/3490099.3511119}.
\end{tcolorbox}

\subsection{Efficiency} \label{subsec:efficiency_survey}

\subsubsection{Efficiency Evaluation}
Researchers use various metrics to evaluate the efficiency of LLM4Code. 
Popular efficiency metrics include the model size, model complexity, latency, memory consumption, etc.
These metrics are usually inter-related. 
For example, a large model typically takes longer to process (i.e., high latency) and consumes more memory to load (i.e., high memory consumption).
We explain each metric as follows.

In this survey, model size is the space needed to store a model, which is usually measured in MB or GB. 
The model size directly affects the memory consumption to load the model. 
Svyatkovskiy et al.~\cite{svyatkovskiy2021fast} suggest that a reasonable upper bound size for an IDE component or an editor plug-in is 50 MB, while a 3 MB model is preferred as it can be deployed even in severely restricted environments (e.g., low-end hardware for teaching students to code).
A model's size is largely determined by the number of parameters in the model and the precision of the parameters.
For example, given the same number of parameters, a model with 8-bit parameters is smaller than a model with 32-bit parameters.

Model complexity is usually measured by FLOPs (floating-point operations)~\cite{compressor}.
Model complexity is related to the model size, but is also affected by other factors such as the model architecture and the type and number of operations in the model.
A model with more FLOPs usually consumes more computing resources and takes longer to process an input, which is reflected in the latency.

Latency is another important metric to evaluate the efficiency of LLM4Code.
Depending on the stages of developing LLM4Code, latency can be further divided into training latency and inference latency.
Training latency is the time required to train a model.
To optimize training latency, researchers have explored methods to reduce the training data size, which we call data efficiency.
Inference latency is the time required to process a single input, which is usually measured in seconds per input or seconds per token.

\subsubsection{Efficiency Enhancement}
Researchers have designed and applied various methods to improve the efficiency of LLM4Code from different perspectives.
For example, Shi et al.~\cite{compressor} conduct the first study on compressing CodeBERT and GraphCodeBERT to 3MB. 
Many studies using different techniques, e.g., quantization, knowledge distillation, and parameter-efficient fine-tuning, are employed to enhance efficiency, which we explain as follows. 

\vspace*{0.2cm}
\noindent \textbf{Model Compression.}
Model compression refers to the techniques that reduce the size of a model and thus optimize other relevant metrics such as latency.
Shi et al.~\cite{compressor} propose a genetic algorithm (GA)-based strategy to find a simplified model and use knowledge distillation to reduce the size of CodeBERT and GraphCodeAttack.
The proposed method can compress pre-trained models to a size (3 MB), which is 160$\times$ smaller than the original size. 
The compressed CodeBERT and GraphCodeBERT are 4.31$\times$ and 4.15$\times$ faster than the original model at inference, respectively. 
They maintain 96.15\% and 97.74\% of the original performance on the vulnerability prediction task.
Shi et al.~\cite{avatar} then present Avatar, which can also compress CodeBERT to 3MB.
The optimized models significantly reduce the energy consumption (up to 184$\times$ less), carbon footprint (up to 157$\times$ less), and inference latency (up to 76$\times$ faster), with a negligible loss in effectiveness (1.67\% on average). 
Su and McMillan~\cite{Su2024} apply knowledge distillation to GPT-3.5 to get a small model for code summarization.
Wei et al.~\cite{10.1145/3611643.3616302} apply quantization to code generation models.
Kaushal et al.~\cite{kaushal2023lord} combine Low Rank Decomposition with quantization.


\vspace*{0.2cm}
\noindent \textbf{Input Optimization.}
As the time complexity of LLM4Code depends on the input length, optimizing input length can improve model efficiency. 
Zhang et al.~\cite{Zhang2022diet} propose DietCode, which can select statements and tokens that receive the most attention weights during pre-training.
DietCode reduces 40\% computational cost in fine-tuning and inferring CodeBERT.
Hidvégi et al.~\cite{hidvégi2024cigar} optimize the prompt to reduce the token cost by 62\% on program repair tasks.
Hussain et al.~\cite{10.1145/3593434.3594236} propose a method that can reduce the length of tokenized inputs efficiently.
Shi et al.~\cite{10.1145/3609437.3609438} simplify the input to LLM4Code and reduce the model training and inference time by 20\% to 40\%.

\vspace*{0.2cm}
\noindent \textbf{Dynamic Inference.}
Dynamic inference manipulates the inference process to improve model efficiency.
Sun et al.~\cite{sun2024neural} find that 54.4\% of tokens can be accurately generated using just the first layer of GPT-2. 
They design a method that can averagely skip 1.7 layers out of 16 layers in the models, leading to an 11.2\% speedup with only a marginal 1.1\% performance reduction.
Sun et al.~\cite{10172653} propose an early-rejection mechanism to turn down low-return prompts that cannot generate helpful completions.
The proposed mechanism helps save 23.3\% of computational cost measured in floating-point operations, and 80.2\% of rejected prompts lead to unhelpful completions.
Grishina et al.~\cite{10.1145/3611643.3616304} build composite representations of code from the early layers of LLM4Code.
Results show that using only 3 out of 12 layers of CodeBERT can provide 3.3$\times$ speedup of fine-tuning and also improve model performance by 2\% on average.
Svyatkovskiy et al.~\cite{svyatkovskiy2021fast} design a two-step method with a modular neural framework for code completion and a reranking module that combines static analysis with granular token encodings. 
They produce a solution that consumes just 6 MB of RAM, computes a single completion in 8 ms, and achieves 90\% accuracy in its top five suggestions.

\vspace*{0.2cm}
\noindent \textbf{Data Efficiency.}
Some efforts are made to improve the data efficiency of LLM4Code.
Hu et al.~\cite{hu2023active} employ active learning to LLM4Code to reduce the training data size.
Studies~\cite{liu2024delving,wang2023adapter} also show that PEFT approaches are beneficial in low-resource scenarios where data is limited.

\vspace*{0.2cm}
\noindent \textbf{Parameter-Efficient Fine-Tuning (PEFT).}
PEFT is a popular method to improve the training stage efficiency.
PEFT approaches only fine-tune a small number of model parameters while freezing most parameters of the pretrained LLMs, thereby greatly decreasing the computational and storage costs.
Popular PEFT approaches include Adapter~\cite{houlsby2019parameter}, LoRA~\cite{hu2022lora}, Prefix tuning~\cite{li-liang-2021-prefix}, and their variants like QLoRA~\cite{dettmers2023qlora}, etc.
Researchers have applied these PEFT methods to LLM4Code on different tasks~\cite{10298587,weyssow2024exploring,liu2024delving,10299938,ayupov2022parameterefficient,saberi2024utilization,10.1145/3540250.3558959,zhuo2024astraios}. 
LoRA is the most popular method, which is used in~\cite{10298587,liu2024delving,weyssow2024exploring,10299938,ayupov2022parameterefficient,zhuo2024astraios}, followed by Adapter~\cite{10298587,liu2024delving,ayupov2022parameterefficient,saberi2024utilization} and Prefix tuning~\cite{10298587,10299938,weyssow2024exploring,10.1145/3540250.3558959}.
Shi et al.~\cite{10.1145/3597926.3598036} find that different layers in LLM4Code have different importance in understanding code. They propose a method to only fine-tune the top two layers to reduce training time cost.


\begin{tcolorbox}[title=\textbf{Findings - Efficiency}, left=2pt, right=2pt,top=2pt,bottom=2pt]
  From the literature, we find that PEFT is becoming increasingly popular to improve the training efficiency of LLM4Code, which demonstrates comparable performance to full fine-tuning and provides benefits in low-resource scenarios.
  Relatively speaking, explorations on other efficiency enhancement strategies (e.g., quantization~\cite{10.1145/3611643.3616302}, model pruning~\cite{gordon2020compressing}) are relatively limited.
  Studies show that there exists a trade-off between efficiency and performance.
  However, how the improvement of efficiency impacts other properties like robustness and security remains unclear and deserves further investigation.
\end{tcolorbox}


\subsection{Usability}

\subsubsection{Usability Evaluation}
The usability is usually evaluated by conducting user studies (including observation, interviews, surveys, etc.) and mining online data relevant to LLM4Code.

\vspace*{0.2cm}
\noindent \textbf{Observation.}
The observational study involves observing the actions of participants in a natural or controlled setting, without any manipulation or intervention by the researcher, to understand how users interact with LLM4Code in practice and evaluate the usability of LLM4Code.
Studies focus on how novice programmers interact with LLM4Code~\cite{10.1145/3617367,10.1145/3544548.3580919,10.1145/3526113.3545659}, showing that LLM4Code can help them write code faster.
Jayagopal et al.~\cite{10.1145/3526113.3545659} focus on the learnability (i.e., first-encounter usability) of five existing code generators.
Nguyen et al.~\cite{nguyen2024beginning} conduct a controlled study of how 120 beginning coders across three academic institutions approach writing and editing prompts.

A series of studies analyze how code generators affect the productivity of developers via observation.
The conclusion is two-fold: LLM4Code can increase productivity but also bring some negative impacts.
Peng et al.~\cite{peng2023impact} recruit 95 professional programmers and ask to implement an HTTP server in JavaScript as quickly as possible.
They find the treated group (with access to Copilot) completed the task 55.8\% faster (95\% confidence interval: 21-89\%) than the control group (without access to Copilot).
However, Vaithilingam et al.~\cite{10.1145/3491101.3519665} observe that participants did face difficulties in understanding, editing, and debugging code snippets generated by Copilot, which hindered their task-solving effectiveness.
Imai~\cite{10.1145/3510454.3522684} finds that Copilot increases productivity as measured by lines of code added, but the quality of code produced is inferior by having more lines of code deleted in the subsequent trial.
Weisz et al.~\cite{10.1145/3490099.3511157} examine whether such imperfect outputs are helpful in the context of Java-to-Python code translation.

Mozannar et al.~\cite{mozannar2023reading} further look into the interaction patterns of programmers when using Copilot.
They propose a taxonomy of common programmer activities, which contains 12 mutually unique activities that programmers perform between consecutive code recommendations.
They studied 21 programmers to analyze the time spent on these activities.
They find that programmers may spend a large fraction of total session time (34.3\%) on just double-checking and editing Copilot suggestions, and spend more than half of the task time on Copilot-related activities, together indicating that introducing Copilot into an IDE can significantly change user behavior.

\vspace*{0.2cm}
\noindent \textbf{Interview.}
Subsequent to observation study, Prather et al.~\cite{10.1145/3617367} conduct interviews to explore students' perceptions of the benefits and dangers of code generators for learning. 
The findings indicate a dichotomy in student attitudes towards tools like Copilot. 
While the majority acknowledges the utility of such tools in facilitating code writing, there is prevalent concern about potential over-reliance and a diminished understanding of the underlying code logic.
Such concerns are also mentioned by educators in other studies~\cite{codex,finnie2022robots}.
Jayagopal et al.~\cite{10.1145/3526113.3545659} conduct a semi-structured interview to understand participants' common behaviors and strategies for using code generators.

\vspace*{0.2cm}
\noindent \textbf{Survey.}
Ziegler et al.~\cite{10.1145/3520312.3534864} analyze 2,631 survey responses from developers using GitHub Copilot.
They draw an observation that the acceptance rate of shown suggestions is a better predictor of perceived productivity than the alternative measures, e.g., persistence of accepted completions in the code.
Liang et al.~\cite{liang2023usability} administer a survey to developers and received responses from a set of 410 developers.
They found that developers are most motivated to use AI programming assistants because they help developers reduce keystrokes, finish programming tasks quickly, and recall syntax, but resonate less with using them to help brainstorm potential solutions.
They also found the most important reasons why developers do not use these tools are because these tools do not output code that addresses certain functional or non-functional requirements and because developers have trouble controlling the tool to generate the desired output.

\vspace*{0.2cm}
\noindent \textbf{Simulation.}
We also notice one study that applies LLM4Code to user interaction data that was previously collected and lets human annotators evaluate how LLM4Code may support the task. 
As this type of study involves no real user interaction, we categorize it as \textit{simulation}.
Druga and Otero~\cite{druga2023scratch} utilize 22 Scratch projects (used for creative family coding) as inputs for GPT-4, which resulted in a pool of 120 creative coding support scenarios.
The authors find that GPT-4 can help families comprehend game code, debug programs, and generate new ideas for future projects.
A following study~\cite{druga2023ai} finds that families found it easier to generate game ideas when prompted with questions by AI assistance and children were more encouraged to write code for novel ideas using AI assistance.

\vspace*{0.2cm}
\noindent \textbf{Mining Online Discussion.}
Zhang et al.~\cite{Zhang_2023} searched and manually collected 169 Stack Overflow posts and 655 GitHub discussions related to the usage of Copilot.
By mining these discussions, they show that (1) The major programming languages used with Copilot are JavaScript and Python, (2) the main IDE used with Copilot is Visual Studio Code, (3) the most commonly used technology with Copilot is Node.js, (4) the leading function implemented by Copilot is data processing, (5) the significant benefit of using Copilot is useful code generation, and (6) the main limitation encountered by practitioners when using Copilot is difficulty in integration.
Li et al.~\cite{li2023always} reveal that while ChatGPT provides swift, clear, and comprehensive replies, enhancing respectfulness compared to Stack Overflow, its reliability is questioned due to an overly confident tone and lack of validation mechanisms akin to Stack Overflow's voting system.
Zhang et al.~\cite{zhou2023concerns} collected data from 476 GitHub issues, 706 GitHub discussions, and 184 Stack Overflow posts, and identified the issues, causes that trigger the issues, and solutions that resolve the issues when using Copilot.
Jin et al.~\cite{jin2024can} analyze the DevGPT dataset~\cite{xiao2023devgpt} that encompasses 29,778 prompts and responses from ChatGPT.
They find that the current practice of using LLM-generated code is typically limited to either demonstrating high-level concepts or providing examples in documentation, rather than to be used as production-ready code.


\subsubsection{Usability Enhancement}
Researchers have identified potential opportunities and design implications to improve the usability of LLM4Code, which we explain as follows.
Jayagopal et al.~\cite{10.1145/3526113.3545659} identified a set of design opportunities to make future coding assistants more learnable.
Li et al.~\cite{li2023always} highlight the importance of consistency, reliability, avoiding overconfidence, and transparency in the design of AI programming assistants.
Jin et al.~\cite{jin2024can} propose to build appropriate trust with LLM4Code, e.g., by providing explanations.
Vaithilingam et al.~\cite{10.1145/3491101.3519665} mention that one way to help users understand the generated code is to provide explanations using inline comments, similar suggestions are also made by~\cite{10.1145/3490099.3511157}.
Prather et al.~\cite{10.1145/3617367} suggest that Copilot's recommendations could be more useful to students (especially novice students) with a selection algorithm that prefers shorter solutions, or filters longer solutions.
Zhou et al.~\cite{zhou2023concerns,Zhang_2023} mention the need for more customization options to allow users to tailor Copilot's behavior to align with their own workflow and integration of Copilot with more IDEs.

Experiments are conducted to show the effectiveness of newly proposed design principles to improve usability. 
Vasconcelos et al.~\cite{vasconcelos2023generation} analyze how uncertainty highlighting can improve the usability of code completion tools.
They find that highlighting tokens with the highest predicted likelihood of being edited leads to faster task completion and more targeted edits, and is subjectively preferred by study participants.
Blinn et al.~\cite{9833110} contribute a conceptual architecture and API to guide programming assistant designers as they confront these integration and human-centered design challenges.
Vaithilingam et al.~\cite{10172834} explore 19 designs and summarize their findings into a set of 5 design principles that can improve usability. 
They build and deploy a new version of IntelliCode with two new inline interfaces in Microsoft Visual Studio 2022 and found that they lead to a significant increase in tool usage.

The usability can also be enhanced from the user side, e.g., by providing clear explanations to prompt the AI programming assistant~\cite{liang2023usability}.
Similarly, Tian et al.~\cite{tian2023chatgpt} find that the models (ChatGPT, Codex, and CodeGen) make more correct predictions when the prompts have a relatively shorter length, suggesting that programmers should provide a clear and concise prompt when using LLM4Code.
For example, novice programmers may benefit from (i) feedback about why a given synthesis run has failed and (ii) interactions that help them understand an incorrect synthesis output and why it was produced.

\begin{tcolorbox}[title=\textbf{Findings - Usability}, left=2pt, right=2pt,top=2pt,bottom=2pt]
  Conclusion regarding whether LLM4Code-based systems can improve productivity is diverging.
  Studies find both positive and negative impacts brought by tools like Copilot, highlighting the space to improve these tools.
  Existing usability studies mainly focus on code completion systems and chatbots, like Copilot and ChatGPT, while attention on LLM4Code for other purposes (like vulnerability detection and code search) is needed. 
  Researchers have proposed a series of design suggestions for improving usability.
  But only a few studies~\cite{9833110,10172834} translate the suggestions and evaluate their effectiveness in practice.
  We believe more studies are required to explore the optimal design of various LLM4Code systems.
\end{tcolorbox}

\section{Challenges and Opportunities}
\label{sec:challenges}
Based on our literature review, we identify challenges and opportunities in enhancing the non-functional properties of LLM4Code, which are organized into three views that can be adopted to develop LLM4Code: data-centric, human-centric, and system-centric views.

\subsection{Data-Centric View}

Our systematic literature review suggests that many issues regarding non-functional requirements are prevalent across LLM4Code of different architectures, which to some extent reveals that the data plays a pivotal role in obtaining high-quality models.
Researchers have pinpointed that certain issues plaguing LLM4Code are, in fact, inherited from the datasets they are trained on~\cite{10006873}. 
For instance, it has been observed that code models tend to memorize and produce portions of the training data, such as license codes or code snippets with vulnerabilities~\cite{yang2023memorzation}.
Furthermore, issues~\cite{10.1145/3551349.3556941} such as class imbalance, mislabeling, and the presence of outdated or irrelevant information in current training sets have been identified. 
Given these considerations, we propose that a data-centric view is crucial in developing LLM4Code, which emphasizes the enhancement of training dataset quality as a fundamental strategy to improve model performance. In light of the challenges mentioned, several research opportunities emerge.

\begin{enumerate}[leftmargin=*]
  \item \textit{Identify High-Quality Training Data.} Efforts are required to obtain high-quality training data. This could involve the development of novel techniques to identify and remove low-quality data points (e.g., duplicated data~\cite{duplicate-code}, vulnerable code~\cite{sandoval2023lost}, mislabeled code~\cite{10.1145/3551349.3556941}) from existing datasets. 
  \item \textit{Data Auditing and Filtering.} Given the fact that poisoned data is generally hard to detect, it is crucial to develop techniques to audit the training data to mitigate potential negative impacts. Other unwanted features should also be identified and filtered, for example, code snippets that contain privacy information~\cite{basak2023secretbench}.
  Conducting thorough reviews of datasets to identify and eliminate potentially poisonous data. This involves detecting data points intentionally designed to mislead or degrade the performance of the model. However, studies show that existing detection methods are not effective in identifying poisoned data~\cite{advdoor,you-see}.
  \item \textit{Data Augmentation.} Implementing techniques to artificially expand the training dataset, thereby increasing model robustness and accuracy~\cite{10123554}. This could involve generating synthetic data examples or applying transformations to existing data points to simulate a broader range of scenarios~\cite{10.1145/3540250.3549162}. Data augmentation can also be used to address class imbalance issues~\cite{10.1145/3551349.3556941}.
  \item \textit{Data Transparency.} The community should advocate for the openness of training data and processes. For example, model developers can provide the training data information and we can further maintain a supply chain of LLM4Code, consisting of dependencies between models and datasets. Such a supply chain allows for traceability. For example, if a dataset is found to be poisoned, we can trace the models that are trained on this dataset. If a model is found to be poisoned, we can trace the datasets that are used to train this model. 
\end{enumerate}

\subsection{Human-Centric View}

LLM4Code have been integrated into developers' daily development workflows. 
In many scenarios, the model is not merely providing outputs or recommendations. 
On one hand, the model continuously interacts with users, such as a code completion tool that needs to provide timely and appropriate suggestions. Users can choose to accept or reject these suggestions, which can influence their code-writing process. 
Inappropriate recommendations might lead to additional time spent on modifications~\cite{10.1145/3617367,10.1145/3544548.3580919,10.1145/3526113.3545659}.
On the other hand, the model needs to provide additional information to assist users in making decisions on how to handle a suggestion. 
For instance, if the model identifies a potential vulnerability in the code, users might need further information to determine the usefulness of the suggestion and whether further action is required.
Therefore, adopting a human-centric view can enhance usability and improve developers' wellbeing. 
This perspective presents several research opportunities:

\begin{enumerate}[leftmargin=*]
  \item \textit{Identify Factors Affecting Usability.} We need to understand the factors that influence the usability of LLM4Code tools so that we can enhance user satisfaction and productivity. How different aspects, e.g., tools' response time, accuracy of code suggestions, relevance of provided information, and the interface's intuitiveness, affect usability remains unclear. Empirical studies, such as user surveys, interviews, and observation of developers using these tools in real-world scenarios, can shed light on these factors. Key areas of focus might include the context in which suggestions are most helpful, how developers interact with suggestions (acceptance, modification, rejection), and any cognitive load or distractions caused by the tool. The goal is to identify patterns and pain points that can inform the development of more user-friendly interfaces and functionalities.
  \item \textit{Enhance Trust.} Various strategies can be explored to enhance the human developers' trust in LLM4Code~\cite{wang2023investigating,lo2023trustworthy}. Enhancing trust in LLM4Code tools through explainable AI (XAI) techniques~\cite{mohammadkhani2023systematic} involves making the AI's decision-making processes transparent and understandable. This can be achieved by employing strategies such as feature importance visualization to highlight influential code parts, providing natural language explanations for the rationale behind suggestions, and offering example-based explanations that draw parallels with previously solved problems. 
  Additionally, integrating confidence scores and uncertainty indicators~\cite{vasconcelos2023generation} with suggestions can help users gauge their reliability, while counterfactual explanations clarify how inputs affect model outputs~\cite{10.1145/3510457.3513081}.
  \item \textit{Personalization.} Personalization for users of LLM4Code~\cite{10.1145/3540250.3558959} involves tailoring the tool's behavior and suggestions to match individual user preferences, coding styles, and project-specific needs. 
  Systems powered by LLM4Code can learn from a user's interaction history, preferred coding patterns, and frequently used libraries or frameworks, enabling the model to offer more relevant and context-aware suggestions.  Another research opportunity is the privacy-preserving personalization methods, such as federated learning~\cite{10123655} or differential privacy~\cite{Behnia2022}, ensuring that users' data remains secure, enhancing user trust and acceptance of personalized recommendations.
  \item \textit{Automated Benchmarking for Usability.} It is important to build a benchmark that can evaluate the usability of LLM4Code tools in interactive environments and can produce estimates of key usability metrics, such as effectiveness, efficiency, and user satisfaction. The inherent challenge lies in the traditional reliance on human participants for usability evaluations, which can be resource-intensive and subjective. However, leveraging recent advancements in AI to simulate human behavior offers a promising solution~\cite{zhang2024android}. By utilizing sophisticated AI models that can mimic developer interactions with coding tools, we can automate the process of usability testing. 
  Incorporating such an AI-driven benchmark into the development lifecycle of LLM4Code tools can significantly enhance our ability to refine these tools in alignment with user needs and preferences, ultimately leading to more intuitive and effective coding environments.
\end{enumerate}

\subsection{System-Centric View}
In practical applications, LLM4Code will be integrated into broader systems and considered 
as an integral part of the overall system. These systems could be commercialized as products, such as paid code auto-completion services, vulnerability detection services, and more. For developers of these systems, there is a need to consider many factors that are crucial for ensuring the system's effectiveness, user-friendliness, and competitive edge in the market.
Although the properties discussed in this document primarily originate from academic research, we believe that considering these properties in real-world settings is very important. Here are some key aspects and research opportunities that need to be taken into account when applying LLM4Code in practice.

\begin{enumerate}[leftmargin=*]
  \item \textit{Build Attack-Resistant System.} Existing research shows that different types of LLM4Code are vulnerable to attacks~\cite{alert,advdoor,yang2023gotcha}, putting the integrity of the system at risk. Therefore, it is crucial to build an attack-resistant system.
  We need to implement comprehensive security measures, e.g., input validation and sanitization, secure and regularly audited training data to avoid learning malicious patterns, and adversarial training to enhance the systems' robustness. 
  Additionally, employing anomaly detection for monitoring unusual activity, strict role-based access control to limit exposure, and robust encryption for data protection are crucial. 
  Regular security audits and updates, along with incident response plans, ensure the system's defenses remain up-to-date and effective.
  \item \textit{AIOps for LLM4Code.} LLM4Code is also a type of AI system, we can apply AIOps to LLM4Code, focusing on versioning of datasets and models, and ensuring efficient operations. This includes implementing version control, adopting continuous integration and deployment (CI/CD) practices for seamless model updates, and deploying real-time monitoring to track model performance. Automated anomaly detection aids in identifying issues promptly, while feedback loops incorporate user insights for continuous model improvement. Efficient update and repair mechanisms, such as automated retraining pipelines, ensure the model remains relevant and effective. 
  \item \textit{Efficiency and Scalability.} Research opportunities to enhance the efficiency and scalability of LLM4Code involve exploring innovative approaches such as model optimization techniques like compression~\cite{compressor} and pruning to reduce computational demands, and employing distributed computing to improve processing times. 
  Efficient data management strategies are crucial for handling large codebases, while adaptive and incremental learning methods allow for continuous model updates without extensive retraining. Hybrid model architectures could offer a balance between deterministic and learning-based approaches, optimizing performance. Additionally, intelligent load balancing can ensure optimal resource utilization, and energy-efficient computing practices are essential for minimizing environmental impact. 
  \item \textit{Legal and Ethical Considerations.} It is important to develop frameworks that ensure these systems of LLM4Code are compliant with copyright laws, protect user privacy, and maintain data security, while also addressing ethical concerns such as bias mitigation and fairness. Research is needed to create mechanisms for copyright compliance in code generation, implement advanced data privacy techniques, and develop algorithms for detecting and correcting biases to prevent discrimination. Ensuring the security of LLM4Code systems against potential exploits, enhancing model transparency for greater user trust, and formulating ethical guidelines for responsible use are also critical. Moreover, establishing legal frameworks to clarify accountability for AI-generated code and ensuring user consent and control over data usage are paramount.
\end{enumerate}


\section{Threats to Validity}
\label{sec:threats}


Selection bias represents a significant threat to the validity of systematic literature reviews, as it can lead to an unrepresentative sample of studies, skewing the review's conclusions. To counteract this bias, we have established clear and reproducible inclusion and exclusion criteria that are meticulously documented in our review protocol. 
This ensures a transparent and objective selection process. 
Moreover, we engage in a comprehensive search on the DBLP database, to capture a broad spectrum of relevant studies including preprints.
By doing so, we aim to minimize the risk of overlooking significant research that can influence our review's findings.

To address the challenge of capturing the most recent studies, which are crucial for maintaining the relevance and timeliness of our review, we implement multiple rounds of snowballing~\cite{snowball1,snowball2}. 
This method involves reviewing the references of selected papers to identify additional relevant publications and repeating this process iteratively. 
The last round of search was conducted on Feb 15, 2024, and we have added the newly found studies to the survey, ensuring that our literature review is comprehensive and up-to-date, reflecting the latest advancements and findings in the field.

\section{Conclusion}
\label{sec:conclusion}

This paper conducts a systematic literature review on the non-functional properties of LLM4Code beyond accuracy. 
We identify 7 important properties, including robustness, security, privacy, explainability, efficiency, privacy, and usability. 
We review 215 relevant papers and analyze their distribution and trends.
We then discuss these non-functional properties in detail, including their definitions, how to evaluate the properties, and the current practices to enhance these properties. For each property, we summarize some key findings and insights. 
We provide three views of LLM4Code development, including data-centric, human-centric, and system-centric views. For each view, we discuss the challenges and potential research opportunities to further improve LLM4Code.

In the future, we plan to revisit the studies relevant to LLM4Code and update the literature review. It is also promising to develop benchmarks and tools to evaluate and enhance the non-functional properties of LLM4Code.

\bibliographystyle{ACM-Reference-Format}
\bibliography{../reference}


\begin{thebibliography}{271}


\ifx \showCODEN    \undefined \def \showCODEN     #1{\unskip}     \fi
\ifx \showDOI      \undefined \def \showDOI       #1{#1}\fi
\ifx \showISBNx    \undefined \def \showISBNx     #1{\unskip}     \fi
\ifx \showISBNxiii \undefined \def \showISBNxiii  #1{\unskip}     \fi
\ifx \showISSN     \undefined \def \showISSN      #1{\unskip}     \fi
\ifx \showLCCN     \undefined \def \showLCCN      #1{\unskip}     \fi
\ifx \shownote     \undefined \def \shownote      #1{#1}          \fi
\ifx \showarticletitle \undefined \def \showarticletitle #1{#1}   \fi
\ifx \showURL      \undefined \def \showURL       {\relax}        \fi
\providecommand\bibfield[2]{#2}
\providecommand\bibinfo[2]{#2}
\providecommand\natexlab[1]{#1}
\providecommand\showeprint[2][]{arXiv:#2}

\bibitem[zli({[n.\,d.]})]%
        {zlibnet}
 \bibinfo{year}{[n.\,d.]}\natexlab{}.
\newblock \bibinfo{title}{A Massively Spiffy Yet Delicately Unobtrusive Compression Library}.
\newblock \bibinfo{howpublished}{\url{https://zlib.net/}}.
\newblock
\newblock
\shownote{Accessed on March 27, 2023}.


\bibitem[owa({[n.\,d.]})]%
        {owasp_benchmark}
 \bibinfo{year}{[n.\,d.]}\natexlab{}.
\newblock \bibinfo{title}{{OWASP Benchmark Project}}.
\newblock \bibinfo{howpublished}{\url{https://owasp.org/www-project-benchmark/}}.
\newblock
\newblock
\shownote{Accessed: 2023-12-19}.


\bibitem[Abuhamad et~al\mbox{.}(2023)]%
        {abuhamad2023shield}
\bibfield{author}{\bibinfo{person}{Mohammed Abuhamad}, \bibinfo{person}{Changhun Jung}, \bibinfo{person}{David Mohaisen}, {and} \bibinfo{person}{DaeHun Nyang}.} \bibinfo{year}{2023}\natexlab{}.
\newblock \bibinfo{title}{SHIELD: Thwarting Code Authorship Attribution}.
\newblock
\newblock


\bibitem[Aghakhani et~al\mbox{.}(2023)]%
        {aghakhani2023trojanpuzzle}
\bibfield{author}{\bibinfo{person}{Hojjat Aghakhani}, \bibinfo{person}{Wei Dai}, \bibinfo{person}{Andre Manoel}, \bibinfo{person}{Xavier Fernandes}, \bibinfo{person}{Anant Kharkar}, \bibinfo{person}{Christopher Kruegel}, \bibinfo{person}{Giovanni Vigna}, \bibinfo{person}{David Evans}, \bibinfo{person}{Ben Zorn}, {and} \bibinfo{person}{Robert Sim}.} \bibinfo{year}{2023}\natexlab{}.
\newblock \showarticletitle{TrojanPuzzle: Covertly Poisoning Code-Suggestion Models}.
\newblock \bibinfo{journal}{\emph{arXiv preprint arXiv:2301.02344}} (\bibinfo{year}{2023}).
\newblock


\bibitem[Ahmad et~al\mbox{.}(2020)]%
        {ahmad-etal-2020-transformer}
\bibfield{author}{\bibinfo{person}{Wasi Ahmad}, \bibinfo{person}{Saikat Chakraborty}, \bibinfo{person}{Baishakhi Ray}, {and} \bibinfo{person}{Kai-Wei Chang}.} \bibinfo{year}{2020}\natexlab{}.
\newblock \showarticletitle{A Transformer-based Approach for Source Code Summarization}. In \bibinfo{booktitle}{\emph{Proceedings of the 58th Annual Meeting of the Association for Computational Linguistics}}. \bibinfo{publisher}{Association for Computational Linguistics}, \bibinfo{address}{Online}, \bibinfo{pages}{4998--5007}.
\newblock
\urldef\tempurl%
\url{https://doi.org/10.18653/v1/2020.acl-main.449}
\showDOI{\tempurl}


\bibitem[Al-Kaswan et~al\mbox{.}(2023)]%
        {alkaswan2023traces}
\bibfield{author}{\bibinfo{person}{Ali Al-Kaswan}, \bibinfo{person}{Maliheh Izadi}, {and} \bibinfo{person}{Arie van Deursen}.} \bibinfo{year}{2023}\natexlab{}.
\newblock \bibinfo{title}{Traces of Memorisation in Large Language Models for Code}.
\newblock
\newblock
\showeprint[arxiv]{2312.11658}~[cs.CR]


\bibitem[Aleithan(2021)]%
        {Reemicse21}
\bibfield{author}{\bibinfo{person}{Reem Aleithan}.} \bibinfo{year}{2021}\natexlab{}.
\newblock \showarticletitle{Explainable Just-in-Time Bug Prediction: Are We There Yet?}. In \bibinfo{booktitle}{\emph{Proceedings of the 43rd International Conference on Software Engineering: Companion Proceedings}} (Virtual Event, Spain) \emph{(\bibinfo{series}{ICSE '21})}. \bibinfo{publisher}{IEEE Press}, \bibinfo{pages}{129–131}.
\newblock
\urldef\tempurl%
\url{https://doi.org/10.1109/ICSE-Companion52605.2021.00056}
\showDOI{\tempurl}


\bibitem[Allal et~al\mbox{.}(2023)]%
        {santacoder}
\bibfield{author}{\bibinfo{person}{Loubna~Ben Allal}, \bibinfo{person}{Raymond Li}, {and} \bibinfo{person}{Denis~Kocetkov et al.}} \bibinfo{year}{2023}\natexlab{}.
\newblock \showarticletitle{SantaCoder: don't reach for the stars!}
\newblock
\showeprint[arxiv]{2301.03988}~[cs.SE]


\bibitem[Allamanis(2019)]%
        {duplicate-code}
\bibfield{author}{\bibinfo{person}{Miltiadis Allamanis}.} \bibinfo{year}{2019}\natexlab{}.
\newblock \showarticletitle{The Adverse Effects of Code Duplication in Machine Learning Models of Code}. In \bibinfo{booktitle}{\emph{Proceedings of the 2019 ACM SIGPLAN International Symposium on New Ideas, New Paradigms, and Reflections on Programming and Software}} (Athens, Greece) \emph{(\bibinfo{series}{Onward! 2019})}. \bibinfo{publisher}{Association for Computing Machinery}, \bibinfo{address}{New York, NY, USA}, \bibinfo{pages}{143–153}.
\newblock
\showISBNx{9781450369954}
\urldef\tempurl%
\url{https://doi.org/10.1145/3359591.3359735}
\showDOI{\tempurl}


\bibitem[Alon et~al\mbox{.}(2019a)]%
        {code2seq}
\bibfield{author}{\bibinfo{person}{Uri Alon}, \bibinfo{person}{Omer Levy}, {and} \bibinfo{person}{Eran Yahav}.} \bibinfo{year}{2019}\natexlab{a}.
\newblock \showarticletitle{code2seq: Generating Sequences from Structured Representations of Code}. In \bibinfo{booktitle}{\emph{International Conference on Learning Representations}}.
\newblock
\urldef\tempurl%
\url{https://openreview.net/forum?id=H1gKYo09tX}
\showURL{%
\tempurl}


\bibitem[Alon et~al\mbox{.}(2019b)]%
        {code2vec}
\bibfield{author}{\bibinfo{person}{Uri Alon}, \bibinfo{person}{Meital Zilberstein}, \bibinfo{person}{Omer Levy}, {and} \bibinfo{person}{Eran Yahav}.} \bibinfo{year}{2019}\natexlab{b}.
\newblock \showarticletitle{Code2vec: Learning Distributed Representations of Code}.
\newblock \bibinfo{journal}{\emph{Proc. ACM Program. Lang.}} \bibinfo{volume}{3}, \bibinfo{number}{POPL}, Article \bibinfo{articleno}{40} (\bibinfo{date}{Jan.} \bibinfo{year}{2019}), \bibinfo{numpages}{29}~pages.
\newblock
\urldef\tempurl%
\url{https://doi.org/10.1145/3290353}
\showDOI{\tempurl}


\bibitem[Alsulami et~al\mbox{.}(2017)]%
        {alsulami_source_2017}
\bibfield{author}{\bibinfo{person}{Bander Alsulami}, \bibinfo{person}{Edwin Dauber}, \bibinfo{person}{Richard Harang}, \bibinfo{person}{Spiros Mancoridis}, {and} \bibinfo{person}{Rachel Greenstadt}.} \bibinfo{year}{2017}\natexlab{}.
\newblock \showarticletitle{Source code authorship attribution using long short-term memory based networks}. In \bibinfo{booktitle}{\emph{Computer {Security} - {ESORICS} 2017}}. \bibinfo{publisher}{Springer Verlag}, \bibinfo{pages}{65--82}.
\newblock
\urldef\tempurl%
\url{https://doi.org/10.1007/978-3-319-66402-6_6}
\showDOI{\tempurl}


\bibitem[Anand et~al\mbox{.}(2021)]%
        {anand2021adversarial}
\bibfield{author}{\bibinfo{person}{Mrinal Anand}, \bibinfo{person}{Pratik Kayal}, {and} \bibinfo{person}{Mayank Singh}.} \bibinfo{year}{2021}\natexlab{}.
\newblock \showarticletitle{Adversarial Robustness of Program Synthesis Models}. In \bibinfo{booktitle}{\emph{Advances in Programming Languages and Neurosymbolic Systems Workshop}}.
\newblock
\urldef\tempurl%
\url{https://openreview.net/forum?id=17C-dfA5X69}
\showURL{%
\tempurl}


\bibitem[Applis et~al\mbox{.}(2021)]%
        {9678706}
\bibfield{author}{\bibinfo{person}{Leonhard Applis}, \bibinfo{person}{Annibale Panichella}, {and} \bibinfo{person}{Arie van Deursen}.} \bibinfo{year}{2021}\natexlab{}.
\newblock \showarticletitle{Assessing Robustness of ML-Based Program Analysis Tools using Metamorphic Program Transformations}. In \bibinfo{booktitle}{\emph{ASE 2021}}. \bibinfo{pages}{1377--1381}.
\newblock
\urldef\tempurl%
\url{https://doi.org/10.1109/ASE51524.2021.9678706}
\showDOI{\tempurl}


\bibitem[Austin et~al\mbox{.}(2021)]%
        {austin2021program}
\bibfield{author}{\bibinfo{person}{Jacob Austin}, \bibinfo{person}{Augustus Odena}, \bibinfo{person}{Maxwell Nye}, \bibinfo{person}{Maarten Bosma}, \bibinfo{person}{Henryk Michalewski}, \bibinfo{person}{David Dohan}, \bibinfo{person}{Ellen Jiang}, \bibinfo{person}{Carrie Cai}, \bibinfo{person}{Michael Terry}, \bibinfo{person}{Quoc Le}, {and} \bibinfo{person}{Charles Sutton}.} \bibinfo{year}{2021}\natexlab{}.
\newblock \bibinfo{title}{Program Synthesis with Large Language Models}.
\newblock
\newblock
\showeprint[arxiv]{2108.07732}~[cs.PL]


\bibitem[Aye and Kaiser(2020)]%
        {aye2020sequence}
\bibfield{author}{\bibinfo{person}{Gareth~Ari Aye} {and} \bibinfo{person}{Gail~E. Kaiser}.} \bibinfo{year}{2020}\natexlab{}.
\newblock \bibinfo{title}{Sequence Model Design for Code Completion in the Modern IDE}.
\newblock
\newblock
\showeprint[arxiv]{2004.05249}~[cs.SE]


\bibitem[Ayupov and Chirkova(2022)]%
        {ayupov2022parameterefficient}
\bibfield{author}{\bibinfo{person}{Shamil Ayupov} {and} \bibinfo{person}{Nadezhda Chirkova}.} \bibinfo{year}{2022}\natexlab{}.
\newblock \bibinfo{title}{Parameter-Efficient Finetuning of Transformers for Source Code}.
\newblock
\newblock
\showeprint[arxiv]{2212.05901}~[cs.CL]


\bibitem[Basak et~al\mbox{.}(2023)]%
        {basak2023secretbench}
\bibfield{author}{\bibinfo{person}{Setu~Kumar Basak}, \bibinfo{person}{Lorenzo Neil}, \bibinfo{person}{Bradley Reaves}, {and} \bibinfo{person}{Laurie Williams}.} \bibinfo{year}{2023}\natexlab{}.
\newblock \showarticletitle{SecretBench: A Dataset of Software Secrets}. In \bibinfo{booktitle}{\emph{Proceedings of the 20th International Conference on Mining Software Repositories}} \emph{(\bibinfo{series}{MSR '23})}. \bibinfo{numpages}{5}~pages.
\newblock


\bibitem[Behnia et~al\mbox{.}(2022)]%
        {Behnia2022}
\bibfield{author}{\bibinfo{person}{Rouzbeh Behnia}, \bibinfo{person}{Mohammadreza~Reza Ebrahimi}, \bibinfo{person}{Jason Pacheco}, {and} \bibinfo{person}{Balaji Padmanabhan}.} \bibinfo{year}{2022}\natexlab{}.
\newblock \showarticletitle{EW-Tune: A Framework for Privately Fine-Tuning Large Language Models with Differential Privacy}. In \bibinfo{booktitle}{\emph{2022 IEEE International Conference on Data Mining Workshops (ICDMW)}}. \bibinfo{publisher}{IEEE}.
\newblock
\urldef\tempurl%
\url{https://doi.org/10.1109/icdmw58026.2022.00078}
\showDOI{\tempurl}


\bibitem[Bielik and Vechev(2020)]%
        {bielik2020adversarial}
\bibfield{author}{\bibinfo{person}{Pavol Bielik} {and} \bibinfo{person}{Martin Vechev}.} \bibinfo{year}{2020}\natexlab{}.
\newblock \showarticletitle{Adversarial robustness for code}. In \bibinfo{booktitle}{\emph{International Conference on Machine Learning}}. PMLR, \bibinfo{pages}{896--907}.
\newblock


\bibitem[Blinn et~al\mbox{.}(2022)]%
        {9833110}
\bibfield{author}{\bibinfo{person}{Andrew Blinn}, \bibinfo{person}{David Moon}, \bibinfo{person}{Eric Griffis}, {and} \bibinfo{person}{Cyrus Omar}.} \bibinfo{year}{2022}\natexlab{}.
\newblock \showarticletitle{An Integrative Human-Centered Architecture for Interactive Programming Assistants}. In \bibinfo{booktitle}{\emph{2022 IEEE Symposium on Visual Languages and Human-Centric Computing (VL/HCC)}}. \bibinfo{pages}{1--5}.
\newblock
\urldef\tempurl%
\url{https://doi.org/10.1109/VL/HCC53370.2022.9833110}
\showDOI{\tempurl}


\bibitem[Caliskan-Islam et~al\mbox{.}(2015)]%
        {10.5555/2831143.2831160}
\bibfield{author}{\bibinfo{person}{Aylin Caliskan-Islam}, \bibinfo{person}{Richard Harang}, \bibinfo{person}{Andrew Liu}, \bibinfo{person}{Arvind Narayanan}, \bibinfo{person}{Clare Voss}, \bibinfo{person}{Fabian Yamaguchi}, {and} \bibinfo{person}{Rachel Greenstadt}.} \bibinfo{year}{2015}\natexlab{}.
\newblock \showarticletitle{De-Anonymizing Programmers via Code Stylometry}. In \bibinfo{booktitle}{\emph{Proceedings of the 24th USENIX Conference on Security Symposium}} (Washington, D.C.) \emph{(\bibinfo{series}{SEC'15})}. \bibinfo{publisher}{USENIX Association}, \bibinfo{address}{USA}, \bibinfo{pages}{255–270}.
\newblock
\showISBNx{9781931971232}


\bibitem[Carlini et~al\mbox{.}(2021)]%
        {carlini21extracting}
\bibfield{author}{\bibinfo{person}{Nicholas Carlini}, \bibinfo{person}{Florian Tramer}, \bibinfo{person}{Eric Wallace}, \bibinfo{person}{Matthew Jagielski}, \bibinfo{person}{Ariel Herbert-Voss}, \bibinfo{person}{Katherine Lee}, \bibinfo{person}{Adam Roberts}, \bibinfo{person}{Tom Brown}, \bibinfo{person}{Dawn Song}, \bibinfo{person}{Ulfar Erlingsson}, \bibinfo{person}{Alina Oprea}, {and} \bibinfo{person}{Colin Raffel}.} \bibinfo{year}{2021}\natexlab{}.
\newblock \showarticletitle{Extracting Training Data from Large Language Models}. In \bibinfo{booktitle}{\emph{USENIX Security Symposium}}.
\newblock


\bibitem[Carlini and Wagner(2017)]%
        {7958570}
\bibfield{author}{\bibinfo{person}{Nicholas Carlini} {and} \bibinfo{person}{David Wagner}.} \bibinfo{year}{2017}\natexlab{}.
\newblock \showarticletitle{Towards Evaluating the Robustness of Neural Networks}. In \bibinfo{booktitle}{\emph{2017 IEEE Symposium on Security and Privacy (SP)}}. \bibinfo{publisher}{IEEE Computer Society}, \bibinfo{address}{Los Alamitos, CA, USA}, \bibinfo{pages}{39--57}.
\newblock
\showISSN{2375-1207}
\urldef\tempurl%
\url{https://doi.org/10.1109/SP.2017.49}
\showDOI{\tempurl}


\bibitem[Chakraborty et~al\mbox{.}(2022a)]%
        {10.1145/3540250.3549162}
\bibfield{author}{\bibinfo{person}{Saikat Chakraborty}, \bibinfo{person}{Toufique Ahmed}, \bibinfo{person}{Yangruibo Ding}, \bibinfo{person}{Premkumar~T. Devanbu}, {and} \bibinfo{person}{Baishakhi Ray}.} \bibinfo{year}{2022}\natexlab{a}.
\newblock \showarticletitle{NatGen: generative pre-training by “naturalizing” source code}. In \bibinfo{booktitle}{\emph{Proceedings of the 30th ACM Joint European Software Engineering Conference and Symposium on the Foundations of Software Engineering}} (<conf-loc>, <city>Singapore</city>, <country>Singapore</country>, </conf-loc>) \emph{(\bibinfo{series}{ESEC/FSE 2022})}. \bibinfo{publisher}{Association for Computing Machinery}, \bibinfo{address}{New York, NY, USA}, \bibinfo{pages}{18–30}.
\newblock
\showISBNx{9781450394130}
\urldef\tempurl%
\url{https://doi.org/10.1145/3540250.3549162}
\showDOI{\tempurl}


\bibitem[Chakraborty et~al\mbox{.}(2022b)]%
        {ReVeal}
\bibfield{author}{\bibinfo{person}{Saikat Chakraborty}, \bibinfo{person}{Rahul Krishna}, \bibinfo{person}{Yangruibo Ding}, {and} \bibinfo{person}{Baishakhi Ray}.} \bibinfo{year}{2022}\natexlab{b}.
\newblock \showarticletitle{Deep Learning Based Vulnerability Detection: Are We There Yet?}
\newblock \bibinfo{journal}{\emph{IEEE Transactions on Software Engineering}} \bibinfo{volume}{48}, \bibinfo{number}{09} (\bibinfo{date}{sep} \bibinfo{year}{2022}), \bibinfo{pages}{3280--3296}.
\newblock
\showISSN{1939-3520}
\urldef\tempurl%
\url{https://doi.org/10.1109/TSE.2021.3087402}
\showDOI{\tempurl}


\bibitem[Cheers et~al\mbox{.}(2021)]%
        {cheers2021evaluating}
\bibfield{author}{\bibinfo{person}{Hayden Cheers}, \bibinfo{person}{Yuqing Lin}, {and} \bibinfo{person}{Shamus~P Smith}.} \bibinfo{year}{2021}\natexlab{}.
\newblock \showarticletitle{Evaluating the robustness of source code plagiarism detection tools to pervasive plagiarism-hiding modifications}.
\newblock \bibinfo{journal}{\emph{Empirical Software Engineering}} \bibinfo{volume}{26}, \bibinfo{number}{5} (\bibinfo{year}{2021}), \bibinfo{pages}{83}.
\newblock


\bibitem[Chen et~al\mbox{.}(2018)]%
        {activation}
\bibfield{author}{\bibinfo{person}{Bryant Chen}, \bibinfo{person}{Wilka Carvalho}, \bibinfo{person}{Nathalie Baracaldo}, \bibinfo{person}{Heiko Ludwig}, \bibinfo{person}{Benjamin Edwards}, \bibinfo{person}{Taesung Lee}, \bibinfo{person}{Ian~M. Molloy}, {and} \bibinfo{person}{Biplav Srivastava}.} \bibinfo{year}{2018}\natexlab{}.
\newblock \showarticletitle{Detecting Backdoor Attacks on Deep Neural Networks by Activation Clustering}.
\newblock \bibinfo{journal}{\emph{CoRR}}  \bibinfo{volume}{abs/1811.03728} (\bibinfo{year}{2018}).
\newblock
\showeprint[arXiv]{1811.03728}
\urldef\tempurl%
\url{http://arxiv.org/abs/1811.03728}
\showURL{%
\tempurl}


\bibitem[Chen et~al\mbox{.}(2021a)]%
        {steal-rl}
\bibfield{author}{\bibinfo{person}{Kangjie Chen}, \bibinfo{person}{Shangwei Guo}, \bibinfo{person}{Tianwei Zhang}, \bibinfo{person}{Xiaofei Xie}, {and} \bibinfo{person}{Yang Liu}.} \bibinfo{year}{2021}\natexlab{a}.
\newblock \showarticletitle{Stealing Deep Reinforcement Learning Models for Fun and Profit}. In \bibinfo{booktitle}{\emph{Proceedings of the 2021 ACM Asia Conference on Computer and Communications Security}} (Virtual Event, Hong Kong) \emph{(\bibinfo{series}{ASIA CCS '21})}. \bibinfo{publisher}{Association for Computing Machinery}, \bibinfo{address}{New York, NY, USA}, \bibinfo{pages}{307–319}.
\newblock
\showISBNx{9781450382878}
\urldef\tempurl%
\url{https://doi.org/10.1145/3433210.3453090}
\showDOI{\tempurl}


\bibitem[Chen et~al\mbox{.}(2021b)]%
        {codex}
\bibfield{author}{\bibinfo{person}{Mark Chen}, \bibinfo{person}{Jerry Tworek}, {and} \bibinfo{person}{Heewoo~Jun et al.}} \bibinfo{year}{2021}\natexlab{b}.
\newblock \showarticletitle{Evaluating Large Language Models Trained on Code}.
\newblock \bibinfo{journal}{\emph{CoRR}} (\bibinfo{year}{2021}).
\newblock


\bibitem[Chen et~al\mbox{.}(2022a)]%
        {9763729}
\bibfield{author}{\bibinfo{person}{Penglong Chen}, \bibinfo{person}{Zhen Li}, \bibinfo{person}{Yu Wen}, {and} \bibinfo{person}{Lili Liu}.} \bibinfo{year}{2022}\natexlab{a}.
\newblock \showarticletitle{Generating Adversarial Source Programs Using Important Tokens-based Structural Transformations}. In \bibinfo{booktitle}{\emph{2022 26th International Conference on Engineering of Complex Computer Systems (ICECCS)}}. \bibinfo{pages}{173--182}.
\newblock
\urldef\tempurl%
\url{https://doi.org/10.1109/ICECCS54210.2022.00029}
\showDOI{\tempurl}


\bibitem[Chen et~al\mbox{.}(2022b)]%
        {chen2022fairness}
\bibfield{author}{\bibinfo{person}{Zhenpeng Chen}, \bibinfo{person}{Jie~M Zhang}, \bibinfo{person}{Max Hort}, \bibinfo{person}{Federica Sarro}, {and} \bibinfo{person}{Mark Harman}.} \bibinfo{year}{2022}\natexlab{b}.
\newblock \showarticletitle{Fairness testing: A comprehensive survey and analysis of trends}.
\newblock  (\bibinfo{year}{2022}).
\newblock


\bibitem[Choi et~al\mbox{.}(2022)]%
        {choi-etal-2022-tabs}
\bibfield{author}{\bibinfo{person}{YunSeok Choi}, \bibinfo{person}{Hyojun Kim}, {and} \bibinfo{person}{Jee-Hyong Lee}.} \bibinfo{year}{2022}\natexlab{}.
\newblock \showarticletitle{{TABS}: Efficient Textual Adversarial Attack for Pre-trained {NL} Code Model Using Semantic Beam Search}. In \bibinfo{booktitle}{\emph{Proceedings of the 2022 Conference on Empirical Methods in Natural Language Processing}}. \bibinfo{publisher}{Association for Computational Linguistics}, \bibinfo{address}{Abu Dhabi, United Arab Emirates}, \bibinfo{pages}{5490--5498}.
\newblock
\urldef\tempurl%
\url{https://doi.org/10.18653/v1/2022.emnlp-main.369}
\showDOI{\tempurl}


\bibitem[Ciniselli et~al\mbox{.}(2022)]%
        {code-model-clone-msr22}
\bibfield{author}{\bibinfo{person}{Matteo Ciniselli}, \bibinfo{person}{Luca Pascarella}, {and} \bibinfo{person}{Gabriele Bavota}.} \bibinfo{year}{2022}\natexlab{}.
\newblock \showarticletitle{To What Extent Do Deep Learning-Based Code Recommenders Generate Predictions by Cloning Code from the Training Set?}. In \bibinfo{booktitle}{\emph{Proceedings of the 19th International Conference on Mining Software Repositories}} (Pittsburgh, Pennsylvania) \emph{(\bibinfo{series}{MSR '22})}. \bibinfo{publisher}{Association for Computing Machinery}, \bibinfo{address}{New York, NY, USA}, \bibinfo{pages}{167–178}.
\newblock
\showISBNx{9781450393034}
\urldef\tempurl%
\url{https://doi.org/10.1145/3524842.3528440}
\showDOI{\tempurl}


\bibitem[Cito et~al\mbox{.}(2022)]%
        {10.1145/3510457.3513081}
\bibfield{author}{\bibinfo{person}{J\"{u}rgen Cito}, \bibinfo{person}{Isil Dillig}, \bibinfo{person}{Vijayaraghavan Murali}, {and} \bibinfo{person}{Satish Chandra}.} \bibinfo{year}{2022}\natexlab{}.
\newblock \showarticletitle{Counterfactual Explanations for Models of Code}. In \bibinfo{booktitle}{\emph{Proceedings of the 44th International Conference on Software Engineering: Software Engineering in Practice}} (Pittsburgh, Pennsylvania) \emph{(\bibinfo{series}{ICSE-SEIP '22})}. \bibinfo{publisher}{Association for Computing Machinery}, \bibinfo{address}{New York, NY, USA}, \bibinfo{pages}{125–134}.
\newblock
\showISBNx{9781450392266}
\urldef\tempurl%
\url{https://doi.org/10.1145/3510457.3513081}
\showDOI{\tempurl}


\bibitem[Cotroneo et~al\mbox{.}(2023)]%
        {cotroneo2023vulnerabilities}
\bibfield{author}{\bibinfo{person}{Domenico Cotroneo}, \bibinfo{person}{Cristina Improta}, \bibinfo{person}{Pietro Liguori}, {and} \bibinfo{person}{Roberto Natella}.} \bibinfo{year}{2023}\natexlab{}.
\newblock \bibinfo{title}{Vulnerabilities in AI Code Generators: Exploring Targeted Data Poisoning Attacks}.
\newblock
\newblock
\showeprint[arxiv]{2308.04451}~[cs.CR]


\bibitem[Dam\'{a}sio et~al\mbox{.}(2023)]%
        {10.1145/3579990.3580012}
\bibfield{author}{\bibinfo{person}{Tha\'{\i}s Dam\'{a}sio}, \bibinfo{person}{Michael Canesche}, \bibinfo{person}{Vin\'{\i}cius Pacheco}, \bibinfo{person}{Marcus Botacin}, \bibinfo{person}{Anderson Faustino~da Silva}, {and} \bibinfo{person}{Fernando~M. Quint\~{a}o Pereira}.} \bibinfo{year}{2023}\natexlab{}.
\newblock \showarticletitle{A Game-Based Framework to Compare Program Classifiers and Evaders}. In \bibinfo{booktitle}{\emph{Proceedings of the 21st ACM/IEEE International Symposium on Code Generation and Optimization}} (Montr\'{e}al, QC, Canada) \emph{(\bibinfo{series}{CGO 2023})}. \bibinfo{publisher}{Association for Computing Machinery}, \bibinfo{address}{New York, NY, USA}, \bibinfo{pages}{108–121}.
\newblock
\showISBNx{9798400701016}
\urldef\tempurl%
\url{https://doi.org/10.1145/3579990.3580012}
\showDOI{\tempurl}


\bibitem[Dettmers et~al\mbox{.}(2023)]%
        {dettmers2023qlora}
\bibfield{author}{\bibinfo{person}{Tim Dettmers}, \bibinfo{person}{Artidoro Pagnoni}, \bibinfo{person}{Ari Holtzman}, {and} \bibinfo{person}{Luke Zettlemoyer}.} \bibinfo{year}{2023}\natexlab{}.
\newblock \bibinfo{title}{QLoRA: Efficient Finetuning of Quantized LLMs}.
\newblock
\newblock


\bibitem[Ding and Samadzadeh(2004)]%
        {DING200449}
\bibfield{author}{\bibinfo{person}{Haibiao Ding} {and} \bibinfo{person}{Mansur~H. Samadzadeh}.} \bibinfo{year}{2004}\natexlab{}.
\newblock \showarticletitle{Extraction of Java program fingerprints for software authorship identification}.
\newblock \bibinfo{journal}{\emph{Journal of Systems and Software}} \bibinfo{volume}{72}, \bibinfo{number}{1} (\bibinfo{year}{2004}), \bibinfo{pages}{49--57}.
\newblock
\showISSN{0164-1212}
\urldef\tempurl%
\url{https://doi.org/10.1016/S0164-1212(03)00049-9}
\showDOI{\tempurl}


\bibitem[Druga and Ko(2023)]%
        {druga2023ai}
\bibfield{author}{\bibinfo{person}{Stefania Druga} {and} \bibinfo{person}{Amy~J. Ko}.} \bibinfo{year}{2023}\natexlab{}.
\newblock \bibinfo{title}{AI Friends: A Design Framework for AI-Powered Creative Programming for Youth}.
\newblock
\newblock
\showeprint[arxiv]{2305.10412}~[cs.HC]


\bibitem[Druga and Otero(2023)]%
        {druga2023scratch}
\bibfield{author}{\bibinfo{person}{Stefania Druga} {and} \bibinfo{person}{Nancy Otero}.} \bibinfo{year}{2023}\natexlab{}.
\newblock \bibinfo{title}{Scratch Copilot Evaluation: Assessing AI-Assisted Creative Coding for Families}.
\newblock
\newblock
\showeprint[arxiv]{2305.10417}~[cs.HC]


\bibitem[Du et~al\mbox{.}(2020)]%
        {10.1145/3427228.3427258}
\bibfield{author}{\bibinfo{person}{Kun Du}, \bibinfo{person}{Hao Yang}, \bibinfo{person}{Yubao Zhang}, \bibinfo{person}{Haixin Duan}, \bibinfo{person}{Haining Wang}, \bibinfo{person}{Shuang Hao}, \bibinfo{person}{Zhou Li}, {and} \bibinfo{person}{Min Yang}.} \bibinfo{year}{2020}\natexlab{}.
\newblock \showarticletitle{Understanding Promotion-as-a-Service on GitHub}. In \bibinfo{booktitle}{\emph{Proceedings of the 36th Annual Computer Security Applications Conference}} \emph{(\bibinfo{series}{ACSAC '20})}. \bibinfo{publisher}{Association for Computing Machinery}, \bibinfo{address}{New York, NY, USA}, \bibinfo{pages}{597–610}.
\newblock
\showISBNx{9781450388580}
\urldef\tempurl%
\url{https://doi.org/10.1145/3427228.3427258}
\showDOI{\tempurl}


\bibitem[Du et~al\mbox{.}(2023)]%
        {du2023extensive}
\bibfield{author}{\bibinfo{person}{Xiaohu Du}, \bibinfo{person}{Ming Wen}, \bibinfo{person}{Zichao Wei}, \bibinfo{person}{Shangwen Wang}, {and} \bibinfo{person}{Hai Jin}.} \bibinfo{year}{2023}\natexlab{}.
\newblock \showarticletitle{An Extensive Study on Adversarial Attack against Pre-trained Models of Code}.
\newblock  (\bibinfo{year}{2023}).
\newblock
\showeprint[arxiv]{2311.07553}~[cs.CR]


\bibitem[Fan et~al\mbox{.}(2023)]%
        {fan2023large}
\bibfield{author}{\bibinfo{person}{Angela Fan}, \bibinfo{person}{Beliz Gokkaya}, \bibinfo{person}{Mark Harman}, \bibinfo{person}{Mitya Lyubarskiy}, \bibinfo{person}{Shubho Sengupta}, \bibinfo{person}{Shin Yoo}, {and} \bibinfo{person}{Jie~M. Zhang}.} \bibinfo{year}{2023}\natexlab{}.
\newblock \bibinfo{title}{Large Language Models for Software Engineering: Survey and Open Problems}.
\newblock
\newblock
\showeprint[arxiv]{2310.03533}~[cs.SE]


\bibitem[Feng et~al\mbox{.}(2022)]%
        {9794113}
\bibfield{author}{\bibinfo{person}{Runhan Feng}, \bibinfo{person}{Ziyang Yan}, \bibinfo{person}{Shiyan Peng}, {and} \bibinfo{person}{Yuanyuan Zhang}.} \bibinfo{year}{2022}\natexlab{}.
\newblock \showarticletitle{Automated Detection of Password Leakage from Public GitHub Repositories}. In \bibinfo{booktitle}{\emph{2022 IEEE/ACM 44th International Conference on Software Engineering (ICSE)}}. \bibinfo{pages}{175--186}.
\newblock
\urldef\tempurl%
\url{https://doi.org/10.1145/3510003.3510150}
\showDOI{\tempurl}


\bibitem[Feng et~al\mbox{.}(2020)]%
        {CodeBERT}
\bibfield{author}{\bibinfo{person}{Zhangyin Feng}, \bibinfo{person}{Daya Guo}, \bibinfo{person}{Duyu Tang}, \bibinfo{person}{Nan Duan}, \bibinfo{person}{Xiaocheng Feng}, \bibinfo{person}{Ming Gong}, \bibinfo{person}{Linjun Shou}, \bibinfo{person}{Bing Qin}, \bibinfo{person}{Ting Liu}, \bibinfo{person}{Daxin Jiang}, {and} \bibinfo{person}{Ming Zhou}.} \bibinfo{year}{2020}\natexlab{}.
\newblock \showarticletitle{{C}ode{BERT}: A Pre-Trained Model for Programming and Natural Languages}. In \bibinfo{booktitle}{\emph{Findings of the Association for Computational Linguistics: EMNLP 2020}}. \bibinfo{publisher}{Association for Computational Linguistics}, \bibinfo{pages}{1536--1547}.
\newblock


\bibitem[Ferretti and Saletta(2021)]%
        {10.1145/3449726.3463222}
\bibfield{author}{\bibinfo{person}{Claudio Ferretti} {and} \bibinfo{person}{Martina Saletta}.} \bibinfo{year}{2021}\natexlab{}.
\newblock \showarticletitle{Deceiving neural source code classifiers: finding adversarial examples with grammatical evolution}. In \bibinfo{booktitle}{\emph{Proceedings of the Genetic and Evolutionary Computation Conference Companion}} (Lille, France) \emph{(\bibinfo{series}{GECCO '21})}. \bibinfo{publisher}{Association for Computing Machinery}, \bibinfo{address}{New York, NY, USA}, \bibinfo{pages}{1889–1897}.
\newblock
\showISBNx{9781450383516}
\urldef\tempurl%
\url{https://doi.org/10.1145/3449726.3463222}
\showDOI{\tempurl}


\bibitem[Finegan-Dollak et~al\mbox{.}(2018)]%
        {finegan-dollak-etal-2018-improving}
\bibfield{author}{\bibinfo{person}{Catherine Finegan-Dollak}, \bibinfo{person}{Jonathan~K. Kummerfeld}, \bibinfo{person}{Li Zhang}, \bibinfo{person}{Karthik Ramanathan}, \bibinfo{person}{Sesh Sadasivam}, \bibinfo{person}{Rui Zhang}, {and} \bibinfo{person}{Dragomir Radev}.} \bibinfo{year}{2018}\natexlab{}.
\newblock \showarticletitle{Improving Text-to-{SQL} Evaluation Methodology}. In \bibinfo{booktitle}{\emph{Proceedings of the 56th Annual Meeting of the Association for Computational Linguistics (Volume 1: Long Papers)}}, \bibfield{editor}{\bibinfo{person}{Iryna Gurevych} {and} \bibinfo{person}{Yusuke Miyao}} (Eds.). \bibinfo{publisher}{Association for Computational Linguistics}, \bibinfo{address}{Melbourne, Australia}, \bibinfo{pages}{351--360}.
\newblock
\urldef\tempurl%
\url{https://doi.org/10.18653/v1/P18-1033}
\showDOI{\tempurl}


\bibitem[Finnie-Ansley et~al\mbox{.}(2022)]%
        {finnie2022robots}
\bibfield{author}{\bibinfo{person}{James Finnie-Ansley}, \bibinfo{person}{Paul Denny}, \bibinfo{person}{Brett~A Becker}, \bibinfo{person}{Andrew Luxton-Reilly}, {and} \bibinfo{person}{James Prather}.} \bibinfo{year}{2022}\natexlab{}.
\newblock \showarticletitle{The robots are coming: Exploring the implications of openai codex on introductory programming}. In \bibinfo{booktitle}{\emph{Proceedings of the 24th Australasian Computing Education Conference}}. \bibinfo{pages}{10--19}.
\newblock


\bibitem[Fitria(2021)]%
        {fitria2021quillbot}
\bibfield{author}{\bibinfo{person}{Tira~Nur Fitria}.} \bibinfo{year}{2021}\natexlab{}.
\newblock \showarticletitle{QuillBot as an online tool: Students’ alternative in paraphrasing and rewriting of English writing}.
\newblock \bibinfo{journal}{\emph{Englisia: Journal of Language, Education, and Humanities}} \bibinfo{volume}{9}, \bibinfo{number}{1} (\bibinfo{year}{2021}), \bibinfo{pages}{183--196}.
\newblock


\bibitem[Frantzeskou et~al\mbox{.}(2006)]%
        {10.1007/0-387-34224-9_59}
\bibfield{author}{\bibinfo{person}{Georgia Frantzeskou}, \bibinfo{person}{Efstathios Stamatatos}, \bibinfo{person}{Stefanos Gritzalis}, {and} \bibinfo{person}{Sokratis Katsikas}.} \bibinfo{year}{2006}\natexlab{}.
\newblock \showarticletitle{Source Code Author Identification Based on N-gram Author Profiles}. In \bibinfo{booktitle}{\emph{Artificial Intelligence Applications and Innovations}}, \bibfield{editor}{\bibinfo{person}{Ilias Maglogiannis}, \bibinfo{person}{Kostas Karpouzis}, {and} \bibinfo{person}{Max Bramer}} (Eds.). \bibinfo{publisher}{Springer US}, \bibinfo{address}{Boston, MA}, \bibinfo{pages}{508--515}.
\newblock
\showISBNx{978-0-387-34224-5}


\bibitem[Fried et~al\mbox{.}(2023)]%
        {fried2023incoder}
\bibfield{author}{\bibinfo{person}{Daniel Fried}, \bibinfo{person}{Armen Aghajanyan}, \bibinfo{person}{Jessy Lin}, \bibinfo{person}{Sida Wang}, \bibinfo{person}{Eric Wallace}, \bibinfo{person}{Freda Shi}, \bibinfo{person}{Ruiqi Zhong}, \bibinfo{person}{Scott Yih}, \bibinfo{person}{Luke Zettlemoyer}, {and} \bibinfo{person}{Mike Lewis}.} \bibinfo{year}{2023}\natexlab{}.
\newblock \showarticletitle{InCoder: A Generative Model for Code Infilling and Synthesis}. In \bibinfo{booktitle}{\emph{The Eleventh International Conference on Learning Representations}}.
\newblock
\urldef\tempurl%
\url{https://openreview.net/forum?id=hQwb-lbM6EL}
\showURL{%
\tempurl}


\bibitem[Ganz et~al\mbox{.}(2023)]%
        {10190498}
\bibfield{author}{\bibinfo{person}{Tom Ganz}, \bibinfo{person}{Philipp Rall}, \bibinfo{person}{Martin Härterich}, {and} \bibinfo{person}{Konrad Rieck}.} \bibinfo{year}{2023}\natexlab{}.
\newblock \showarticletitle{Hunting for Truth: Analyzing Explanation Methods in Learning-based Vulnerability Discovery}. In \bibinfo{booktitle}{\emph{2023 IEEE 8th European Symposium on Security and Privacy (EuroS\&P)}}. \bibinfo{pages}{524--541}.
\newblock
\urldef\tempurl%
\url{https://doi.org/10.1109/EuroSP57164.2023.00038}
\showDOI{\tempurl}


\bibitem[Gao et~al\mbox{.}(2023)]%
        {10.1145/3591227}
\bibfield{author}{\bibinfo{person}{Fengjuan Gao}, \bibinfo{person}{Yu Wang}, {and} \bibinfo{person}{Ke Wang}.} \bibinfo{year}{2023}\natexlab{}.
\newblock \showarticletitle{Discrete Adversarial Attack to Models of Code}.
\newblock \bibinfo{journal}{\emph{Proc. ACM Program. Lang.}} \bibinfo{volume}{7}, \bibinfo{number}{PLDI}, Article \bibinfo{articleno}{113} (\bibinfo{date}{jun} \bibinfo{year}{2023}), \bibinfo{numpages}{24}~pages.
\newblock
\urldef\tempurl%
\url{https://doi.org/10.1145/3591227}
\showDOI{\tempurl}


\bibitem[Goodfellow et~al\mbox{.}(2015)]%
        {FGSM}
\bibfield{author}{\bibinfo{person}{Ian Goodfellow}, \bibinfo{person}{Jonathon Shlens}, {and} \bibinfo{person}{Christian Szegedy}.} \bibinfo{year}{2015}\natexlab{}.
\newblock \showarticletitle{Explaining and Harnessing Adversarial Examples}. In \bibinfo{booktitle}{\emph{International Conference on Learning Representations}}.
\newblock
\urldef\tempurl%
\url{http://arxiv.org/abs/1412.6572}
\showURL{%
\tempurl}


\bibitem[Gordon et~al\mbox{.}(2020)]%
        {gordon2020compressing}
\bibfield{author}{\bibinfo{person}{Mitchell~A Gordon}, \bibinfo{person}{Kevin Duh}, {and} \bibinfo{person}{Nicholas Andrews}.} \bibinfo{year}{2020}\natexlab{}.
\newblock \showarticletitle{Compressing bert: Studying the effects of weight pruning on transfer learning}.
\newblock \bibinfo{journal}{\emph{arXiv preprint arXiv:2002.08307}} (\bibinfo{year}{2020}).
\newblock


\bibitem[Grishina et~al\mbox{.}(2023)]%
        {10.1145/3611643.3616304}
\bibfield{author}{\bibinfo{person}{Anastasiia Grishina}, \bibinfo{person}{Max Hort}, {and} \bibinfo{person}{Leon Moonen}.} \bibinfo{year}{2023}\natexlab{}.
\newblock \showarticletitle{The EarlyBIRD Catches the Bug: On Exploiting Early Layers of Encoder Models for More Efficient Code Classification}. In \bibinfo{booktitle}{\emph{Proceedings of the 31st ACM Joint European Software Engineering Conference and Symposium on the Foundations of Software Engineering}} \emph{(\bibinfo{series}{ESEC/FSE 2023})}. \bibinfo{publisher}{Association for Computing Machinery}, \bibinfo{address}{New York, NY, USA}, \bibinfo{pages}{895–907}.
\newblock
\showISBNx{9798400703270}
\urldef\tempurl%
\url{https://doi.org/10.1145/3611643.3616304}
\showDOI{\tempurl}


\bibitem[Guo et~al\mbox{.}(2021)]%
        {GraphCodeBERT}
\bibfield{author}{\bibinfo{person}{Daya Guo}, \bibinfo{person}{Shuo Ren}, \bibinfo{person}{Shuai Lu}, \bibinfo{person}{Zhangyin Feng}, \bibinfo{person}{Duyu Tang}, \bibinfo{person}{Shujie Liu}, \bibinfo{person}{Long Zhou}, \bibinfo{person}{Nan Duan}, \bibinfo{person}{Alexey Svyatkovskiy}, \bibinfo{person}{Shengyu~Fu andz Michele~Tufano}, \bibinfo{person}{Shao~Kun Deng}, \bibinfo{person}{Colin~B. Clement}, \bibinfo{person}{Dawn Drain}, \bibinfo{person}{Neel Sundaresan}, \bibinfo{person}{Jian Yin}, \bibinfo{person}{Daxin Jiang}, {and} \bibinfo{person}{Ming Zhou}.} \bibinfo{year}{2021}\natexlab{}.
\newblock \showarticletitle{GraphCodeBERT: Pre-training Code Representations with Data Flow}. In \bibinfo{booktitle}{\emph{9th International Conference on Learning Representations, {ICLR} 2021, Virtual Event, Austria, May 3-7, 2021}}.
\newblock


\bibitem[Hajipour et~al\mbox{.}(2023)]%
        {hajipour2023systematically}
\bibfield{author}{\bibinfo{person}{Hossein Hajipour}, \bibinfo{person}{Keno Hassler}, \bibinfo{person}{Thorsten Holz}, \bibinfo{person}{Lea Schönherr}, {and} \bibinfo{person}{Mario Fritz}.} \bibinfo{year}{2023}\natexlab{}.
\newblock \bibinfo{title}{CodeLMSec Benchmark: Systematically Evaluating and Finding Security Vulnerabilities in Black-Box Code Language Models}.
\newblock
\newblock
\showeprint[arxiv]{2302.04012}~[cs.CR]


\bibitem[He and Vechev(2023)]%
        {he2023large}
\bibfield{author}{\bibinfo{person}{Jingxuan He} {and} \bibinfo{person}{Martin Vechev}.} \bibinfo{year}{2023}\natexlab{}.
\newblock \bibinfo{title}{Large Language Models for Code: Security Hardening and Adversarial Testing}.
\newblock
\newblock
\showeprint[arxiv]{2302.05319}~[cs.CR]


\bibitem[Hellendoorn and Sawant(2021)]%
        {10.1145/3501261}
\bibfield{author}{\bibinfo{person}{Vincent~J. Hellendoorn} {and} \bibinfo{person}{Anand~Ashok Sawant}.} \bibinfo{year}{2021}\natexlab{}.
\newblock \showarticletitle{The Growing Cost of Deep Learning for Source Code}.
\newblock \bibinfo{journal}{\emph{Commun. ACM}} \bibinfo{volume}{65}, \bibinfo{number}{1} (\bibinfo{date}{dec} \bibinfo{year}{2021}), \bibinfo{pages}{31–33}.
\newblock
\showISSN{0001-0782}
\urldef\tempurl%
\url{https://doi.org/10.1145/3501261}
\showDOI{\tempurl}


\bibitem[Hendrycks et~al\mbox{.}(2021)]%
        {apps}
\bibfield{author}{\bibinfo{person}{Dan Hendrycks}, \bibinfo{person}{Steven Basart}, \bibinfo{person}{Saurav Kadavath}, \bibinfo{person}{Mantas Mazeika}, \bibinfo{person}{Akul Arora}, \bibinfo{person}{Ethan Guo}, \bibinfo{person}{Collin Burns}, \bibinfo{person}{Samir Puranik}, \bibinfo{person}{Horace He}, \bibinfo{person}{Dawn Song}, {and} \bibinfo{person}{Jacob Steinhardt}.} \bibinfo{year}{2021}\natexlab{}.
\newblock \bibinfo{title}{Measuring Coding Challenge Competence With APPS}.
\newblock
\newblock
\showeprint[arxiv]{2105.09938}~[cs.SE]


\bibitem[Henkel et~al\mbox{.}(2022)]%
        {9825895}
\bibfield{author}{\bibinfo{person}{Jordan Henkel}, \bibinfo{person}{Goutham Ramakrishnan}, \bibinfo{person}{Zi Wang}, \bibinfo{person}{Aws Albarghouthi}, \bibinfo{person}{Somesh Jha}, {and} \bibinfo{person}{Thomas Reps}.} \bibinfo{year}{2022}\natexlab{}.
\newblock \showarticletitle{Semantic Robustness of Models of Source Code}. In \bibinfo{booktitle}{\emph{2022 IEEE International Conference on Software Analysis, Evolution and Reengineering (SANER)}}. \bibinfo{pages}{526--537}.
\newblock
\urldef\tempurl%
\url{https://doi.org/10.1109/SANER53432.2022.00070}
\showDOI{\tempurl}


\bibitem[Hidvégi et~al\mbox{.}(2024)]%
        {hidvégi2024cigar}
\bibfield{author}{\bibinfo{person}{Dávid Hidvégi}, \bibinfo{person}{Khashayar Etemadi}, \bibinfo{person}{Sofia Bobadilla}, {and} \bibinfo{person}{Martin Monperrus}.} \bibinfo{year}{2024}\natexlab{}.
\newblock \bibinfo{title}{CigaR: Cost-efficient Program Repair with LLMs}.
\newblock
\newblock


\bibitem[Hooda et~al\mbox{.}(2024)]%
        {hooda2024large}
\bibfield{author}{\bibinfo{person}{Ashish Hooda}, \bibinfo{person}{Mihai Christodorescu}, \bibinfo{person}{Miltiadis Allamanis}, \bibinfo{person}{Aaron Wilson}, \bibinfo{person}{Kassem Fawaz}, {and} \bibinfo{person}{Somesh Jha}.} \bibinfo{year}{2024}\natexlab{}.
\newblock \bibinfo{title}{Do Large Code Models Understand Programming Concepts? A Black-box Approach}.
\newblock
\newblock
\showeprint[arxiv]{2402.05980}~[cs.SE]


\bibitem[Hou et~al\mbox{.}(2023)]%
        {codellm_survey}
\bibfield{author}{\bibinfo{person}{Xinyi Hou}, \bibinfo{person}{Yanjie Zhao}, \bibinfo{person}{Yue Liu}, \bibinfo{person}{Zhou Yang}, \bibinfo{person}{Kailong Wang}, \bibinfo{person}{Li Li}, \bibinfo{person}{Xiapu Luo}, \bibinfo{person}{David Lo}, \bibinfo{person}{John Grundy}, {and} \bibinfo{person}{Haoyu Wang}.} \bibinfo{year}{2023}\natexlab{}.
\newblock \bibinfo{title}{Large Language Models for Software Engineering: A Systematic Literature Review}.
\newblock
\newblock
\showeprint[arxiv]{2308.10620}~[cs.SE]


\bibitem[Houlsby et~al\mbox{.}(2019)]%
        {houlsby2019parameter}
\bibfield{author}{\bibinfo{person}{Neil Houlsby}, \bibinfo{person}{Andrei Giurgiu}, \bibinfo{person}{Stanislaw Jastrzebski}, \bibinfo{person}{Bruna Morrone}, \bibinfo{person}{Quentin De~Laroussilhe}, \bibinfo{person}{Andrea Gesmundo}, \bibinfo{person}{Mona Attariyan}, {and} \bibinfo{person}{Sylvain Gelly}.} \bibinfo{year}{2019}\natexlab{}.
\newblock \showarticletitle{Parameter-efficient transfer learning for NLP}. In \bibinfo{booktitle}{\emph{International Conference on Machine Learning}}. PMLR, \bibinfo{pages}{2790--2799}.
\newblock


\bibitem[Hu et~al\mbox{.}(2022)]%
        {hu2022lora}
\bibfield{author}{\bibinfo{person}{Edward~J Hu}, \bibinfo{person}{yelong shen}, \bibinfo{person}{Phillip Wallis}, \bibinfo{person}{Zeyuan Allen-Zhu}, \bibinfo{person}{Yuanzhi Li}, \bibinfo{person}{Shean Wang}, \bibinfo{person}{Lu Wang}, {and} \bibinfo{person}{Weizhu Chen}.} \bibinfo{year}{2022}\natexlab{}.
\newblock \showarticletitle{Lo{RA}: Low-Rank Adaptation of Large Language Models}. In \bibinfo{booktitle}{\emph{International Conference on Learning Representations}}.
\newblock
\urldef\tempurl%
\url{https://openreview.net/forum?id=nZeVKeeFYf9}
\showURL{%
\tempurl}


\bibitem[Hu et~al\mbox{.}(2023)]%
        {hu2023active}
\bibfield{author}{\bibinfo{person}{Qiang Hu}, \bibinfo{person}{Yuejun Guo}, \bibinfo{person}{Xiaofei Xie}, \bibinfo{person}{Maxime Cordy}, \bibinfo{person}{Lei Ma}, \bibinfo{person}{Mike Papadakis}, {and} \bibinfo{person}{Yves~Le Traon}.} \bibinfo{year}{2023}\natexlab{}.
\newblock \bibinfo{title}{Active Code Learning: Benchmarking Sample-Efficient Training of Code Models}.
\newblock
\newblock
\showeprint[arxiv]{2306.01250}~[cs.SE]


\bibitem[Hu et~al\mbox{.}(2018)]%
        {10.5555/3304889.3304975}
\bibfield{author}{\bibinfo{person}{Xing Hu}, \bibinfo{person}{Ge Li}, \bibinfo{person}{Xin Xia}, \bibinfo{person}{David Lo}, \bibinfo{person}{Shuai Lu}, {and} \bibinfo{person}{Zhi Jin}.} \bibinfo{year}{2018}\natexlab{}.
\newblock \showarticletitle{Summarizing Source Code with Transferred API Knowledge}. In \bibinfo{booktitle}{\emph{Proceedings of the 27th International Joint Conference on Artificial Intelligence}} (Stockholm, Sweden) \emph{(\bibinfo{series}{IJCAI'18})}. \bibinfo{publisher}{AAAI Press}, \bibinfo{pages}{2269–2275}.
\newblock
\showISBNx{9780999241127}


\bibitem[Huang et~al\mbox{.}(2023)]%
        {huang2023away}
\bibfield{author}{\bibinfo{person}{Yizhan Huang}, \bibinfo{person}{Yichen Li}, \bibinfo{person}{Weibin Wu}, \bibinfo{person}{Jianping Zhang}, {and} \bibinfo{person}{Michael~R. Lyu}.} \bibinfo{year}{2023}\natexlab{}.
\newblock \bibinfo{title}{Do Not Give Away My Secrets: Uncovering the Privacy Issue of Neural Code Completion Tools}.
\newblock
\newblock
\showeprint[arxiv]{2309.07639}~[cs.CR]


\bibitem[Huber et~al\mbox{.}(2023)]%
        {huber2023look}
\bibfield{author}{\bibinfo{person}{Dominik Huber}, \bibinfo{person}{Matteo Paltenghi}, {and} \bibinfo{person}{Michael Pradel}.} \bibinfo{year}{2023}\natexlab{}.
\newblock \bibinfo{title}{Where to Look When Repairing Code? Comparing the Attention of Neural Models and Developers}.
\newblock
\newblock
\showeprint[arxiv]{2305.07287}~[cs.SE]


\bibitem[Husain et~al\mbox{.}(2019)]%
        {husain2019codesearchnet}
\bibfield{author}{\bibinfo{person}{Hamel Husain}, \bibinfo{person}{Ho-Hsiang Wu}, \bibinfo{person}{Tiferet Gazit}, \bibinfo{person}{Miltiadis Allamanis}, {and} \bibinfo{person}{Marc Brockschmidt}.} \bibinfo{year}{2019}\natexlab{}.
\newblock \showarticletitle{{CodeSearchNet} challenge: Evaluating the state of semantic code search}.
\newblock \bibinfo{journal}{\emph{arXiv preprint arXiv:1909.09436}} (\bibinfo{year}{2019}).
\newblock


\bibitem[Hussain et~al\mbox{.}(2023b)]%
        {hussain2023occlusionbased}
\bibfield{author}{\bibinfo{person}{Aftab Hussain}, \bibinfo{person}{Md~Rafiqul~Islam Rabin}, \bibinfo{person}{Toufique Ahmed}, \bibinfo{person}{Mohammad~Amin Alipour}, {and} \bibinfo{person}{Bowen Xu}.} \bibinfo{year}{2023}\natexlab{b}.
\newblock \bibinfo{title}{Occlusion-based Detection of Trojan-triggering Inputs in Large Language Models of Code}.
\newblock
\newblock
\showeprint[arxiv]{2312.04004}~[cs.SE]


\bibitem[Hussain et~al\mbox{.}(2023c)]%
        {hussain2023survey}
\bibfield{author}{\bibinfo{person}{Aftab Hussain}, \bibinfo{person}{Md~Rafiqul~Islam Rabin}, \bibinfo{person}{Toufique Ahmed}, \bibinfo{person}{Navid Ayoobi}, \bibinfo{person}{Bowen Xu}, \bibinfo{person}{Prem Devanbu}, {and} \bibinfo{person}{Mohammad~Amin Alipour}.} \bibinfo{year}{2023}\natexlab{c}.
\newblock \bibinfo{title}{A Survey of Trojans in Neural Models of Source Code: Taxonomy and Techniques}.
\newblock
\newblock
\showeprint[arxiv]{2305.03803}~[cs.SE]


\bibitem[Hussain et~al\mbox{.}(2023d)]%
        {hussain2023trojanedcm}
\bibfield{author}{\bibinfo{person}{Aftab Hussain}, \bibinfo{person}{Md~Rafiqul~Islam Rabin}, {and} \bibinfo{person}{Mohammad~Amin Alipour}.} \bibinfo{year}{2023}\natexlab{d}.
\newblock \bibinfo{title}{TrojanedCM: A Repository of Trojaned Large Language Models of Code}.
\newblock
\newblock
\showeprint[arxiv]{2311.14850}~[cs.SE]


\bibitem[Hussain et~al\mbox{.}(2023a)]%
        {10.1145/3593434.3594236}
\bibfield{author}{\bibinfo{person}{Yasir Hussain}, \bibinfo{person}{Zhiqiu Huang}, \bibinfo{person}{Yu Zhou}, \bibinfo{person}{Izhar~Ahmed Khan}, \bibinfo{person}{Nasrullah Khan}, {and} \bibinfo{person}{Muhammad~Zahid Abbas}.} \bibinfo{year}{2023}\natexlab{a}.
\newblock \showarticletitle{Optimized Tokenization Process for Open-Vocabulary Code Completion: An Empirical Study}. In \bibinfo{booktitle}{\emph{Proceedings of the 27th International Conference on Evaluation and Assessment in Software Engineering}} (Oulu, Finland) \emph{(\bibinfo{series}{EASE '23})}. \bibinfo{publisher}{Association for Computing Machinery}, \bibinfo{address}{New York, NY, USA}, \bibinfo{pages}{398–405}.
\newblock
\showISBNx{9798400700446}
\urldef\tempurl%
\url{https://doi.org/10.1145/3593434.3594236}
\showDOI{\tempurl}


\bibitem[Imai(2022)]%
        {10.1145/3510454.3522684}
\bibfield{author}{\bibinfo{person}{Saki Imai}.} \bibinfo{year}{2022}\natexlab{}.
\newblock \showarticletitle{Is GitHub Copilot a Substitute for Human Pair-Programming? An Empirical Study}. In \bibinfo{booktitle}{\emph{Proceedings of the ACM/IEEE 44th International Conference on Software Engineering: Companion Proceedings}} (Pittsburgh, Pennsylvania) \emph{(\bibinfo{series}{ICSE '22})}. \bibinfo{publisher}{Association for Computing Machinery}, \bibinfo{address}{New York, NY, USA}, \bibinfo{pages}{319–321}.
\newblock
\showISBNx{9781450392235}
\urldef\tempurl%
\url{https://doi.org/10.1145/3510454.3522684}
\showDOI{\tempurl}


\bibitem[Improta et~al\mbox{.}(2023)]%
        {improta2023enhancing}
\bibfield{author}{\bibinfo{person}{Cristina Improta}, \bibinfo{person}{Pietro Liguori}, \bibinfo{person}{Roberto Natella}, \bibinfo{person}{Bojan Cukic}, {and} \bibinfo{person}{Domenico Cotroneo}.} \bibinfo{year}{2023}\natexlab{}.
\newblock \bibinfo{title}{Enhancing Robustness of AI Offensive Code Generators via Data Augmentation}.
\newblock
\newblock
\showeprint[arxiv]{2306.05079}~[cs.LG]


\bibitem[Jalali and Wohlin(2012)]%
        {snowball1}
\bibfield{author}{\bibinfo{person}{Samireh Jalali} {and} \bibinfo{person}{Claes Wohlin}.} \bibinfo{year}{2012}\natexlab{}.
\newblock \showarticletitle{Systematic literature studies: Database searches vs. backward snowballing}. In \bibinfo{booktitle}{\emph{Proceedings of the 2012 ACM-IEEE International Symposium on Empirical Software Engineering and Measurement}}. \bibinfo{pages}{29--38}.
\newblock
\urldef\tempurl%
\url{https://doi.org/10.1145/2372251.2372257}
\showDOI{\tempurl}


\bibitem[Jayagopal et~al\mbox{.}(2022)]%
        {10.1145/3526113.3545659}
\bibfield{author}{\bibinfo{person}{Dhanya Jayagopal}, \bibinfo{person}{Justin Lubin}, {and} \bibinfo{person}{Sarah~E. Chasins}.} \bibinfo{year}{2022}\natexlab{}.
\newblock \showarticletitle{Exploring the Learnability of Program Synthesizers by Novice Programmers}. In \bibinfo{booktitle}{\emph{Proceedings of the 35th Annual ACM Symposium on User Interface Software and Technology}} (Bend, OR, USA) \emph{(\bibinfo{series}{UIST '22})}. \bibinfo{publisher}{Association for Computing Machinery}, \bibinfo{address}{New York, NY, USA}, Article \bibinfo{articleno}{64}, \bibinfo{numpages}{15}~pages.
\newblock
\showISBNx{9781450393201}
\urldef\tempurl%
\url{https://doi.org/10.1145/3526113.3545659}
\showDOI{\tempurl}


\bibitem[Jha and Reddy(2023)]%
        {codeattack}
\bibfield{author}{\bibinfo{person}{Akshita Jha} {and} \bibinfo{person}{Chandan~K. Reddy}.} \bibinfo{year}{2023}\natexlab{}.
\newblock \showarticletitle{CodeAttack: code-based adversarial attacks for pre-trained programming language models}. In \bibinfo{booktitle}{\emph{Proceedings of the Thirty-Seventh AAAI Conference on Artificial Intelligence and Thirty-Fifth Conference on Innovative Applications of Artificial Intelligence and Thirteenth Symposium on Educational Advances in Artificial Intelligence}} \emph{(\bibinfo{series}{AAAI'23/IAAI'23/EAAI'23})}. \bibinfo{publisher}{AAAI Press}, Article \bibinfo{articleno}{1670}, \bibinfo{numpages}{9}~pages.
\newblock
\showISBNx{978-1-57735-880-0}
\urldef\tempurl%
\url{https://doi.org/10.1609/aaai.v37i12.26739}
\showDOI{\tempurl}


\bibitem[Ji et~al\mbox{.}(2023)]%
        {ji2023benchmarking}
\bibfield{author}{\bibinfo{person}{Zhenlan Ji}, \bibinfo{person}{Pingchuan Ma}, \bibinfo{person}{Zongjie Li}, {and} \bibinfo{person}{Shuai Wang}.} \bibinfo{year}{2023}\natexlab{}.
\newblock \bibinfo{title}{Benchmarking and Explaining Large Language Model-based Code Generation: A Causality-Centric Approach}.
\newblock
\newblock
\showeprint[arxiv]{2310.06680}~[cs.SE]


\bibitem[Ji et~al\mbox{.}(2022)]%
        {ji2022unlearnable}
\bibfield{author}{\bibinfo{person}{Zhenlan Ji}, \bibinfo{person}{Pingchuan Ma}, {and} \bibinfo{person}{Shuai Wang}.} \bibinfo{year}{2022}\natexlab{}.
\newblock \showarticletitle{Unlearnable Examples: Protecting Open-Source Software from Unauthorized Neural Code Learning.}. In \bibinfo{booktitle}{\emph{SEKE}}. \bibinfo{pages}{525--530}.
\newblock


\bibitem[Jia et~al\mbox{.}(2023)]%
        {10123554}
\bibfield{author}{\bibinfo{person}{Jinghan Jia}, \bibinfo{person}{Shashank Srikant}, \bibinfo{person}{Tamara Mitrovska}, \bibinfo{person}{Chuang Gan}, \bibinfo{person}{Shiyu Chang}, \bibinfo{person}{Sijia Liu}, {and} \bibinfo{person}{Una-May O'Reilly}.} \bibinfo{year}{2023}\natexlab{}.
\newblock \showarticletitle{ClawSAT: Towards Both Robust and Accurate Code Models}. In \bibinfo{booktitle}{\emph{2023 IEEE International Conference on Software Analysis, Evolution and Reengineering (SANER)}}. \bibinfo{publisher}{IEEE Computer Society}, \bibinfo{address}{Los Alamitos, CA, USA}, \bibinfo{pages}{212--223}.
\newblock
\urldef\tempurl%
\url{https://doi.org/10.1109/SANER56733.2023.00029}
\showDOI{\tempurl}


\bibitem[Jiarpakdee et~al\mbox{.}(2022)]%
        {9044387}
\bibfield{author}{\bibinfo{person}{Jirayus Jiarpakdee}, \bibinfo{person}{Chakkrit~Kla Tantithamthavorn}, \bibinfo{person}{Hoa~Khanh Dam}, {and} \bibinfo{person}{John Grundy}.} \bibinfo{year}{2022}\natexlab{}.
\newblock \showarticletitle{An Empirical Study of Model-Agnostic Techniques for Defect Prediction Models}.
\newblock \bibinfo{journal}{\emph{IEEE Transactions on Software Engineering}} \bibinfo{volume}{48}, \bibinfo{number}{1} (\bibinfo{year}{2022}), \bibinfo{pages}{166--185}.
\newblock
\urldef\tempurl%
\url{https://doi.org/10.1109/TSE.2020.2982385}
\showDOI{\tempurl}


\bibitem[Jin et~al\mbox{.}(2024)]%
        {jin2024can}
\bibfield{author}{\bibinfo{person}{Kailun Jin}, \bibinfo{person}{Chung-Yu Wang}, \bibinfo{person}{Hung~Viet Pham}, {and} \bibinfo{person}{Hadi Hemmati}.} \bibinfo{year}{2024}\natexlab{}.
\newblock \showarticletitle{Can ChatGPT Support Developers? An Empirical Evaluation of Large Language Models for Code Generation}.
\newblock \bibinfo{journal}{\emph{arXiv preprint arXiv:2402.11702}} (\bibinfo{year}{2024}).
\newblock


\bibitem[Jungwirth et~al\mbox{.}(2023)]%
        {10174120}
\bibfield{author}{\bibinfo{person}{Gerhard Jungwirth}, \bibinfo{person}{Aakanksha Saha}, \bibinfo{person}{Michael Schröder}, \bibinfo{person}{Tobias Fiebig}, \bibinfo{person}{Martina Lindorfer}, {and} \bibinfo{person}{Jürgen Cito}.} \bibinfo{year}{2023}\natexlab{}.
\newblock \showarticletitle{Connecting the .dotfiles: Checked-In Secret Exposure with Extra (Lateral Movement) Steps}. In \bibinfo{booktitle}{\emph{2023 IEEE/ACM 20th International Conference on Mining Software Repositories (MSR)}}. \bibinfo{pages}{322--333}.
\newblock
\urldef\tempurl%
\url{https://doi.org/10.1109/MSR59073.2023.00051}
\showDOI{\tempurl}


\bibitem[Kaushal et~al\mbox{.}(2023)]%
        {kaushal2023lord}
\bibfield{author}{\bibinfo{person}{Ayush Kaushal}, \bibinfo{person}{Tejas Vaidhya}, {and} \bibinfo{person}{Irina Rish}.} \bibinfo{year}{2023}\natexlab{}.
\newblock \bibinfo{title}{LORD: Low Rank Decomposition Of Monolingual Code LLMs For One-Shot Compression}.
\newblock
\newblock
\showeprint[arxiv]{2309.14021}~[cs.CL]


\bibitem[Kazemitabaar et~al\mbox{.}(2023)]%
        {10.1145/3544548.3580919}
\bibfield{author}{\bibinfo{person}{Majeed Kazemitabaar}, \bibinfo{person}{Justin Chow}, \bibinfo{person}{Carl Ka~To Ma}, \bibinfo{person}{Barbara~J. Ericson}, \bibinfo{person}{David Weintrop}, {and} \bibinfo{person}{Tovi Grossman}.} \bibinfo{year}{2023}\natexlab{}.
\newblock \showarticletitle{Studying the Effect of AI Code Generators on Supporting Novice Learners in Introductory Programming}. In \bibinfo{booktitle}{\emph{Proceedings of the 2023 CHI Conference on Human Factors in Computing Systems}} (Hamburg, Germany) \emph{(\bibinfo{series}{CHI '23})}. \bibinfo{publisher}{Association for Computing Machinery}, \bibinfo{address}{New York, NY, USA}, Article \bibinfo{articleno}{455}, \bibinfo{numpages}{23}~pages.
\newblock
\showISBNx{9781450394215}
\urldef\tempurl%
\url{https://doi.org/10.1145/3544548.3580919}
\showDOI{\tempurl}


\bibitem[Kitchenham et~al\mbox{.}(2023)]%
        {9772383}
\bibfield{author}{\bibinfo{person}{Barbara Kitchenham}, \bibinfo{person}{Lech Madeyski}, {and} \bibinfo{person}{David Budgen}.} \bibinfo{year}{2023}\natexlab{}.
\newblock \showarticletitle{SEGRESS: Software Engineering Guidelines for REporting Secondary Studies}.
\newblock \bibinfo{journal}{\emph{IEEE Transactions on Software Engineering}} \bibinfo{volume}{49}, \bibinfo{number}{03} (\bibinfo{date}{mar} \bibinfo{year}{2023}), \bibinfo{pages}{1273--1298}.
\newblock
\showISSN{1939-3520}
\urldef\tempurl%
\url{https://doi.org/10.1109/TSE.2022.3174092}
\showDOI{\tempurl}


\bibitem[Kothari et~al\mbox{.}(2007)]%
        {4151691}
\bibfield{author}{\bibinfo{person}{Jay Kothari}, \bibinfo{person}{Maxim Shevertalov}, \bibinfo{person}{Edward Stehle}, {and} \bibinfo{person}{Spiros Mancoridis}.} \bibinfo{year}{2007}\natexlab{}.
\newblock \showarticletitle{A Probabilistic Approach to Source Code Authorship Identification}. In \bibinfo{booktitle}{\emph{Fourth International Conference on Information Technology (ITNG'07)}}. \bibinfo{pages}{243--248}.
\newblock
\urldef\tempurl%
\url{https://doi.org/10.1109/ITNG.2007.17}
\showDOI{\tempurl}


\bibitem[Kou et~al\mbox{.}(2023)]%
        {kou2023model}
\bibfield{author}{\bibinfo{person}{Bonan Kou}, \bibinfo{person}{Shengmai Chen}, \bibinfo{person}{Zhijie Wang}, \bibinfo{person}{Lei Ma}, {and} \bibinfo{person}{Tianyi Zhang}.} \bibinfo{year}{2023}\natexlab{}.
\newblock \bibinfo{title}{Is Model Attention Aligned with Human Attention? An Empirical Study on Large Language Models for Code Generation}.
\newblock
\newblock
\showeprint[arxiv]{2306.01220}~[cs.SE]


\bibitem[Le-Cong et~al\mbox{.}(2024)]%
        {lecong2024evaluating}
\bibfield{author}{\bibinfo{person}{Thanh Le-Cong}, \bibinfo{person}{Dat Nguyen}, \bibinfo{person}{Bach Le}, {and} \bibinfo{person}{Toby Murray}.} \bibinfo{year}{2024}\natexlab{}.
\newblock \bibinfo{title}{Evaluating Program Repair with Semantic-Preserving Transformations: A Naturalness Assessment}.
\newblock
\newblock
\showeprint[arxiv]{2402.11892}~[cs.SE]


\bibitem[Ledel and Herbold(2022)]%
        {ledel2022studying}
\bibfield{author}{\bibinfo{person}{Benjamin Ledel} {and} \bibinfo{person}{Steffen Herbold}.} \bibinfo{year}{2022}\natexlab{}.
\newblock \bibinfo{title}{Studying the explanations for the automated prediction of bug and non-bug issues using LIME and SHAP}.
\newblock
\newblock
\showeprint[arxiv]{2209.07623}~[cs.SE]


\bibitem[Lee et~al\mbox{.}(2024)]%
        {lee2024wrote}
\bibfield{author}{\bibinfo{person}{Taehyun Lee}, \bibinfo{person}{Seokhee Hong}, \bibinfo{person}{Jaewoo Ahn}, \bibinfo{person}{Ilgee Hong}, \bibinfo{person}{Hwaran Lee}, \bibinfo{person}{Sangdoo Yun}, \bibinfo{person}{Jamin Shin}, {and} \bibinfo{person}{Gunhee Kim}.} \bibinfo{year}{2024}\natexlab{}.
\newblock \bibinfo{title}{Who Wrote this Code? Watermarking for Code Generation}.
\newblock
\newblock
\showeprint[arxiv]{2305.15060}~[cs.CL]


\bibitem[Li et~al\mbox{.}(2024)]%
        {li2024resilient}
\bibfield{author}{\bibinfo{person}{Boquan Li}, \bibinfo{person}{Mengdi Zhang}, \bibinfo{person}{Peixin Zhang}, \bibinfo{person}{Jun Sun}, {and} \bibinfo{person}{Xingmei Wang}.} \bibinfo{year}{2024}\natexlab{}.
\newblock \bibinfo{title}{Resilient Watermarking for LLM-Generated Codes}.
\newblock
\newblock
\showeprint[arxiv]{2402.07518}~[cs.CR]


\bibitem[Li et~al\mbox{.}(2019)]%
        {Li_2019}
\bibfield{author}{\bibinfo{person}{Jinfeng Li}, \bibinfo{person}{Shouling Ji}, \bibinfo{person}{Tianyu Du}, \bibinfo{person}{Bo Li}, {and} \bibinfo{person}{Ting Wang}.} \bibinfo{year}{2019}\natexlab{}.
\newblock \showarticletitle{{TextBugger}: Generating Adversarial Text Against Real-world Applications}. In \bibinfo{booktitle}{\emph{Proceedings 2019 Network and Distributed System Security Symposium}}. \bibinfo{publisher}{Internet Society}.
\newblock
\urldef\tempurl%
\url{https://doi.org/10.14722/ndss.2019.23138}
\showDOI{\tempurl}


\bibitem[Li et~al\mbox{.}(2022b)]%
        {CodePoisoner}
\bibfield{author}{\bibinfo{person}{Jia Li}, \bibinfo{person}{Zhuo Li}, \bibinfo{person}{Huangzhao Zhang}, \bibinfo{person}{Ge Li}, \bibinfo{person}{Zhi Jin}, \bibinfo{person}{Xing Hu}, {and} \bibinfo{person}{Xin Xia}.} \bibinfo{year}{2022}\natexlab{b}.
\newblock \bibinfo{title}{Poison Attack and Defense on Deep Source Code Processing Models}.
\newblock
\newblock
\urldef\tempurl%
\url{https://doi.org/10.48550/ARXIV.2210.17029}
\showDOI{\tempurl}


\bibitem[Li et~al\mbox{.}(2023d)]%
        {li2023always}
\bibfield{author}{\bibinfo{person}{Jiachen Li}, \bibinfo{person}{Elizabeth Mynatt}, \bibinfo{person}{Varun Mishra}, {and} \bibinfo{person}{Jonathan Bell}.} \bibinfo{year}{2023}\natexlab{d}.
\newblock \bibinfo{title}{"Always Nice and Confident, Sometimes wrong": Developer's Experiences Engaging Generative AI Chatbots Versus Human-Powered Q\&A Platforms}.
\newblock
\newblock
\showeprint[arxiv]{2309.13684}~[cs.HC]


\bibitem[Li et~al\mbox{.}(2023a)]%
        {starcoder}
\bibfield{author}{\bibinfo{person}{Raymond Li}, \bibinfo{person}{Loubna~Ben Allal}, {and} \bibinfo{person}{Yangtian~Zi et al.}} \bibinfo{year}{2023}\natexlab{a}.
\newblock \bibinfo{title}{StarCoder: may the source be with you!}
\newblock
\newblock
\showeprint[arxiv]{2305.06161}~[cs.CL]


\bibitem[Li and Liang(2021)]%
        {li-liang-2021-prefix}
\bibfield{author}{\bibinfo{person}{Xiang~Lisa Li} {and} \bibinfo{person}{Percy Liang}.} \bibinfo{year}{2021}\natexlab{}.
\newblock \showarticletitle{Prefix-Tuning: Optimizing Continuous Prompts for Generation}. In \bibinfo{booktitle}{\emph{Proceedings of the 59th Annual Meeting of the Association for Computational Linguistics and the 11th International Joint Conference on Natural Language Processing (Volume 1: Long Papers)}}, \bibfield{editor}{\bibinfo{person}{Chengqing Zong}, \bibinfo{person}{Fei Xia}, \bibinfo{person}{Wenjie Li}, {and} \bibinfo{person}{Roberto Navigli}} (Eds.). \bibinfo{publisher}{Association for Computational Linguistics}, \bibinfo{address}{Online}, \bibinfo{pages}{4582--4597}.
\newblock
\urldef\tempurl%
\url{https://doi.org/10.18653/v1/2021.acl-long.353}
\showDOI{\tempurl}


\bibitem[Li et~al\mbox{.}(2023c)]%
        {li-etal-2023-multi-target}
\bibfield{author}{\bibinfo{person}{Yanzhou Li}, \bibinfo{person}{Shangqing Liu}, \bibinfo{person}{Kangjie Chen}, \bibinfo{person}{Xiaofei Xie}, \bibinfo{person}{Tianwei Zhang}, {and} \bibinfo{person}{Yang Liu}.} \bibinfo{year}{2023}\natexlab{c}.
\newblock \showarticletitle{Multi-target Backdoor Attacks for Code Pre-trained Models}. In \bibinfo{booktitle}{\emph{Proceedings of the 61st Annual Meeting of the Association for Computational Linguistics (Volume 1: Long Papers)}}. \bibinfo{publisher}{Association for Computational Linguistics}, \bibinfo{address}{Toronto, Canada}, \bibinfo{pages}{7236--7254}.
\newblock
\urldef\tempurl%
\url{https://doi.org/10.18653/v1/2023.acl-long.399}
\showDOI{\tempurl}


\bibitem[Li et~al\mbox{.}(2022c)]%
        {li2022closer}
\bibfield{author}{\bibinfo{person}{Yaoxian Li}, \bibinfo{person}{Shiyi Qi}, \bibinfo{person}{Cuiyun Gao}, \bibinfo{person}{Yun Peng}, \bibinfo{person}{David Lo}, \bibinfo{person}{Zenglin Xu}, {and} \bibinfo{person}{Michael~R. Lyu}.} \bibinfo{year}{2022}\natexlab{c}.
\newblock \bibinfo{title}{A Closer Look into Transformer-Based Code Intelligence Through Code Transformation: Challenges and Opportunities}.
\newblock
\newblock
\showeprint[arxiv]{2207.04285}~[cs.SE]


\bibitem[Li et~al\mbox{.}(2022d)]%
        {li-etal-2022-semantic}
\bibfield{author}{\bibinfo{person}{Yiyang Li}, \bibinfo{person}{Hongqiu Wu}, {and} \bibinfo{person}{Hai Zhao}.} \bibinfo{year}{2022}\natexlab{d}.
\newblock \showarticletitle{Semantic-Preserving Adversarial Code Comprehension}. In \bibinfo{booktitle}{\emph{Proceedings of the 29th International Conference on Computational Linguistics}}. \bibinfo{publisher}{International Committee on Computational Linguistics}, \bibinfo{address}{Gyeongju, Republic of Korea}, \bibinfo{pages}{3017--3028}.
\newblock
\urldef\tempurl%
\url{https://aclanthology.org/2022.coling-1.267}
\showURL{%
\tempurl}


\bibitem[Li et~al\mbox{.}(2023g)]%
        {10232920}
\bibfield{author}{\bibinfo{person}{Yao Li}, \bibinfo{person}{Tao Zhang}, \bibinfo{person}{Xiapu Luo}, \bibinfo{person}{Haipeng Cai}, \bibinfo{person}{Sen Fang}, {and} \bibinfo{person}{Dawei Yuan}.} \bibinfo{year}{2023}\natexlab{g}.
\newblock \showarticletitle{Do Pretrained Language Models Indeed Understand Software Engineering Tasks?}
\newblock \bibinfo{journal}{\emph{IEEE Transactions on Software Engineering}} \bibinfo{volume}{49}, \bibinfo{number}{10} (\bibinfo{date}{oct} \bibinfo{year}{2023}), \bibinfo{pages}{4639--4655}.
\newblock
\showISSN{1939-3520}
\urldef\tempurl%
\url{https://doi.org/10.1109/TSE.2023.3308952}
\showDOI{\tempurl}


\bibitem[Li et~al\mbox{.}(2022a)]%
        {RoPGen}
\bibfield{author}{\bibinfo{person}{Zhen Li}, \bibinfo{person}{Guenevere~(Qian) Chen}, \bibinfo{person}{Chen Chen}, \bibinfo{person}{Yayi Zou}, {and} \bibinfo{person}{Shouhuai Xu}.} \bibinfo{year}{2022}\natexlab{a}.
\newblock \showarticletitle{RoPGen: Towards Robust Code Authorship Attribution via Automatic Coding Style Transformation}. In \bibinfo{booktitle}{\emph{Proceedings of the 44th International Conference on Software Engineering}} (Pittsburgh, Pennsylvania) \emph{(\bibinfo{series}{ICSE '22})}. \bibinfo{publisher}{Association for Computing Machinery}, \bibinfo{address}{New York, NY, USA}, \bibinfo{pages}{1906–1918}.
\newblock
\showISBNx{9781450392211}
\urldef\tempurl%
\url{https://doi.org/10.1145/3510003.3510181}
\showDOI{\tempurl}


\bibitem[Li et~al\mbox{.}(2023b)]%
        {10.1016/j.future.2022.12.030}
\bibfield{author}{\bibinfo{person}{Zhen Li}, \bibinfo{person}{Xiang Huang}, \bibinfo{person}{Yangrui Li}, {and} \bibinfo{person}{Guenevere Chen}.} \bibinfo{year}{2023}\natexlab{b}.
\newblock \showarticletitle{A Comparative Study of Adversarial Training Methods for Neural Models of Source Code}.
\newblock \bibinfo{journal}{\emph{Future Gener. Comput. Syst.}} \bibinfo{volume}{142}, \bibinfo{number}{C} (\bibinfo{date}{may} \bibinfo{year}{2023}), \bibinfo{pages}{165–181}.
\newblock
\showISSN{0167-739X}
\urldef\tempurl%
\url{https://doi.org/10.1016/j.future.2022.12.030}
\showDOI{\tempurl}


\bibitem[Li et~al\mbox{.}(2023e)]%
        {10.1145/3551349.3556941}
\bibfield{author}{\bibinfo{person}{Zhong Li}, \bibinfo{person}{Minxue Pan}, \bibinfo{person}{Yu Pei}, \bibinfo{person}{Tian Zhang}, \bibinfo{person}{Linzhang Wang}, {and} \bibinfo{person}{Xuandong Li}.} \bibinfo{year}{2023}\natexlab{e}.
\newblock \showarticletitle{Robust Learning of Deep Predictive Models from Noisy and Imbalanced Software Engineering Datasets}. In \bibinfo{booktitle}{\emph{Proceedings of the 37th IEEE/ACM International Conference on Automated Software Engineering}} (Rochester, MI, USA) \emph{(\bibinfo{series}{ASE '22})}. Article \bibinfo{articleno}{86}, \bibinfo{numpages}{13}~pages.
\newblock
\showISBNx{9781450394758}
\urldef\tempurl%
\url{https://doi.org/10.1145/3551349.3556941}
\showDOI{\tempurl}


\bibitem[Li et~al\mbox{.}(2023f)]%
        {10.1145/3576915.3623120}
\bibfield{author}{\bibinfo{person}{Zongjie Li}, \bibinfo{person}{Chaozheng Wang}, \bibinfo{person}{Shuai Wang}, {and} \bibinfo{person}{Cuiyun Gao}.} \bibinfo{year}{2023}\natexlab{f}.
\newblock \showarticletitle{Protecting Intellectual Property of Large Language Model-Based Code Generation APIs via Watermarks}. In \bibinfo{booktitle}{\emph{Proceedings of the 2023 ACM SIGSAC Conference on Computer and Communications Security}} \emph{(\bibinfo{series}{CCS '23})}. \bibinfo{publisher}{Association for Computing Machinery}, \bibinfo{address}{New York, NY, USA}, \bibinfo{pages}{2336–2350}.
\newblock
\showISBNx{9798400700507}
\urldef\tempurl%
\url{https://doi.org/10.1145/3576915.3623120}
\showDOI{\tempurl}


\bibitem[Li et~al\mbox{.}(2023h)]%
        {10298318}
\bibfield{author}{\bibinfo{person}{Zhen Li}, \bibinfo{person}{Ruqian Zhang}, \bibinfo{person}{Deqing Zou}, \bibinfo{person}{Ning Wang}, \bibinfo{person}{Yating Li}, \bibinfo{person}{Shouhuai Xu}, \bibinfo{person}{Chen Chen}, {and} \bibinfo{person}{Hai Jin}.} \bibinfo{year}{2023}\natexlab{h}.
\newblock \showarticletitle{Robin: A Novel Method to Produce Robust Interpreters for Deep Learning-Based Code Classifiers}. In \bibinfo{booktitle}{\emph{2023 38th IEEE/ACM International Conference on Automated Software Engineering (ASE)}}. \bibinfo{publisher}{IEEE Computer Society}, \bibinfo{address}{Los Alamitos, CA, USA}, \bibinfo{pages}{27--39}.
\newblock
\urldef\tempurl%
\url{https://doi.org/10.1109/ASE56229.2023.00164}
\showDOI{\tempurl}


\bibitem[Li et~al\mbox{.}(2022e)]%
        {SySeVR}
\bibfield{author}{\bibinfo{person}{Zhen Li}, \bibinfo{person}{Deqing Zou}, \bibinfo{person}{Shouhuai Xu}, \bibinfo{person}{Hai Jin}, \bibinfo{person}{Yawei Zhu}, {and} \bibinfo{person}{Zhaoxuan Chen}.} \bibinfo{year}{2022}\natexlab{e}.
\newblock \showarticletitle{SySeVR: A Framework for Using Deep Learning to Detect Software Vulnerabilities}.
\newblock \bibinfo{journal}{\emph{IEEE Transactions on Dependable and Secure Computing}} \bibinfo{volume}{19}, \bibinfo{number}{04} (\bibinfo{date}{jul} \bibinfo{year}{2022}), \bibinfo{pages}{2244--2258}.
\newblock
\showISSN{1941-0018}
\urldef\tempurl%
\url{https://doi.org/10.1109/TDSC.2021.3051525}
\showDOI{\tempurl}


\bibitem[Liang et~al\mbox{.}(2024)]%
        {liang2023usability}
\bibfield{author}{\bibinfo{person}{Jenny~T. Liang}, \bibinfo{person}{Chenyang Yang}, {and} \bibinfo{person}{Brad~A. Myers}.} \bibinfo{year}{2024}\natexlab{}.
\newblock \showarticletitle{A Large-Scale Survey on the Usability of AI Programming Assistants: Successes and Challenges}. In \bibinfo{booktitle}{\emph{2024 IEEE/ACM 46th International Conference on Software Engineering (ICSE)}}. \bibinfo{publisher}{IEEE Computer Society}, \bibinfo{address}{Los Alamitos, CA, USA}, \bibinfo{pages}{605--617}.
\newblock
\showISSN{1558-1225}
\urldef\tempurl%
\url{https://doi.ieeecomputersociety.org/}
\showURL{%
\tempurl}


\bibitem[Liguori et~al\mbox{.}(2021)]%
        {9700297}
\bibfield{author}{\bibinfo{person}{Pietro Liguori}, \bibinfo{person}{Erfan Al-Hossami}, \bibinfo{person}{Vittorio Orbinato}, \bibinfo{person}{Roberto Natella}, \bibinfo{person}{Samira Shaikh}, \bibinfo{person}{Domenico Cotroneo}, {and} \bibinfo{person}{Bojan Cukic}.} \bibinfo{year}{2021}\natexlab{}.
\newblock \showarticletitle{EVIL: Exploiting Software via Natural Language}. In \bibinfo{booktitle}{\emph{2021 IEEE 32nd International Symposium on Software Reliability Engineering (ISSRE)}}. \bibinfo{publisher}{IEEE Computer Society}, \bibinfo{address}{Los Alamitos, CA, USA}, \bibinfo{pages}{321--332}.
\newblock
\urldef\tempurl%
\url{https://doi.org/10.1109/ISSRE52982.2021.00042}
\showDOI{\tempurl}


\bibitem[Liguori et~al\mbox{.}(2023)]%
        {10.1145/3528588.3528653}
\bibfield{author}{\bibinfo{person}{Pietro Liguori}, \bibinfo{person}{Cristina Improta}, \bibinfo{person}{Simona De~Vivo}, \bibinfo{person}{Roberto Natella}, \bibinfo{person}{Bojan Cukic}, {and} \bibinfo{person}{Domenico Cotroneo}.} \bibinfo{year}{2023}\natexlab{}.
\newblock \showarticletitle{Can NMT Understand Me? Towards Perturbation-Based Evaluation of NMT Models for Code Generation}. In \bibinfo{booktitle}{\emph{Proceedings of the 1st International Workshop on Natural Language-Based Software Engineering}} (Pittsburgh, Pennsylvania) \emph{(\bibinfo{series}{NLBSE '22})}. \bibinfo{publisher}{Association for Computing Machinery}, \bibinfo{address}{New York, NY, USA}, \bibinfo{pages}{59–66}.
\newblock
\showISBNx{9781450393430}
\urldef\tempurl%
\url{https://doi.org/10.1145/3528588.3528653}
\showDOI{\tempurl}


\bibitem[Lipton(2018)]%
        {10.1145/3236386.3241340}
\bibfield{author}{\bibinfo{person}{Zachary~C. Lipton}.} \bibinfo{year}{2018}\natexlab{}.
\newblock \showarticletitle{The Mythos of Model Interpretability: In Machine Learning, the Concept of Interpretability is Both Important and Slippery.}
\newblock \bibinfo{journal}{\emph{Queue}} \bibinfo{volume}{16}, \bibinfo{number}{3} (\bibinfo{date}{jun} \bibinfo{year}{2018}), \bibinfo{pages}{31–57}.
\newblock
\showISSN{1542-7730}
\urldef\tempurl%
\url{https://doi.org/10.1145/3236386.3241340}
\showDOI{\tempurl}


\bibitem[Liu et~al\mbox{.}(2023a)]%
        {10298587}
\bibfield{author}{\bibinfo{person}{Jiaxing Liu}, \bibinfo{person}{Chaofeng Sha}, {and} \bibinfo{person}{Xin Peng}.} \bibinfo{year}{2023}\natexlab{a}.
\newblock \showarticletitle{An Empirical Study of Parameter-Efficient Fine-Tuning Methods for Pre-Trained Code Models}. In \bibinfo{booktitle}{\emph{2023 38th IEEE/ACM International Conference on Automated Software Engineering (ASE)}}. \bibinfo{publisher}{IEEE Computer Society}, \bibinfo{address}{Los Alamitos, CA, USA}, \bibinfo{pages}{397--408}.
\newblock
\urldef\tempurl%
\url{https://doi.org/10.1109/ASE56229.2023.00125}
\showDOI{\tempurl}


\bibitem[Liu et~al\mbox{.}(2021)]%
        {9454564}
\bibfield{author}{\bibinfo{person}{Qianjun Liu}, \bibinfo{person}{Shouling Ji}, \bibinfo{person}{Changchang Liu}, {and} \bibinfo{person}{Chunming Wu}.} \bibinfo{year}{2021}\natexlab{}.
\newblock \showarticletitle{A Practical Black-Box Attack on Source Code Authorship Identification Classifiers}.
\newblock \bibinfo{journal}{\emph{IEEE Transactions on Information Forensics and Security}}  \bibinfo{volume}{16} (\bibinfo{year}{2021}), \bibinfo{pages}{3620--3633}.
\newblock
\urldef\tempurl%
\url{https://doi.org/10.1109/TIFS.2021.3080507}
\showDOI{\tempurl}


\bibitem[Liu et~al\mbox{.}(2024a)]%
        {liu2024delving}
\bibfield{author}{\bibinfo{person}{Shuo Liu}, \bibinfo{person}{Jacky Keung}, \bibinfo{person}{Zhen Yang}, \bibinfo{person}{Fang Liu}, \bibinfo{person}{Qilin Zhou}, {and} \bibinfo{person}{Yihan Liao}.} \bibinfo{year}{2024}\natexlab{a}.
\newblock \bibinfo{title}{Delving into Parameter-Efficient Fine-Tuning in Code Change Learning: An Empirical Study}.
\newblock
\newblock
\showeprint[arxiv]{2402.06247}~[cs.SE]


\bibitem[Liu et~al\mbox{.}(2023b)]%
        {Liuyue2023}
\bibfield{author}{\bibinfo{person}{Yue Liu}, \bibinfo{person}{Chakkrit Tantithamthavorn}, \bibinfo{person}{Yonghui Liu}, {and} \bibinfo{person}{Li Li}.} \bibinfo{year}{2023}\natexlab{b}.
\newblock \showarticletitle{{On the Reliability and Explainability of Automated Code Generation Approaches}}.
\newblock  \bibinfo{volume}{1}, \bibinfo{number}{1} (\bibinfo{year}{2023}), \bibinfo{pages}{1--20}.
\newblock
\showeprint[arxiv]{2302.09587}
\urldef\tempurl%
\url{http://arxiv.org/abs/2302.09587}
\showURL{%
\tempurl}


\bibitem[Liu et~al\mbox{.}(2024b)]%
        {10.1145/3641540}
\bibfield{author}{\bibinfo{person}{Yue Liu}, \bibinfo{person}{Chakkrit Tantithamthavorn}, \bibinfo{person}{Yonghui Liu}, {and} \bibinfo{person}{Li Li}.} \bibinfo{year}{2024}\natexlab{b}.
\newblock \showarticletitle{On the Reliability and Explainability of Language Models for Program Generation}.
\newblock \bibinfo{journal}{\emph{ACM Trans. Softw. Eng. Methodol.}} (\bibinfo{date}{jan} \bibinfo{year}{2024}).
\newblock
\showISSN{1049-331X}
\urldef\tempurl%
\url{https://doi.org/10.1145/3641540}
\showDOI{\tempurl}
\newblock
\shownote{Just Accepted}.


\bibitem[Lo(2023)]%
        {lo2023trustworthy}
\bibfield{author}{\bibinfo{person}{David Lo}.} \bibinfo{year}{2023}\natexlab{}.
\newblock \bibinfo{title}{Trustworthy and Synergistic Artificial Intelligence for Software Engineering: Vision and Roadmaps}.
\newblock
\newblock
\showeprint[arxiv]{2309.04142}~[cs.SE]


\bibitem[Lu et~al\mbox{.}(2023)]%
        {10299938}
\bibfield{author}{\bibinfo{person}{Junyi Lu}, \bibinfo{person}{Lei Yu}, \bibinfo{person}{Xiaojia Li}, \bibinfo{person}{Li Yang}, {and} \bibinfo{person}{Chun Zuo}.} \bibinfo{year}{2023}\natexlab{}.
\newblock \showarticletitle{LLaMA-Reviewer: Advancing Code Review Automation with Large Language Models through Parameter-Efficient Fine-Tuning}. In \bibinfo{booktitle}{\emph{2023 IEEE 34th International Symposium on Software Reliability Engineering (ISSRE)}}. \bibinfo{publisher}{IEEE Computer Society}, \bibinfo{address}{Los Alamitos, CA, USA}, \bibinfo{pages}{647--658}.
\newblock
\urldef\tempurl%
\url{https://doi.org/10.1109/ISSRE59848.2023.00026}
\showDOI{\tempurl}


\bibitem[Lu et~al\mbox{.}(2021)]%
        {CodeXGLUE}
\bibfield{author}{\bibinfo{person}{Shuai Lu}, \bibinfo{person}{Daya Guo}, \bibinfo{person}{Shuo Ren}, \bibinfo{person}{Junjie Huang}, \bibinfo{person}{Alexey Svyatkovskiy}, \bibinfo{person}{Ambrosio Blanco}, \bibinfo{person}{Colin~B. Clement}, \bibinfo{person}{Dawn Drain}, \bibinfo{person}{Daxin Jiang}, \bibinfo{person}{Duyu Tang}, \bibinfo{person}{Ge Li}, \bibinfo{person}{Lidong Zhou}, \bibinfo{person}{Linjun Shou}, \bibinfo{person}{Long Zhou}, \bibinfo{person}{Michele Tufano}, \bibinfo{person}{Ming Gong}, \bibinfo{person}{Ming Zhou}, \bibinfo{person}{Nan Duan}, \bibinfo{person}{Neel Sundaresan}, \bibinfo{person}{Shao~Kun Deng}, \bibinfo{person}{Shengyu Fu}, {and} \bibinfo{person}{Shujie Liu}.} \bibinfo{year}{2021}\natexlab{}.
\newblock \showarticletitle{CodeXGLUE: {A} Machine Learning Benchmark Dataset for Code Understanding and Generation}.
\newblock \bibinfo{journal}{\emph{CoRR}} (\bibinfo{year}{2021}).
\newblock


\bibitem[Lundberg and Lee(2017)]%
        {SHAP}
\bibfield{author}{\bibinfo{person}{Scott~M Lundberg} {and} \bibinfo{person}{Su-In Lee}.} \bibinfo{year}{2017}\natexlab{}.
\newblock \showarticletitle{A unified approach to interpreting model predictions}.
\newblock \bibinfo{journal}{\emph{NeruIPS}}  \bibinfo{volume}{30} (\bibinfo{year}{2017}).
\newblock


\bibitem[Ma et~al\mbox{.}(2023a)]%
        {ma2023code2}
\bibfield{author}{\bibinfo{person}{Wanlun Ma}, \bibinfo{person}{Yiliao Song}, \bibinfo{person}{Minhui Xue}, \bibinfo{person}{Sheng Wen}, {and} \bibinfo{person}{Yang Xiang}.} \bibinfo{year}{2023}\natexlab{a}.
\newblock \bibinfo{title}{The ``code'' of Ethics:A Holistic Audit of AI Code Generators}.
\newblock
\newblock


\bibitem[Ma et~al\mbox{.}(2023b)]%
        {ma2023code}
\bibfield{author}{\bibinfo{person}{Wei Ma}, \bibinfo{person}{Mengjie Zhao}, \bibinfo{person}{Xiaofei Xie}, \bibinfo{person}{Qiang Hu}, \bibinfo{person}{Shangqing Liu}, \bibinfo{person}{Jie Zhang}, \bibinfo{person}{Wenhan Wang}, {and} \bibinfo{person}{Yang Liu}.} \bibinfo{year}{2023}\natexlab{b}.
\newblock \bibinfo{title}{Are Code Pre-trained Models Powerful to Learn Code Syntax and Semantics?}
\newblock
\newblock
\showeprint[arxiv]{2212.10017}~[cs.SE]


\bibitem[Majdinasab et~al\mbox{.}(2024)]%
        {majdinasab2024trained}
\bibfield{author}{\bibinfo{person}{Vahid Majdinasab}, \bibinfo{person}{Amin Nikanjam}, {and} \bibinfo{person}{Foutse Khomh}.} \bibinfo{year}{2024}\natexlab{}.
\newblock \bibinfo{title}{Trained Without My Consent: Detecting Code Inclusion In Language Models Trained on Code}.
\newblock
\newblock
\showeprint[arxiv]{2402.09299}~[cs.SE]


\bibitem[Mastropaolo et~al\mbox{.}(2023)]%
        {10.1109/ICSE48619.2023.00181}
\bibfield{author}{\bibinfo{person}{Antonio Mastropaolo}, \bibinfo{person}{Luca Pascarella}, \bibinfo{person}{Emanuela Guglielmi}, \bibinfo{person}{Matteo Ciniselli}, \bibinfo{person}{Simone Scalabrino}, \bibinfo{person}{Rocco Oliveto}, {and} \bibinfo{person}{Gabriele Bavota}.} \bibinfo{year}{2023}\natexlab{}.
\newblock \showarticletitle{On the Robustness of Code Generation Techniques: An Empirical Study on GitHub Copilot}. In \bibinfo{booktitle}{\emph{Proceedings of the 45th International Conference on Software Engineering}} (Melbourne, Victoria, Australia) \emph{(\bibinfo{series}{ICSE '23})}. \bibinfo{publisher}{IEEE Press}, \bibinfo{pages}{2149–2160}.
\newblock
\showISBNx{9781665457019}
\urldef\tempurl%
\url{https://doi.org/10.1109/ICSE48619.2023.00181}
\showDOI{\tempurl}


\bibitem[Matyukhina et~al\mbox{.}(2019)]%
        {Matyukhina2019AdversarialAA}
\bibfield{author}{\bibinfo{person}{Alina Matyukhina}, \bibinfo{person}{Natalia Stakhanova}, \bibinfo{person}{Mila Dalla~Preda}, {and} \bibinfo{person}{Celine Perley}.} \bibinfo{year}{2019}\natexlab{}.
\newblock \showarticletitle{Adversarial Authorship Attribution in Open-Source Projects}. In \bibinfo{booktitle}{\emph{Proceedings of the Ninth ACM Conference on Data and Application Security and Privacy}} (Richardson, Texas, USA) \emph{(\bibinfo{series}{CODASPY '19})}. \bibinfo{publisher}{Association for Computing Machinery}, \bibinfo{address}{New York, NY, USA}, \bibinfo{pages}{291–302}.
\newblock
\showISBNx{9781450360999}
\urldef\tempurl%
\url{https://doi.org/10.1145/3292006.3300032}
\showDOI{\tempurl}


\bibitem[Mercuri et~al\mbox{.}(2023)]%
        {a16100478}
\bibfield{author}{\bibinfo{person}{Valeria Mercuri}, \bibinfo{person}{Martina Saletta}, {and} \bibinfo{person}{Claudio Ferretti}.} \bibinfo{year}{2023}\natexlab{}.
\newblock \showarticletitle{Evolutionary Approaches for Adversarial Attacks on Neural Source Code Classifiers}.
\newblock \bibinfo{journal}{\emph{Algorithms}} \bibinfo{volume}{16}, \bibinfo{number}{10} (\bibinfo{year}{2023}).
\newblock
\showISSN{1999-4893}
\urldef\tempurl%
\url{https://doi.org/10.3390/a16100478}
\showDOI{\tempurl}


\bibitem[Meta({[n.\,d.]})]%
        {codellama}
\bibfield{author}{\bibinfo{person}{Meta}.} \bibinfo{year}{[n.\,d.]}\natexlab{}.
\newblock \bibinfo{title}{Code Llama}.
\newblock
\newblock
\urldef\tempurl%
\url{https://ai.meta.com/blog/code-llama-large-language-model-coding/}
\showURL{%
\tempurl}


\bibitem[Metropolis et~al\mbox{.}(1953)]%
        {metropolis1953equation}
\bibfield{author}{\bibinfo{person}{Nicholas Metropolis}, \bibinfo{person}{Arianna~W Rosenbluth}, \bibinfo{person}{Marshall~N Rosenbluth}, \bibinfo{person}{Augusta~H Teller}, {and} \bibinfo{person}{Edward Teller}.} \bibinfo{year}{1953}\natexlab{}.
\newblock \showarticletitle{Equation of state calculations by fast computing machines}.
\newblock \bibinfo{journal}{\emph{The journal of chemical physics}} \bibinfo{volume}{21}, \bibinfo{number}{6} (\bibinfo{year}{1953}), \bibinfo{pages}{1087--1092}.
\newblock


\bibitem[Mohammadkhani et~al\mbox{.}(2023a)]%
        {mohammadkhani2023systematic}
\bibfield{author}{\bibinfo{person}{Ahmad~Haji Mohammadkhani}, \bibinfo{person}{Nitin~Sai Bommi}, \bibinfo{person}{Mariem Daboussi}, \bibinfo{person}{Onkar Sabnis}, \bibinfo{person}{Chakkrit Tantithamthavorn}, {and} \bibinfo{person}{Hadi Hemmati}.} \bibinfo{year}{2023}\natexlab{a}.
\newblock \bibinfo{title}{A Systematic Literature Review of Explainable AI for Software Engineering}.
\newblock
\newblock
\showeprint[arxiv]{2302.06065}~[cs.SE]


\bibitem[Mohammadkhani et~al\mbox{.}(2022)]%
        {Mohammadkhani2022}
\bibfield{author}{\bibinfo{person}{Ahmad~Haji Mohammadkhani}, \bibinfo{person}{Chakkrit Tantithamthavorn}, {and} \bibinfo{person}{Hadi Hemmati}.} \bibinfo{year}{2022}\natexlab{}.
\newblock \showarticletitle{{Explainable AI for Pre-Trained Code Models: What Do They Learn? When They Do Not Work?}}
\newblock  (\bibinfo{year}{2022}).
\newblock
\showeprint[arxiv]{2211.12821}
\urldef\tempurl%
\url{http://arxiv.org/abs/2211.12821}
\showURL{%
\tempurl}


\bibitem[Mohammadkhani et~al\mbox{.}(2023b)]%
        {10356671}
\bibfield{author}{\bibinfo{person}{Ahmad~Haji Mohammadkhani}, \bibinfo{person}{Chakkrit Tantithamthavorn}, {and} \bibinfo{person}{Hadi Hemmatif}.} \bibinfo{year}{2023}\natexlab{b}.
\newblock \showarticletitle{Explaining Transformer-based Code Models: What Do They Learn? When They Do Not Work?}. In \bibinfo{booktitle}{\emph{2023 IEEE 23rd International Working Conference on Source Code Analysis and Manipulation (SCAM)}}. \bibinfo{publisher}{IEEE Computer Society}, \bibinfo{address}{Los Alamitos, CA, USA}, \bibinfo{pages}{96--106}.
\newblock
\urldef\tempurl%
\url{https://doi.org/10.1109/SCAM59687.2023.00020}
\showDOI{\tempurl}


\bibitem[Mou et~al\mbox{.}(2016)]%
        {OJClone}
\bibfield{author}{\bibinfo{person}{Lili Mou}, \bibinfo{person}{Ge Li}, \bibinfo{person}{Lu Zhang}, \bibinfo{person}{Tao Wang}, {and} \bibinfo{person}{Zhi Jin}.} \bibinfo{year}{2016}\natexlab{}.
\newblock \showarticletitle{Convolutional Neural Networks over Tree Structures for Programming Language Processing}. In \bibinfo{booktitle}{\emph{Proceedings of the Thirtieth AAAI Conference on Artificial Intelligence}} (Phoenix, Arizona) \emph{(\bibinfo{series}{AAAI'16})}. \bibinfo{publisher}{AAAI Press}, \bibinfo{pages}{1287–1293}.
\newblock


\bibitem[Mozannar et~al\mbox{.}(2023)]%
        {mozannar2023reading}
\bibfield{author}{\bibinfo{person}{Hussein Mozannar}, \bibinfo{person}{Gagan Bansal}, \bibinfo{person}{Adam Fourney}, {and} \bibinfo{person}{Eric Horvitz}.} \bibinfo{year}{2023}\natexlab{}.
\newblock \bibinfo{title}{When to Show a Suggestion? Integrating Human Feedback in AI-Assisted Programming}.
\newblock
\newblock
\showeprint[arxiv]{2306.04930}~[cs.HC]


\bibitem[Na et~al\mbox{.}(2023)]%
        {na-etal-2023-dip}
\bibfield{author}{\bibinfo{person}{CheolWon Na}, \bibinfo{person}{YunSeok Choi}, {and} \bibinfo{person}{Jee-Hyong Lee}.} \bibinfo{year}{2023}\natexlab{}.
\newblock \showarticletitle{{DIP}: Dead code Insertion based Black-box Attack for Programming Language Model}. In \bibinfo{booktitle}{\emph{Proceedings of the 61st Annual Meeting of the Association for Computational Linguistics (Volume 1: Long Papers)}}. \bibinfo{publisher}{Association for Computational Linguistics}, \bibinfo{address}{Toronto, Canada}, \bibinfo{pages}{7777--7791}.
\newblock
\urldef\tempurl%
\url{https://doi.org/10.18653/v1/2023.acl-long.430}
\showDOI{\tempurl}


\bibitem[Naik et~al\mbox{.}(2018)]%
        {naik-etal-2018-stress}
\bibfield{author}{\bibinfo{person}{Aakanksha Naik}, \bibinfo{person}{Abhilasha Ravichander}, \bibinfo{person}{Norman Sadeh}, \bibinfo{person}{Carolyn Rose}, {and} \bibinfo{person}{Graham Neubig}.} \bibinfo{year}{2018}\natexlab{}.
\newblock \showarticletitle{Stress Test Evaluation for Natural Language Inference}. In \bibinfo{booktitle}{\emph{Proceedings of the 27th International Conference on Computational Linguistics}}. \bibinfo{publisher}{Association for Computational Linguistics}, \bibinfo{address}{Santa Fe, New Mexico, USA}, \bibinfo{pages}{2340--2353}.
\newblock
\urldef\tempurl%
\url{https://aclanthology.org/C18-1198}
\showURL{%
\tempurl}


\bibitem[Nguyen et~al\mbox{.}(2021)]%
        {coffee}
\bibfield{author}{\bibinfo{person}{Phuong~T. Nguyen}, \bibinfo{person}{Claudio Di~Sipio}, \bibinfo{person}{Juri Di~Rocco}, \bibinfo{person}{Massimiliano Di~Penta}, {and} \bibinfo{person}{Davide Di~Ruscio}.} \bibinfo{year}{2021}\natexlab{}.
\newblock \showarticletitle{Adversarial Attacks to API Recommender Systems: Time to Wake Up and Smell the Coffee?}. In \bibinfo{booktitle}{\emph{2021 36th IEEE/ACM International Conference on Automated Software Engineering (ASE)}}. \bibinfo{pages}{253--265}.
\newblock
\urldef\tempurl%
\url{https://doi.org/10.1109/ASE51524.2021.9678946}
\showDOI{\tempurl}


\bibitem[Nguyen et~al\mbox{.}(2024)]%
        {nguyen2024beginning}
\bibfield{author}{\bibinfo{person}{Sydney Nguyen}, \bibinfo{person}{Hannah~McLean Babe}, \bibinfo{person}{Yangtian Zi}, \bibinfo{person}{Arjun Guha}, \bibinfo{person}{Carolyn~Jane Anderson}, {and} \bibinfo{person}{Molly~Q Feldman}.} \bibinfo{year}{2024}\natexlab{}.
\newblock \bibinfo{title}{How Beginning Programmers and Code LLMs (Mis)read Each Other}.
\newblock
\newblock
\showeprint[arxiv]{2401.15232}~[cs.HC]


\bibitem[Nguyen et~al\mbox{.}(2023)]%
        {nguyen2023adversarial}
\bibfield{author}{\bibinfo{person}{Thanh-Dat Nguyen}, \bibinfo{person}{Zhou Yang}, \bibinfo{person}{Xuan Bach~D. Le}, \bibinfo{person}{Patanamon}, \bibinfo{person}{Thongtanunam}, {and} \bibinfo{person}{David Lo}.} \bibinfo{year}{2023}\natexlab{}.
\newblock \bibinfo{title}{Adversarial Attacks on Code Models with Discriminative Graph Patterns}.
\newblock
\newblock
\showeprint[arxiv]{2308.11161}~[cs.SE]


\bibitem[Nguyen-Duc et~al\mbox{.}(2023)]%
        {nguyenduc2023generative}
\bibfield{author}{\bibinfo{person}{Anh Nguyen-Duc}, \bibinfo{person}{Beatriz Cabrero-Daniel}, \bibinfo{person}{Adam Przybylek}, \bibinfo{person}{Chetan Arora}, \bibinfo{person}{Dron Khanna}, \bibinfo{person}{Tomas Herda}, \bibinfo{person}{Usman Rafiq}, \bibinfo{person}{Jorge Melegati}, \bibinfo{person}{Eduardo Guerra}, \bibinfo{person}{Kai-Kristian Kemell}, \bibinfo{person}{Mika Saari}, \bibinfo{person}{Zheying Zhang}, \bibinfo{person}{Huy Le}, \bibinfo{person}{Tho Quan}, {and} \bibinfo{person}{Pekka Abrahamsson}.} \bibinfo{year}{2023}\natexlab{}.
\newblock \bibinfo{title}{Generative Artificial Intelligence for Software Engineering -- A Research Agenda}.
\newblock
\newblock
\showeprint[arxiv]{2310.18648}~[cs.SE]


\bibitem[Nijkamp et~al\mbox{.}(2023)]%
        {codegen}
\bibfield{author}{\bibinfo{person}{Erik Nijkamp}, \bibinfo{person}{Bo Pang}, \bibinfo{person}{Hiroaki Hayashi}, \bibinfo{person}{Lifu Tu}, \bibinfo{person}{Huan Wang}, \bibinfo{person}{Yingbo Zhou}, \bibinfo{person}{Silvio Savarese}, {and} \bibinfo{person}{Caiming Xiong}.} \bibinfo{year}{2023}\natexlab{}.
\newblock \showarticletitle{CodeGen: An Open Large Language Model for Code with Multi-Turn Program Synthesis}. In \bibinfo{booktitle}{\emph{The Eleventh International Conference on Learning Representations}}.
\newblock


\bibitem[Niu et~al\mbox{.}(2023a)]%
        {niu2023empirical}
\bibfield{author}{\bibinfo{person}{Changan Niu}, \bibinfo{person}{Chuanyi Li}, \bibinfo{person}{Vincent Ng}, \bibinfo{person}{Dongxiao Chen}, \bibinfo{person}{Jidong Ge}, {and} \bibinfo{person}{Bin Luo}.} \bibinfo{year}{2023}\natexlab{a}.
\newblock \bibinfo{title}{An Empirical Comparison of Pre-Trained Models of Source Code}.
\newblock
\newblock
\showeprint[arxiv]{2302.04026}~[cs.SE]


\bibitem[Niu et~al\mbox{.}(2023b)]%
        {291327}
\bibfield{author}{\bibinfo{person}{Liang Niu}, \bibinfo{person}{Shujaat Mirza}, \bibinfo{person}{Zayd Maradni}, {and} \bibinfo{person}{Christina P{\"o}pper}.} \bibinfo{year}{2023}\natexlab{b}.
\newblock \showarticletitle{{CodexLeaks}: Privacy Leaks from Code Generation Language Models in {GitHub} Copilot}. In \bibinfo{booktitle}{\emph{32nd USENIX Security Symposium (USENIX Security 23)}}. \bibinfo{publisher}{USENIX Association}, \bibinfo{address}{Anaheim, CA}, \bibinfo{pages}{2133--2150}.
\newblock
\showISBNx{978-1-939133-37-3}


\bibitem[Oh et~al\mbox{.}(2023)]%
        {oh2023poisoned}
\bibfield{author}{\bibinfo{person}{Sanghak Oh}, \bibinfo{person}{Kiho Lee}, \bibinfo{person}{Seonhye Park}, \bibinfo{person}{Doowon Kim}, {and} \bibinfo{person}{Hyoungshick Kim}.} \bibinfo{year}{2023}\natexlab{}.
\newblock \bibinfo{title}{Poisoned ChatGPT Finds Work for Idle Hands: Exploring Developers' Coding Practices with Insecure Suggestions from Poisoned AI Models}.
\newblock
\newblock
\showeprint[arxiv]{2312.06227}~[cs.CR]


\bibitem[Palacio et~al\mbox{.}(2023)]%
        {palacio2023evaluating}
\bibfield{author}{\bibinfo{person}{David~N Palacio}, \bibinfo{person}{Alejandro Velasco}, \bibinfo{person}{Daniel Rodriguez-Cardenas}, \bibinfo{person}{Kevin Moran}, {and} \bibinfo{person}{Denys Poshyvanyk}.} \bibinfo{year}{2023}\natexlab{}.
\newblock \bibinfo{title}{Evaluating and Explaining Large Language Models for Code Using Syntactic Structures}.
\newblock
\newblock
\showeprint[arxiv]{2308.03873}~[cs.SE]


\bibitem[Paltenghi et~al\mbox{.}(2022)]%
        {paltenghi2022extracting}
\bibfield{author}{\bibinfo{person}{Matteo Paltenghi}, \bibinfo{person}{Rahul Pandita}, \bibinfo{person}{Austin~Z. Henley}, {and} \bibinfo{person}{Albert Ziegler}.} \bibinfo{year}{2022}\natexlab{}.
\newblock \bibinfo{title}{Extracting Meaningful Attention on Source Code: An Empirical Study of Developer and Neural Model Code Exploration}.
\newblock
\newblock
\showeprint[arxiv]{2210.05506}~[cs.SE]


\bibitem[Paltenghi and Pradel(2021)]%
        {9678712}
\bibfield{author}{\bibinfo{person}{Matteo Paltenghi} {and} \bibinfo{person}{Michael Pradel}.} \bibinfo{year}{2021}\natexlab{}.
\newblock \showarticletitle{Thinking Like a Developer? Comparing the Attention of Humans with Neural Models of Code}. In \bibinfo{booktitle}{\emph{2021 36th IEEE/ACM International Conference on Automated Software Engineering (ASE)}}. \bibinfo{pages}{867--879}.
\newblock
\urldef\tempurl%
\url{https://doi.org/10.1109/ASE51524.2021.9678712}
\showDOI{\tempurl}


\bibitem[Papernot et~al\mbox{.}(2016)]%
        {papernot2016transferability}
\bibfield{author}{\bibinfo{person}{Nicolas Papernot}, \bibinfo{person}{Patrick McDaniel}, {and} \bibinfo{person}{Ian Goodfellow}.} \bibinfo{year}{2016}\natexlab{}.
\newblock \showarticletitle{Transferability in machine learning: from phenomena to black-box attacks using adversarial samples}.
\newblock \bibinfo{journal}{\emph{arXiv preprint arXiv:1605.07277}} (\bibinfo{year}{2016}).
\newblock


\bibitem[Pearce et~al\mbox{.}(2022)]%
        {DBLP:conf/sp/PearceA0DK22}
\bibfield{author}{\bibinfo{person}{Hammond Pearce}, \bibinfo{person}{Baleegh Ahmad}, \bibinfo{person}{Benjamin Tan}, \bibinfo{person}{Brendan Dolan{-}Gavitt}, {and} \bibinfo{person}{Ramesh Karri}.} \bibinfo{year}{2022}\natexlab{}.
\newblock \showarticletitle{Asleep at the Keyboard? Assessing the Security of GitHub Copilot's Code Contributions}. In \bibinfo{booktitle}{\emph{43rd {IEEE} Symposium on Security and Privacy, {SP} 2022, San Francisco, CA, USA, May 22-26, 2022}}. \bibinfo{publisher}{{IEEE}}, \bibinfo{pages}{754--768}.
\newblock
\urldef\tempurl%
\url{https://doi.org/10.1109/SP46214.2022.9833571}
\showDOI{\tempurl}


\bibitem[Peng et~al\mbox{.}(2023)]%
        {peng2023impact}
\bibfield{author}{\bibinfo{person}{Sida Peng}, \bibinfo{person}{Eirini Kalliamvakou}, \bibinfo{person}{Peter Cihon}, {and} \bibinfo{person}{Mert Demirer}.} \bibinfo{year}{2023}\natexlab{}.
\newblock \bibinfo{title}{The Impact of AI on Developer Productivity: Evidence from GitHub Copilot}.
\newblock
\newblock
\showeprint[arxiv]{2302.06590}~[cs.SE]


\bibitem[Polosukhin and Skidanov(2018)]%
        {polosukhin2018neural}
\bibfield{author}{\bibinfo{person}{Illia Polosukhin} {and} \bibinfo{person}{Alexander Skidanov}.} \bibinfo{year}{2018}\natexlab{}.
\newblock \bibinfo{title}{Neural Program Search: Solving Programming Tasks from Description and Examples}.
\newblock
\newblock
\showeprint[arxiv]{1802.04335}~[cs.AI]


\bibitem[Pornprasit et~al\mbox{.}(2022)]%
        {10.1109/ASE51524.2021.9678763}
\bibfield{author}{\bibinfo{person}{Chanathip Pornprasit}, \bibinfo{person}{Chakkrit Tantithamthavorn}, \bibinfo{person}{Jirayus Jiarpakdee}, \bibinfo{person}{Michael Fu}, {and} \bibinfo{person}{Patanamon Thongtanunam}.} \bibinfo{year}{2022}\natexlab{}.
\newblock \showarticletitle{PyExplainer: Explaining the Predictions of Just-in-Time Defect Models}. In \bibinfo{booktitle}{\emph{Proceedings of the 36th IEEE/ACM International Conference on Automated Software Engineering}} (Melbourne, Australia) \emph{(\bibinfo{series}{ASE '21})}. \bibinfo{publisher}{IEEE Press}, \bibinfo{pages}{407–418}.
\newblock
\showISBNx{9781665403375}
\urldef\tempurl%
\url{https://doi.org/10.1109/ASE51524.2021.9678763}
\showDOI{\tempurl}


\bibitem[Pour et~al\mbox{.}(2021)]%
        {9438605}
\bibfield{author}{\bibinfo{person}{Maryam~Vahdat Pour}, \bibinfo{person}{Zhuo Li}, \bibinfo{person}{Lei Ma}, {and} \bibinfo{person}{Hadi Hemmati}.} \bibinfo{year}{2021}\natexlab{}.
\newblock \showarticletitle{A Search-Based Testing Framework for Deep Neural Networks of Source Code Embedding}. In \bibinfo{booktitle}{\emph{14th {IEEE} Conference on Software Testing, Verification and Validation, {ICST} 2021, Porto de Galinhas, Brazil, April 12-16, 2021}}. \bibinfo{publisher}{{IEEE}}.
\newblock


\bibitem[Prather et~al\mbox{.}(2023)]%
        {10.1145/3617367}
\bibfield{author}{\bibinfo{person}{James Prather}, \bibinfo{person}{Brent~N. Reeves}, \bibinfo{person}{Paul Denny}, \bibinfo{person}{Brett~A. Becker}, \bibinfo{person}{Juho Leinonen}, \bibinfo{person}{Andrew Luxton-Reilly}, \bibinfo{person}{Garrett Powell}, \bibinfo{person}{James Finnie-Ansley}, {and} \bibinfo{person}{Eddie~Antonio Santos}.} \bibinfo{year}{2023}\natexlab{}.
\newblock \showarticletitle{“It's Weird That It Knows What I Want”: Usability and Interactions with Copilot for Novice Programmers}.
\newblock \bibinfo{journal}{\emph{ACM Trans. Comput.-Hum. Interact.}} (\bibinfo{date}{aug} \bibinfo{year}{2023}).
\newblock
\showISSN{1073-0516}
\urldef\tempurl%
\url{https://doi.org/10.1145/3617367}
\showDOI{\tempurl}
\newblock
\shownote{Just Accepted}.


\bibitem[Qi et~al\mbox{.}(2021)]%
        {qi-etal-2021-onion}
\bibfield{author}{\bibinfo{person}{Fanchao Qi}, \bibinfo{person}{Yangyi Chen}, \bibinfo{person}{Mukai Li}, \bibinfo{person}{Yuan Yao}, \bibinfo{person}{Zhiyuan Liu}, {and} \bibinfo{person}{Maosong Sun}.} \bibinfo{year}{2021}\natexlab{}.
\newblock \showarticletitle{{ONION}: A Simple and Effective Defense Against Textual Backdoor Attacks}. In \bibinfo{booktitle}{\emph{Proceedings of the 2021 Conference on Empirical Methods in Natural Language Processing}}. \bibinfo{publisher}{Association for Computational Linguistics}, \bibinfo{address}{Online and Punta Cana, Dominican Republic}, \bibinfo{pages}{9558--9566}.
\newblock
\urldef\tempurl%
\url{https://doi.org/10.18653/v1/2021.emnlp-main.752}
\showDOI{\tempurl}


\bibitem[Qi et~al\mbox{.}(2023)]%
        {qi2023badcs}
\bibfield{author}{\bibinfo{person}{Shiyi Qi}, \bibinfo{person}{Yuanhang Yang}, \bibinfo{person}{Shuzhzeng Gao}, \bibinfo{person}{Cuiyun Gao}, {and} \bibinfo{person}{Zenglin Xu}.} \bibinfo{year}{2023}\natexlab{}.
\newblock \bibinfo{title}{BadCS: A Backdoor Attack Framework for Code search}.
\newblock
\newblock


\bibitem[Quiring et~al\mbox{.}(2019)]%
        {10.5555/3361338.3361372}
\bibfield{author}{\bibinfo{person}{Erwin Quiring}, \bibinfo{person}{Alwin Maier}, {and} \bibinfo{person}{Konrad Rieck}.} \bibinfo{year}{2019}\natexlab{}.
\newblock \showarticletitle{Misleading Authorship Attribution of Source Code Using Adversarial Learning}. In \bibinfo{booktitle}{\emph{Proceedings of the 28th USENIX Conference on Security Symposium}} (Santa Clara, CA, USA) \emph{(\bibinfo{series}{SEC'19})}. \bibinfo{publisher}{USENIX Association}, \bibinfo{address}{USA}, \bibinfo{pages}{479–496}.
\newblock
\showISBNx{9781939133069}


\bibitem[Rabin and Alipour(2021)]%
        {rabin2021evaluation}
\bibfield{author}{\bibinfo{person}{Md~Rafiqul~Islam Rabin} {and} \bibinfo{person}{Mohammad~Amin Alipour}.} \bibinfo{year}{2021}\natexlab{}.
\newblock \bibinfo{title}{Evaluation of Generalizability of Neural Program Analyzers under Semantic-Preserving Transformations}.
\newblock
\newblock
\showeprint[arxiv]{2004.07313}~[cs.SE]


\bibitem[Rabin and Alipour(2022)]%
        {RABIN2022100432}
\bibfield{author}{\bibinfo{person}{Md~Rafiqul~Islam Rabin} {and} \bibinfo{person}{Mohammad~Amin Alipour}.} \bibinfo{year}{2022}\natexlab{}.
\newblock \showarticletitle{FeatureExtractor: A tool for extracting key input features of code intelligence models}.
\newblock \bibinfo{journal}{\emph{Software Impacts}}  \bibinfo{volume}{14} (\bibinfo{year}{2022}), \bibinfo{pages}{100432}.
\newblock
\showISSN{2665-9638}
\urldef\tempurl%
\url{https://doi.org/10.1016/j.simpa.2022.100432}
\showDOI{\tempurl}


\bibitem[Rabin et~al\mbox{.}(2021a)]%
        {RABIN2021106552}
\bibfield{author}{\bibinfo{person}{Md~Rafiqul~Islam Rabin}, \bibinfo{person}{Nghi~D.Q. Bui}, \bibinfo{person}{Ke Wang}, \bibinfo{person}{Yijun Yu}, \bibinfo{person}{Lingxiao Jiang}, {and} \bibinfo{person}{Mohammad~Amin Alipour}.} \bibinfo{year}{2021}\natexlab{a}.
\newblock \showarticletitle{On the generalizability of Neural Program Models with respect to semantic-preserving program transformations}.
\newblock \bibinfo{journal}{\emph{Information and Software Technology}}  \bibinfo{volume}{135} (\bibinfo{year}{2021}), \bibinfo{pages}{106552}.
\newblock
\showISSN{0950-5849}
\urldef\tempurl%
\url{https://doi.org/10.1016/j.infsof.2021.106552}
\showDOI{\tempurl}


\bibitem[Rabin et~al\mbox{.}(2021b)]%
        {rabin2021generalizability}
\bibfield{author}{\bibinfo{person}{Md~Rafiqul~Islam Rabin}, \bibinfo{person}{Nghi~DQ Bui}, \bibinfo{person}{Ke Wang}, \bibinfo{person}{Yijun Yu}, \bibinfo{person}{Lingxiao Jiang}, {and} \bibinfo{person}{Mohammad~Amin Alipour}.} \bibinfo{year}{2021}\natexlab{b}.
\newblock \showarticletitle{On the generalizability of Neural Program Models with respect to semantic-preserving program transformations}.
\newblock \bibinfo{journal}{\emph{Information and Software Technology}}  \bibinfo{volume}{135} (\bibinfo{year}{2021}), \bibinfo{pages}{106552}.
\newblock


\bibitem[Rabin et~al\mbox{.}(2021c)]%
        {10.1145/3468264.3468539}
\bibfield{author}{\bibinfo{person}{Md~Rafiqul~Islam Rabin}, \bibinfo{person}{Vincent~J. Hellendoorn}, {and} \bibinfo{person}{Mohammad~Amin Alipour}.} \bibinfo{year}{2021}\natexlab{c}.
\newblock \showarticletitle{Understanding neural code intelligence through program simplification}. In \bibinfo{booktitle}{\emph{Proceedings of the 29th ACM Joint Meeting on European Software Engineering Conference and Symposium on the Foundations of Software Engineering}} (Athens, Greece) \emph{(\bibinfo{series}{ESEC/FSE 2021})}. \bibinfo{publisher}{Association for Computing Machinery}, \bibinfo{address}{New York, NY, USA}, \bibinfo{pages}{441–452}.
\newblock
\showISBNx{9781450385626}
\urldef\tempurl%
\url{https://doi.org/10.1145/3468264.3468539}
\showDOI{\tempurl}


\bibitem[Rabin et~al\mbox{.}(2023)]%
        {RABIN2023107066}
\bibfield{author}{\bibinfo{person}{Md~Rafiqul~Islam Rabin}, \bibinfo{person}{Aftab Hussain}, \bibinfo{person}{Mohammad~Amin Alipour}, {and} \bibinfo{person}{Vincent~J. Hellendoorn}.} \bibinfo{year}{2023}\natexlab{}.
\newblock \showarticletitle{Memorization and generalization in neural code intelligence models}.
\newblock \bibinfo{journal}{\emph{Information and Software Technology}}  \bibinfo{volume}{153} (\bibinfo{year}{2023}), \bibinfo{pages}{107066}.
\newblock
\showISSN{0950-5849}
\urldef\tempurl%
\url{https://doi.org/10.1016/j.infsof.2022.107066}
\showDOI{\tempurl}


\bibitem[Rabin et~al\mbox{.}(2019)]%
        {rabin2019testing}
\bibfield{author}{\bibinfo{person}{Md~Rafiqul~Islam Rabin}, \bibinfo{person}{Ke Wang}, {and} \bibinfo{person}{Mohammad~Amin Alipour}.} \bibinfo{year}{2019}\natexlab{}.
\newblock \bibinfo{title}{Testing Neural Program Analyzers}.
\newblock
\newblock
\showeprint[arxiv]{1908.10711}~[cs.LG]


\bibitem[Ramakrishnan and Albarghouthi(2022)]%
        {codebackdoor}
\bibfield{author}{\bibinfo{person}{Goutham Ramakrishnan} {and} \bibinfo{person}{Aws Albarghouthi}.} \bibinfo{year}{2022}\natexlab{}.
\newblock \showarticletitle{Backdoors in Neural Models of Source Code}. In \bibinfo{booktitle}{\emph{2022 26th International Conference on Pattern Recognition (ICPR)}}. \bibinfo{publisher}{IEEE Computer Society}, \bibinfo{address}{Los Alamitos, CA, USA}, \bibinfo{pages}{2892--2899}.
\newblock
\urldef\tempurl%
\url{https://doi.org/10.1109/ICPR56361.2022.9956690}
\showDOI{\tempurl}


\bibitem[Raychev et~al\mbox{.}(2016)]%
        {py150}
\bibfield{author}{\bibinfo{person}{Veselin Raychev}, \bibinfo{person}{Pavol Bielik}, {and} \bibinfo{person}{Martin Vechev}.} \bibinfo{year}{2016}\natexlab{}.
\newblock \showarticletitle{Probabilistic Model for Code with Decision Trees}. In \bibinfo{booktitle}{\emph{Proceedings of the 2016 ACM SIGPLAN International Conference on Object-Oriented Programming, Systems, Languages, and Applications}} (Amsterdam, Netherlands) \emph{(\bibinfo{series}{OOPSLA 2016})}. \bibinfo{publisher}{Association for Computing Machinery}, \bibinfo{address}{New York, NY, USA}, \bibinfo{pages}{731–747}.
\newblock
\showISBNx{9781450344449}
\urldef\tempurl%
\url{https://doi.org/10.1145/2983990.2984041}
\showDOI{\tempurl}


\bibitem[Ren et~al\mbox{.}(2019)]%
        {10.1145/3324916}
\bibfield{author}{\bibinfo{person}{Xiaoxue Ren}, \bibinfo{person}{Zhenchang Xing}, \bibinfo{person}{Xin Xia}, \bibinfo{person}{David Lo}, \bibinfo{person}{Xinyu Wang}, {and} \bibinfo{person}{John Grundy}.} \bibinfo{year}{2019}\natexlab{}.
\newblock \showarticletitle{Neural Network-Based Detection of Self-Admitted Technical Debt: From Performance to Explainability}.
\newblock \bibinfo{journal}{\emph{ACM Trans. Softw. Eng. Methodol.}} \bibinfo{volume}{28}, \bibinfo{number}{3}, Article \bibinfo{articleno}{15} (\bibinfo{date}{jul} \bibinfo{year}{2019}), \bibinfo{numpages}{45}~pages.
\newblock
\showISSN{1049-331X}
\urldef\tempurl%
\url{https://doi.org/10.1145/3324916}
\showDOI{\tempurl}


\bibitem[Ribeiro et~al\mbox{.}(2016)]%
        {LIME}
\bibfield{author}{\bibinfo{person}{Marco~Tulio Ribeiro}, \bibinfo{person}{Sameer Singh}, {and} \bibinfo{person}{Carlos Guestrin}.} \bibinfo{year}{2016}\natexlab{}.
\newblock \showarticletitle{"Why Should I Trust You?": Explaining the Predictions of Any Classifier}. In \bibinfo{booktitle}{\emph{Proceedings of the 22nd ACM SIGKDD International Conference on Knowledge Discovery and Data Mining}} (San Francisco, California, USA) \emph{(\bibinfo{series}{KDD '16})}. \bibinfo{publisher}{Association for Computing Machinery}, \bibinfo{address}{New York, NY, USA}, \bibinfo{pages}{1135–1144}.
\newblock
\showISBNx{9781450342322}
\urldef\tempurl%
\url{https://doi.org/10.1145/2939672.2939778}
\showDOI{\tempurl}


\bibitem[Rodriguez-Cardenas et~al\mbox{.}(2023)]%
        {10336302}
\bibfield{author}{\bibinfo{person}{Daniel Rodriguez-Cardenas}, \bibinfo{person}{David~N. Palacio}, \bibinfo{person}{Dipin Khati}, \bibinfo{person}{Henry Burke}, {and} \bibinfo{person}{Denys Poshyvanyk}.} \bibinfo{year}{2023}\natexlab{}.
\newblock \showarticletitle{Benchmarking Causal Study to Interpret Large Language Models for Source Code}. In \bibinfo{booktitle}{\emph{2023 IEEE International Conference on Software Maintenance and Evolution (ICSME)}}. \bibinfo{publisher}{IEEE Computer Society}, \bibinfo{address}{Los Alamitos, CA, USA}, \bibinfo{pages}{329--334}.
\newblock
\urldef\tempurl%
\url{https://doi.org/10.1109/ICSME58846.2023.00040}
\showDOI{\tempurl}


\bibitem[Roy et~al\mbox{.}(2022)]%
        {9978217}
\bibfield{author}{\bibinfo{person}{Saumendu Roy}, \bibinfo{person}{Gabriel Laberge}, \bibinfo{person}{Banani Roy}, \bibinfo{person}{Foutse Khomh}, \bibinfo{person}{Amin Nikanjam}, {and} \bibinfo{person}{Saikat Mondal}.} \bibinfo{year}{2022}\natexlab{}.
\newblock \showarticletitle{Why Don’t XAI Techniques Agree? Characterizing the Disagreements Between Post-hoc Explanations of Defect Predictions}. In \bibinfo{booktitle}{\emph{2022 IEEE International Conference on Software Maintenance and Evolution (ICSME)}}. \bibinfo{publisher}{IEEE Computer Society}, \bibinfo{address}{Los Alamitos, CA, USA}, \bibinfo{pages}{444--448}.
\newblock
\urldef\tempurl%
\url{https://doi.org/10.1109/ICSME55016.2022.00056}
\showDOI{\tempurl}


\bibitem[Saad and Sharma(2023)]%
        {saad2023naturalness}
\bibfield{author}{\bibinfo{person}{Mootez Saad} {and} \bibinfo{person}{Tushar Sharma}.} \bibinfo{year}{2023}\natexlab{}.
\newblock \bibinfo{title}{Naturalness of Attention: Revisiting Attention in Code Language Models}.
\newblock
\newblock
\showeprint[arxiv]{2311.13508}~[cs.SE]


\bibitem[Saberi et~al\mbox{.}(2024)]%
        {saberi2024utilization}
\bibfield{author}{\bibinfo{person}{Iman Saberi}, \bibinfo{person}{Fatemeh Fard}, {and} \bibinfo{person}{Fuxiang Chen}.} \bibinfo{year}{2024}\natexlab{}.
\newblock \bibinfo{title}{Utilization of Pre-trained Language Model for Adapter-based Knowledge Transfer in Software Engineering}.
\newblock
\newblock
\showeprint[arxiv]{2307.08540}~[cs.SE]


\bibitem[Sandoval et~al\mbox{.}(2023)]%
        {sandoval2023lost}
\bibfield{author}{\bibinfo{person}{Gustavo Sandoval}, \bibinfo{person}{Hammond Pearce}, \bibinfo{person}{Teo Nys}, \bibinfo{person}{Ramesh Karri}, \bibinfo{person}{Siddharth Garg}, {and} \bibinfo{person}{Brendan Dolan-Gavitt}.} \bibinfo{year}{2023}\natexlab{}.
\newblock \showarticletitle{Lost at C: A User Study on the Security Implications of Large Language Model Code Assistants}. In \bibinfo{booktitle}{\emph{32nd USENIX Security Symposium (USENIX Security 23)}}. \bibinfo{publisher}{USENIX Association}, \bibinfo{address}{Anaheim, CA}, \bibinfo{pages}{2205--2222}.
\newblock
\showISBNx{978-1-939133-37-3}
\urldef\tempurl%
\url{https://www.usenix.org/conference/usenixsecurity23/presentation/sandoval}
\showURL{%
\tempurl}


\bibitem[Schuster et~al\mbox{.}(2021)]%
        {263874}
\bibfield{author}{\bibinfo{person}{Roei Schuster}, \bibinfo{person}{Congzheng Song}, \bibinfo{person}{Eran Tromer}, {and} \bibinfo{person}{Vitaly Shmatikov}.} \bibinfo{year}{2021}\natexlab{}.
\newblock \showarticletitle{You Autocomplete Me: Poisoning Vulnerabilities in Neural Code Completion}. In \bibinfo{booktitle}{\emph{30th USENIX Security Symposium (USENIX Security 21)}}. \bibinfo{publisher}{USENIX Association}, \bibinfo{pages}{1559--1575}.
\newblock
\showISBNx{978-1-939133-24-3}


\bibitem[Sharma et~al\mbox{.}(2022)]%
        {10.1145/3524610.3527921}
\bibfield{author}{\bibinfo{person}{Rishab Sharma}, \bibinfo{person}{Fuxiang Chen}, \bibinfo{person}{Fatemeh Fard}, {and} \bibinfo{person}{David Lo}.} \bibinfo{year}{2022}\natexlab{}.
\newblock \showarticletitle{An Exploratory Study on Code Attention in BERT}. In \bibinfo{booktitle}{\emph{Proceedings of the 30th IEEE/ACM International Conference on Program Comprehension}} (Virtual Event) \emph{(\bibinfo{series}{ICPC '22})}. \bibinfo{publisher}{Association for Computing Machinery}, \bibinfo{address}{New York, NY, USA}, \bibinfo{pages}{437–448}.
\newblock
\showISBNx{9781450392983}
\urldef\tempurl%
\url{https://doi.org/10.1145/3524610.3527921}
\showDOI{\tempurl}


\bibitem[She et~al\mbox{.}(2023)]%
        {she2023pitfalls}
\bibfield{author}{\bibinfo{person}{Xinyu She}, \bibinfo{person}{Yue Liu}, \bibinfo{person}{Yanjie Zhao}, \bibinfo{person}{Yiling He}, \bibinfo{person}{Li Li}, \bibinfo{person}{Chakkrit Tantithamthavorn}, \bibinfo{person}{Zhan Qin}, {and} \bibinfo{person}{Haoyu Wang}.} \bibinfo{year}{2023}\natexlab{}.
\newblock \bibinfo{title}{Pitfalls in Language Models for Code Intelligence: A Taxonomy and Survey}.
\newblock
\newblock
\showeprint[arxiv]{2310.17903}~[cs.SE]


\bibitem[Shi et~al\mbox{.}(2023d)]%
        {10.1145/3609437.3609438}
\bibfield{author}{\bibinfo{person}{Chaoxuan Shi}, \bibinfo{person}{Tingwei Zhu}, \bibinfo{person}{Tian Zhang}, \bibinfo{person}{Jun Pang}, {and} \bibinfo{person}{Minxue Pan}.} \bibinfo{year}{2023}\natexlab{d}.
\newblock \showarticletitle{Structural-semantics Guided Program Simplification for Understanding Neural Code Intelligence Models}. In \bibinfo{booktitle}{\emph{Proceedings of the 14th Asia-Pacific Symposium on Internetware}} \emph{(\bibinfo{series}{Internetware '23})}. \bibinfo{publisher}{Association for Computing Machinery}, \bibinfo{address}{New York, NY, USA}, \bibinfo{pages}{1–11}.
\newblock
\showISBNx{9798400708947}
\urldef\tempurl%
\url{https://doi.org/10.1145/3609437.3609438}
\showDOI{\tempurl}


\bibitem[Shi et~al\mbox{.}(2023a)]%
        {10.1145/3597926.3598036}
\bibfield{author}{\bibinfo{person}{Ensheng Shi}, \bibinfo{person}{Yanlin Wang}, \bibinfo{person}{Hongyu Zhang}, \bibinfo{person}{Lun Du}, \bibinfo{person}{Shi Han}, \bibinfo{person}{Dongmei Zhang}, {and} \bibinfo{person}{Hongbin Sun}.} \bibinfo{year}{2023}\natexlab{a}.
\newblock \showarticletitle{Towards Efficient Fine-Tuning of Pre-trained Code Models: An Experimental Study and Beyond}. In \bibinfo{booktitle}{\emph{Proceedings of the 32nd ACM SIGSOFT International Symposium on Software Testing and Analysis}} \emph{(\bibinfo{series}{ISSTA 2023})}. \bibinfo{publisher}{Association for Computing Machinery}, \bibinfo{address}{New York, NY, USA}, \bibinfo{pages}{39–51}.
\newblock
\showISBNx{9798400702211}
\urldef\tempurl%
\url{https://doi.org/10.1145/3597926.3598036}
\showDOI{\tempurl}


\bibitem[Shi et~al\mbox{.}(2023b)]%
        {avatar}
\bibfield{author}{\bibinfo{person}{Jieke Shi}, \bibinfo{person}{Zhou Yang}, \bibinfo{person}{Hong~Jin Kang}, \bibinfo{person}{Bowen Xu}, \bibinfo{person}{Junda He}, {and} \bibinfo{person}{David Lo}.} \bibinfo{year}{2023}\natexlab{b}.
\newblock \showarticletitle{Smaller, Faster, Greener: Compressing Pre-trained Code Models via Surrogate-Assisted Optimization}.
\newblock \bibinfo{journal}{\emph{arXiv preprint arXiv:2309.04076}} (\bibinfo{year}{2023}).
\newblock


\bibitem[Shi et~al\mbox{.}(2023c)]%
        {compressor}
\bibfield{author}{\bibinfo{person}{Jieke Shi}, \bibinfo{person}{Zhou Yang}, \bibinfo{person}{Bowen Xu}, \bibinfo{person}{Hong~Jin Kang}, {and} \bibinfo{person}{David Lo}.} \bibinfo{year}{2023}\natexlab{c}.
\newblock \showarticletitle{Compressing Pre-Trained Models of Code into 3 MB} \emph{(\bibinfo{series}{ASE '22})}. \bibinfo{publisher}{Association for Computing Machinery}, \bibinfo{address}{New York, NY, USA}, Article \bibinfo{articleno}{24}, \bibinfo{numpages}{12}~pages.
\newblock
\showISBNx{9781450394758}
\urldef\tempurl%
\url{https://doi.org/10.1145/3551349.3556964}
\showDOI{\tempurl}


\bibitem[Shin et~al\mbox{.}(2021)]%
        {shin2021explainable}
\bibfield{author}{\bibinfo{person}{Jiho Shin}, \bibinfo{person}{Reem Aleithan}, \bibinfo{person}{Jaechang Nam}, \bibinfo{person}{Junjie Wang}, {and} \bibinfo{person}{Song Wang}.} \bibinfo{year}{2021}\natexlab{}.
\newblock \bibinfo{title}{Explainable Software Defect Prediction: Are We There Yet?}
\newblock
\newblock


\bibitem[Shirafuji et~al\mbox{.}(2023)]%
        {shirafuji2023exploring}
\bibfield{author}{\bibinfo{person}{Atsushi Shirafuji}, \bibinfo{person}{Yutaka Watanobe}, \bibinfo{person}{Takumi Ito}, \bibinfo{person}{Makoto Morishita}, \bibinfo{person}{Yuki Nakamura}, \bibinfo{person}{Yusuke Oda}, {and} \bibinfo{person}{Jun Suzuki}.} \bibinfo{year}{2023}\natexlab{}.
\newblock \bibinfo{title}{Exploring the Robustness of Large Language Models for Solving Programming Problems}.
\newblock
\newblock
\showeprint[arxiv]{2306.14583}~[cs.CL]


\bibitem[Shiri et~al\mbox{.}(2023)]%
        {shiri2023paraphrasing}
\bibfield{author}{\bibinfo{person}{Fatemeh Shiri}, \bibinfo{person}{Terry~Yue Zhuo}, \bibinfo{person}{Zhuang Li}, \bibinfo{person}{Van Nguyen}, \bibinfo{person}{Shirui Pan}, \bibinfo{person}{Weiqing Wang}, \bibinfo{person}{Reza Haffari}, {and} \bibinfo{person}{Yuan-Fang Li}.} \bibinfo{year}{2023}\natexlab{}.
\newblock \bibinfo{title}{Paraphrasing Techniques for Maritime QA system}.
\newblock
\newblock
\showeprint[arxiv]{2203.10854}~[cs.CL]


\bibitem[Siddiq et~al\mbox{.}(2022)]%
        {10006873}
\bibfield{author}{\bibinfo{person}{Mohammed~Latif Siddiq}, \bibinfo{person}{Shafayat~H. Majumder}, \bibinfo{person}{Maisha~R. Mim}, \bibinfo{person}{Sourov Jajodia}, {and} \bibinfo{person}{Joanna C.~S. Santos}.} \bibinfo{year}{2022}\natexlab{}.
\newblock \showarticletitle{An Empirical Study of Code Smells in Transformer-based Code Generation Techniques}. In \bibinfo{booktitle}{\emph{2022 IEEE 22nd International Working Conference on Source Code Analysis and Manipulation (SCAM)}}. \bibinfo{pages}{71--82}.
\newblock
\urldef\tempurl%
\url{https://doi.org/10.1109/SCAM55253.2022.00014}
\showDOI{\tempurl}


\bibitem[Simko et~al\mbox{.}(2018)]%
        {simko2018recognizing}
\bibfield{author}{\bibinfo{person}{Lucy Simko}, \bibinfo{person}{Luke Zettlemoyer}, {and} \bibinfo{person}{Tadayoshi Kohno}.} \bibinfo{year}{2018}\natexlab{}.
\newblock \showarticletitle{Recognizing and Imitating Programmer Style: Adversaries in Program Authorship Attribution.}
\newblock \bibinfo{journal}{\emph{Proc. Priv. Enhancing Technol.}} \bibinfo{volume}{2018}, \bibinfo{number}{1} (\bibinfo{year}{2018}), \bibinfo{pages}{127--144}.
\newblock


\bibitem[Song and Ding(2023)]%
        {10197074}
\bibfield{author}{\bibinfo{person}{Leo Song} {and} \bibinfo{person}{Steven~H.H. Ding}.} \bibinfo{year}{2023}\natexlab{}.
\newblock \showarticletitle{Milo: Attacking Deep Pre-trained Model for Programming Languages Tasks with Anti-analysis Code Obfuscation}. In \bibinfo{booktitle}{\emph{COMPSAC}}. \bibinfo{pages}{586--594}.
\newblock
\urldef\tempurl%
\url{https://doi.org/10.1109/COMPSAC57700.2023.00084}
\showDOI{\tempurl}


\bibitem[Springer et~al\mbox{.}(2021)]%
        {springer2021strata}
\bibfield{author}{\bibinfo{person}{Jacob~M. Springer}, \bibinfo{person}{Bryn~Marie Reinstadler}, {and} \bibinfo{person}{Una-May O'Reilly}.} \bibinfo{year}{2021}\natexlab{}.
\newblock \bibinfo{title}{STRATA: Simple, Gradient-Free Attacks for Models of Code}.
\newblock
\newblock


\bibitem[Srikant et~al\mbox{.}(2021)]%
        {Epresentation2021}
\bibfield{author}{\bibinfo{person}{Shashank Srikant}, \bibinfo{person}{Sijia Liu}, \bibinfo{person}{Tamara Mitrovska}, \bibinfo{person}{Shiyu Chang}, \bibinfo{person}{Quanfu Fan}, \bibinfo{person}{Gaoyuan Zhang}, {and} \bibinfo{person}{Una{-}May O'Reilly}.} \bibinfo{year}{2021}\natexlab{}.
\newblock \showarticletitle{{Generating Adversarial Computer Programs using Optimized Obfuscations}}.
\newblock \bibinfo{journal}{\emph{ICLR}}  \bibinfo{volume}{16} (\bibinfo{year}{2021}), \bibinfo{pages}{209--226}.
\newblock


\bibitem[Staniak and Biecek(2018)]%
        {breakdown}
\bibfield{author}{\bibinfo{person}{Mateusz Staniak} {and} \bibinfo{person}{Przemyslaw Biecek}.} \bibinfo{year}{2018}\natexlab{}.
\newblock \showarticletitle{Explanations of model predictions with live and breakDown packages}.
\newblock  (\bibinfo{year}{2018}).
\newblock


\bibitem[Su and McMillan(2024)]%
        {Su2024}
\bibfield{author}{\bibinfo{person}{Chia-Yi Su} {and} \bibinfo{person}{Collin McMillan}.} \bibinfo{year}{2024}\natexlab{}.
\newblock \showarticletitle{Distilled GPT for source code summarization}.
\newblock \bibinfo{journal}{\emph{Automated Software Engineering}} \bibinfo{volume}{31}, \bibinfo{number}{1} (\bibinfo{year}{2024}), \bibinfo{pages}{22}.
\newblock
\showISSN{1573-7535}
\urldef\tempurl%
\url{https://doi.org/10.1007/s10515-024-00421-4}
\showDOI{\tempurl}


\bibitem[Sun et~al\mbox{.}(2022b)]%
        {10.1145/3490099.3511119}
\bibfield{author}{\bibinfo{person}{Jiao Sun}, \bibinfo{person}{Q.~Vera Liao}, \bibinfo{person}{Michael Muller}, \bibinfo{person}{Mayank Agarwal}, \bibinfo{person}{Stephanie Houde}, \bibinfo{person}{Kartik Talamadupula}, {and} \bibinfo{person}{Justin~D. Weisz}.} \bibinfo{year}{2022}\natexlab{b}.
\newblock \showarticletitle{Investigating Explainability of Generative AI for Code through Scenario-Based Design}. In \bibinfo{booktitle}{\emph{27th International Conference on Intelligent User Interfaces}} (Helsinki, Finland) \emph{(\bibinfo{series}{IUI '22})}. \bibinfo{publisher}{Association for Computing Machinery}, \bibinfo{address}{New York, NY, USA}, \bibinfo{pages}{212–228}.
\newblock
\showISBNx{9781450391443}
\urldef\tempurl%
\url{https://doi.org/10.1145/3490099.3511119}
\showDOI{\tempurl}


\bibitem[Sun et~al\mbox{.}(2023a)]%
        {Sun2023backdoor}
\bibfield{author}{\bibinfo{person}{Weisong Sun}, \bibinfo{person}{Yuchen Chen}, \bibinfo{person}{Guanhong Tao}, \bibinfo{person}{Chunrong Fang}, \bibinfo{person}{Xiangyu Zhang}, \bibinfo{person}{Quanjun Zhang}, {and} \bibinfo{person}{Bin Luo}.} \bibinfo{year}{2023}\natexlab{a}.
\newblock \showarticletitle{Backdooring Neural Code Search}. In \bibinfo{booktitle}{\emph{Proceedings of the 61st Annual Meeting of the Association for Computational Linguistics (Volume 1: Long Papers)}}. \bibinfo{publisher}{Association for Computational Linguistics}, \bibinfo{address}{Toronto, Canada}, \bibinfo{pages}{9692--9708}.
\newblock
\urldef\tempurl%
\url{https://doi.org/10.18653/v1/2023.acl-long.540}
\showDOI{\tempurl}


\bibitem[Sun et~al\mbox{.}(2023b)]%
        {CodeMark}
\bibfield{author}{\bibinfo{person}{Zhensu Sun}, \bibinfo{person}{Xiaoning Du}, \bibinfo{person}{Fu Song}, {and} \bibinfo{person}{Li Li}.} \bibinfo{year}{2023}\natexlab{b}.
\newblock \showarticletitle{CodeMark: Imperceptible Watermarking for Code Datasets against Neural Code Completion Models}. In \bibinfo{booktitle}{\emph{Proceedings of the 31st ACM Joint European Software Engineering Conference and Symposium on the Foundations of Software Engineering}} \emph{(\bibinfo{series}{ESEC/FSE 2023})}. \bibinfo{publisher}{Association for Computing Machinery}, \bibinfo{address}{New York, NY, USA}, \bibinfo{pages}{1561–1572}.
\newblock
\showISBNx{9798400703270}
\urldef\tempurl%
\url{https://doi.org/10.1145/3611643.3616297}
\showDOI{\tempurl}


\bibitem[Sun et~al\mbox{.}(2022a)]%
        {CoProtector}
\bibfield{author}{\bibinfo{person}{Zhensu Sun}, \bibinfo{person}{Xiaoning Du}, \bibinfo{person}{Fu Song}, \bibinfo{person}{Mingze Ni}, {and} \bibinfo{person}{Li Li}.} \bibinfo{year}{2022}\natexlab{a}.
\newblock \showarticletitle{CoProtector: Protect Open-Source Code against Unauthorized Training Usage with Data Poisoning}. In \bibinfo{booktitle}{\emph{Proceedings of the ACM Web Conference 2022}} (Virtual Event, Lyon, France) \emph{(\bibinfo{series}{WWW '22})}. \bibinfo{publisher}{Association for Computing Machinery}, \bibinfo{address}{New York, NY, USA}, \bibinfo{pages}{652–660}.
\newblock
\showISBNx{9781450390965}
\urldef\tempurl%
\url{https://doi.org/10.1145/3485447.3512225}
\showDOI{\tempurl}


\bibitem[Sun et~al\mbox{.}(2024)]%
        {sun2024neural}
\bibfield{author}{\bibinfo{person}{Zhensu Sun}, \bibinfo{person}{Xiaoning Du}, \bibinfo{person}{Fu Song}, \bibinfo{person}{Shangwen Wang}, {and} \bibinfo{person}{Li Li}.} \bibinfo{year}{2024}\natexlab{}.
\newblock \bibinfo{title}{When Neural Code Completion Models Size up the Situation: Attaining Cheaper and Faster Completion through Dynamic Model Inference}.
\newblock
\newblock
\showeprint[arxiv]{2401.09964}~[cs.SE]


\bibitem[Sun et~al\mbox{.}(2023c)]%
        {10172653}
\bibfield{author}{\bibinfo{person}{Zhensu Sun}, \bibinfo{person}{Xiaoning Du}, \bibinfo{person}{Fu Song}, \bibinfo{person}{Shangwen Wang}, \bibinfo{person}{Mingze Ni}, {and} \bibinfo{person}{Li Li}.} \bibinfo{year}{2023}\natexlab{c}.
\newblock \showarticletitle{Don't Complete It! Preventing Unhelpful Code Completion for Productive and Sustainable Neural Code Completion Systems}. In \bibinfo{booktitle}{\emph{2023 IEEE/ACM 45th International Conference on Software Engineering: Companion Proceedings (ICSE-Companion)}}. \bibinfo{pages}{324--325}.
\newblock
\urldef\tempurl%
\url{https://doi.org/10.1109/ICSE-Companion58688.2023.00089}
\showDOI{\tempurl}


\bibitem[Svajlenko et~al\mbox{.}(2014)]%
        {BigCloneBench}
\bibfield{author}{\bibinfo{person}{Jeffrey Svajlenko}, \bibinfo{person}{Judith~F. Islam}, \bibinfo{person}{Iman Keivanloo}, \bibinfo{person}{Chanchal~K. Roy}, {and} \bibinfo{person}{Mohammad~Mamun Mia}.} \bibinfo{year}{2014}\natexlab{}.
\newblock \showarticletitle{Towards a Big Data Curated Benchmark of Inter-project Code Clones}. In \bibinfo{booktitle}{\emph{2014 ICSME}}. \bibinfo{pages}{476--480}.
\newblock
\urldef\tempurl%
\url{https://doi.org/10.1109/ICSME.2014.77}
\showDOI{\tempurl}


\bibitem[Svyatkovskiy et~al\mbox{.}(2021)]%
        {svyatkovskiy2021fast}
\bibfield{author}{\bibinfo{person}{Alexey Svyatkovskiy}, \bibinfo{person}{Sebastian Lee}, \bibinfo{person}{Anna Hadjitofi}, \bibinfo{person}{Maik Riechert}, \bibinfo{person}{Juliana~Vicente Franco}, {and} \bibinfo{person}{Miltiadis Allamanis}.} \bibinfo{year}{2021}\natexlab{}.
\newblock \showarticletitle{Fast and memory-efficient neural code completion}. In \bibinfo{booktitle}{\emph{2021 IEEE/ACM 18th International Conference on Mining Software Repositories (MSR)}}. IEEE, \bibinfo{pages}{329--340}.
\newblock


\bibitem[Tian et~al\mbox{.}(2023b)]%
        {tian2023chatgpt}
\bibfield{author}{\bibinfo{person}{Haoye Tian}, \bibinfo{person}{Weiqi Lu}, \bibinfo{person}{Tsz~On Li}, \bibinfo{person}{Xunzhu Tang}, \bibinfo{person}{Shing-Chi Cheung}, \bibinfo{person}{Jacques Klein}, {and} \bibinfo{person}{Tegawendé~F. Bissyandé}.} \bibinfo{year}{2023}\natexlab{b}.
\newblock \bibinfo{title}{Is ChatGPT the Ultimate Programming Assistant -- How far is it?}
\newblock
\newblock
\showeprint[arxiv]{2304.11938}~[cs.SE]


\bibitem[Tian et~al\mbox{.}(2021)]%
        {9724884}
\bibfield{author}{\bibinfo{person}{Junfeng Tian}, \bibinfo{person}{Chenxin Wang}, \bibinfo{person}{Zhen Li}, {and} \bibinfo{person}{Yu Wen}.} \bibinfo{year}{2021}\natexlab{}.
\newblock \showarticletitle{Generating Adversarial Examples of Source Code Classification Models via Q-Learning-Based Markov Decision Process}. In \bibinfo{booktitle}{\emph{2021 IEEE 21st International Conference on Software Quality, Reliability and Security (QRS)}}. \bibinfo{pages}{807--818}.
\newblock
\urldef\tempurl%
\url{https://doi.org/10.1109/QRS54544.2021.00090}
\showDOI{\tempurl}


\bibitem[Tian et~al\mbox{.}(2023a)]%
        {coda}
\bibfield{author}{\bibinfo{person}{Zhao Tian}, \bibinfo{person}{Junjie Chen}, {and} \bibinfo{person}{Zhi Jin}.} \bibinfo{year}{2023}\natexlab{a}.
\newblock \bibinfo{title}{Code Difference Guided Adversarial Example Generation for Deep Code Models}.
\newblock , \bibinfo{numpages}{850-862}~pages.
\newblock


\bibitem[Tran et~al\mbox{.}(2018)]%
        {spectral}
\bibfield{author}{\bibinfo{person}{Brandon Tran}, \bibinfo{person}{Jerry Li}, {and} \bibinfo{person}{Aleksander Madry}.} \bibinfo{year}{2018}\natexlab{}.
\newblock \showarticletitle{Spectral Signatures in Backdoor Attacks}. In \bibinfo{booktitle}{\emph{Advances in Neural Information Processing Systems}}, \bibfield{editor}{\bibinfo{person}{S.~Bengio}, \bibinfo{person}{H.~Wallach}, \bibinfo{person}{H.~Larochelle}, \bibinfo{person}{K.~Grauman}, \bibinfo{person}{N.~Cesa-Bianchi}, {and} \bibinfo{person}{R.~Garnett}} (Eds.), Vol.~\bibinfo{volume}{31}. \bibinfo{publisher}{Curran Associates, Inc.}
\newblock


\bibitem[Troshin and Chirkova(2022)]%
        {troshin-chirkova-2022-probing}
\bibfield{author}{\bibinfo{person}{Sergey Troshin} {and} \bibinfo{person}{Nadezhda Chirkova}.} \bibinfo{year}{2022}\natexlab{}.
\newblock \showarticletitle{Probing Pretrained Models of Source Codes}. In \bibinfo{booktitle}{\emph{Proceedings of the Fifth BlackboxNLP Workshop on Analyzing and Interpreting Neural Networks for NLP}}, \bibfield{editor}{\bibinfo{person}{Jasmijn Bastings}, \bibinfo{person}{Yonatan Belinkov}, \bibinfo{person}{Yanai Elazar}, \bibinfo{person}{Dieuwke Hupkes}, \bibinfo{person}{Naomi Saphra}, {and} \bibinfo{person}{Sarah Wiegreffe}} (Eds.). \bibinfo{publisher}{Association for Computational Linguistics}, \bibinfo{address}{Abu Dhabi, United Arab Emirates (Hybrid)}, \bibinfo{pages}{371--383}.
\newblock
\urldef\tempurl%
\url{https://doi.org/10.18653/v1/2022.blackboxnlp-1.31}
\showDOI{\tempurl}


\bibitem[Tufano et~al\mbox{.}(2019)]%
        {10.1145/3340544}
\bibfield{author}{\bibinfo{person}{Michele Tufano}, \bibinfo{person}{Cody Watson}, \bibinfo{person}{Gabriele Bavota}, \bibinfo{person}{Massimiliano~Di Penta}, \bibinfo{person}{Martin White}, {and} \bibinfo{person}{Denys Poshyvanyk}.} \bibinfo{year}{2019}\natexlab{}.
\newblock \showarticletitle{An Empirical Study on Learning Bug-Fixing Patches in the Wild via Neural Machine Translation}.
\newblock \bibinfo{journal}{\emph{ACM Trans. Softw. Eng. Methodol.}} \bibinfo{volume}{28}, \bibinfo{number}{4}, Article \bibinfo{articleno}{19} (\bibinfo{date}{sep} \bibinfo{year}{2019}), \bibinfo{numpages}{29}~pages.
\newblock
\showISSN{1049-331X}
\urldef\tempurl%
\url{https://doi.org/10.1145/3340544}
\showDOI{\tempurl}


\bibitem[Vaithilingam et~al\mbox{.}(2023)]%
        {10172834}
\bibfield{author}{\bibinfo{person}{Priyan Vaithilingam}, \bibinfo{person}{Elena~L. Glassman}, \bibinfo{person}{Peter Groenwegen}, \bibinfo{person}{Sumit Gulwani}, \bibinfo{person}{Austin~Z. Henley}, \bibinfo{person}{Rohan Malpani}, \bibinfo{person}{David Pugh}, \bibinfo{person}{Arjun Radhakrishna}, \bibinfo{person}{Gustavo Soares}, \bibinfo{person}{Joey Wang}, {and} \bibinfo{person}{Aaron Yim}.} \bibinfo{year}{2023}\natexlab{}.
\newblock \showarticletitle{Towards More Effective AI-Assisted Programming: A Systematic Design Exploration to Improve Visual Studio IntelliCode’s User Experience}. In \bibinfo{booktitle}{\emph{2023 IEEE/ACM 45th International Conference on Software Engineering: Software Engineering in Practice (ICSE-SEIP)}}. \bibinfo{pages}{185--195}.
\newblock
\urldef\tempurl%
\url{https://doi.org/10.1109/ICSE-SEIP58684.2023.00022}
\showDOI{\tempurl}


\bibitem[Vaithilingam et~al\mbox{.}(2022)]%
        {10.1145/3491101.3519665}
\bibfield{author}{\bibinfo{person}{Priyan Vaithilingam}, \bibinfo{person}{Tianyi Zhang}, {and} \bibinfo{person}{Elena~L. Glassman}.} \bibinfo{year}{2022}\natexlab{}.
\newblock \showarticletitle{Expectation vs. Experience: Evaluating the Usability of Code Generation Tools Powered by Large Language Models}. In \bibinfo{booktitle}{\emph{Extended Abstracts of the 2022 CHI Conference on Human Factors in Computing Systems}} (New Orleans, LA, USA) \emph{(\bibinfo{series}{CHI EA '22})}. \bibinfo{publisher}{Association for Computing Machinery}, \bibinfo{address}{New York, NY, USA}, Article \bibinfo{articleno}{332}, \bibinfo{numpages}{7}~pages.
\newblock
\showISBNx{9781450391566}
\urldef\tempurl%
\url{https://doi.org/10.1145/3491101.3519665}
\showDOI{\tempurl}


\bibitem[Vasconcelos et~al\mbox{.}(2023)]%
        {vasconcelos2023generation}
\bibfield{author}{\bibinfo{person}{Helena Vasconcelos}, \bibinfo{person}{Gagan Bansal}, \bibinfo{person}{Adam Fourney}, \bibinfo{person}{Q.~Vera Liao}, {and} \bibinfo{person}{Jennifer~Wortman Vaughan}.} \bibinfo{year}{2023}\natexlab{}.
\newblock \bibinfo{title}{Generation Probabilities Are Not Enough: Exploring the Effectiveness of Uncertainty Highlighting in AI-Powered Code Completions}.
\newblock
\newblock
\showeprint[arxiv]{2302.07248}~[cs.HC]


\bibitem[Wan et~al\mbox{.}(2022a)]%
        {you-see}
\bibfield{author}{\bibinfo{person}{Yao Wan}, \bibinfo{person}{Shijie Zhang}, \bibinfo{person}{Hongyu Zhang}, \bibinfo{person}{Yulei Sui}, \bibinfo{person}{Guandong Xu}, \bibinfo{person}{Dezhong Yao}, \bibinfo{person}{Hai Jin}, {and} \bibinfo{person}{Lichao Sun}.} \bibinfo{year}{2022}\natexlab{a}.
\newblock \showarticletitle{You See What I Want You to See: Poisoning Vulnerabilities in Neural Code Search}. In \bibinfo{booktitle}{\emph{Proceedings of the 30th ACM Joint European Software Engineering Conference and Symposium on the Foundations of Software Engineering}} (Singapore, Singapore) \emph{(\bibinfo{series}{ESEC/FSE 2022})}. \bibinfo{publisher}{Association for Computing Machinery}, \bibinfo{address}{New York, NY, USA}, \bibinfo{pages}{1233–1245}.
\newblock
\showISBNx{9781450394130}
\urldef\tempurl%
\url{https://doi.org/10.1145/3540250.3549153}
\showDOI{\tempurl}


\bibitem[Wan et~al\mbox{.}(2022b)]%
        {10.1145/3510003.3510050}
\bibfield{author}{\bibinfo{person}{Yao Wan}, \bibinfo{person}{Wei Zhao}, \bibinfo{person}{Hongyu Zhang}, \bibinfo{person}{Yulei Sui}, \bibinfo{person}{Guandong Xu}, {and} \bibinfo{person}{Hai Jin}.} \bibinfo{year}{2022}\natexlab{b}.
\newblock \showarticletitle{What Do They Capture? A Structural Analysis of Pre-Trained Language Models for Source Code}. In \bibinfo{booktitle}{\emph{Proceedings of the 44th International Conference on Software Engineering}} (Pittsburgh, Pennsylvania) \emph{(\bibinfo{series}{ICSE '22})}. \bibinfo{publisher}{Association for Computing Machinery}, \bibinfo{address}{New York, NY, USA}, \bibinfo{pages}{2377–2388}.
\newblock
\showISBNx{9781450392211}
\urldef\tempurl%
\url{https://doi.org/10.1145/3510003.3510050}
\showDOI{\tempurl}


\bibitem[Wang et~al\mbox{.}(2021c)]%
        {wang2021adversarial}
\bibfield{author}{\bibinfo{person}{Boxin Wang}, \bibinfo{person}{Chejian Xu}, \bibinfo{person}{Shuohang Wang}, \bibinfo{person}{Zhe Gan}, \bibinfo{person}{Yu Cheng}, \bibinfo{person}{Jianfeng Gao}, \bibinfo{person}{Ahmed~Hassan Awadallah}, {and} \bibinfo{person}{Bo Li}.} \bibinfo{year}{2021}\natexlab{c}.
\newblock \showarticletitle{Adversarial {GLUE}: A Multi-Task Benchmark for Robustness Evaluation of Language Models}. In \bibinfo{booktitle}{\emph{Thirty-fifth Conference on Neural Information Processing Systems Datasets and Benchmarks Track (Round 2)}}.
\newblock
\urldef\tempurl%
\url{https://openreview.net/forum?id=GF9cSKI3A_q}
\showURL{%
\tempurl}


\bibitem[Wang et~al\mbox{.}(2023a)]%
        {wang2023adapter}
\bibfield{author}{\bibinfo{person}{Deze Wang}, \bibinfo{person}{Boxing Chen}, \bibinfo{person}{Shanshan Li}, \bibinfo{person}{Wei Luo}, \bibinfo{person}{Shaoliang Peng}, \bibinfo{person}{Wei Dong}, {and} \bibinfo{person}{Xiangke Liao}.} \bibinfo{year}{2023}\natexlab{a}.
\newblock \bibinfo{title}{One Adapter for All Programming Languages? Adapter Tuning for Code Search and Summarization}.
\newblock
\newblock
\showeprint[arxiv]{2303.15822}~[cs.SE]


\bibitem[Wang et~al\mbox{.}(2023b)]%
        {wang2023investigating}
\bibfield{author}{\bibinfo{person}{Ruotong Wang}, \bibinfo{person}{Ruijia Cheng}, \bibinfo{person}{Denae Ford}, {and} \bibinfo{person}{Thomas Zimmermann}.} \bibinfo{year}{2023}\natexlab{b}.
\newblock \bibinfo{title}{Investigating and Designing for Trust in AI-powered Code Generation Tools}.
\newblock
\newblock
\showeprint[arxiv]{2305.11248}~[cs.HC]


\bibitem[Wang et~al\mbox{.}(2023c)]%
        {wang2023-recode}
\bibfield{author}{\bibinfo{person}{Shiqi Wang}, \bibinfo{person}{Zheng Li}, \bibinfo{person}{Haifeng Qian}, \bibinfo{person}{Chenghao Yang}, \bibinfo{person}{Zijian Wang}, \bibinfo{person}{Mingyue Shang}, \bibinfo{person}{Varun Kumar}, \bibinfo{person}{Samson Tan}, \bibinfo{person}{Baishakhi Ray}, \bibinfo{person}{Parminder Bhatia}, \bibinfo{person}{Ramesh Nallapati}, \bibinfo{person}{Murali~Krishna Ramanathan}, \bibinfo{person}{Dan Roth}, {and} \bibinfo{person}{Bing Xiang}.} \bibinfo{year}{2023}\natexlab{c}.
\newblock \showarticletitle{{R}e{C}ode: Robustness Evaluation of Code Generation Models}. In \bibinfo{booktitle}{\emph{Proceedings of the 61st Annual Meeting of the Association for Computational Linguistics (Volume 1: Long Papers)}}. \bibinfo{publisher}{Association for Computational Linguistics}, \bibinfo{address}{Toronto, Canada}, \bibinfo{pages}{13818--13843}.
\newblock
\urldef\tempurl%
\url{https://doi.org/10.18653/v1/2023.acl-long.773}
\showDOI{\tempurl}


\bibitem[Wang et~al\mbox{.}(2021a)]%
        {10.1145/3324884.3416583}
\bibfield{author}{\bibinfo{person}{Xin Wang}, \bibinfo{person}{Jin Liu}, \bibinfo{person}{Li Li}, \bibinfo{person}{Xiao Chen}, \bibinfo{person}{Xiao Liu}, {and} \bibinfo{person}{Hao Wu}.} \bibinfo{year}{2021}\natexlab{a}.
\newblock \showarticletitle{Detecting and Explaining Self-Admitted Technical Debts with Attention-Based Neural Networks}. In \bibinfo{booktitle}{\emph{Proceedings of the 35th IEEE/ACM International Conference on Automated Software Engineering}} (Virtual Event, Australia) \emph{(\bibinfo{series}{ASE '20})}. \bibinfo{publisher}{Association for Computing Machinery}, \bibinfo{address}{New York, NY, USA}, \bibinfo{pages}{871–882}.
\newblock
\showISBNx{9781450367684}
\urldef\tempurl%
\url{https://doi.org/10.1145/3324884.3416583}
\showDOI{\tempurl}


\bibitem[Wang et~al\mbox{.}(2022)]%
        {wang2022robust}
\bibfield{author}{\bibinfo{person}{Yizhen Wang}, \bibinfo{person}{Mohannad Alhanahnah}, \bibinfo{person}{Xiaozhu Meng}, \bibinfo{person}{Ke Wang}, \bibinfo{person}{Mihai Christodorescu}, {and} \bibinfo{person}{Somesh Jha}.} \bibinfo{year}{2022}\natexlab{}.
\newblock \showarticletitle{Robust learning against relational adversaries}.
\newblock \bibinfo{journal}{\emph{Advances in Neural Information Processing Systems}}  \bibinfo{volume}{35} (\bibinfo{year}{2022}), \bibinfo{pages}{16246--16260}.
\newblock


\bibitem[Wang and Wang(2023)]%
        {wang2023demystifying}
\bibfield{author}{\bibinfo{person}{Yu Wang} {and} \bibinfo{person}{Ke Wang}.} \bibinfo{year}{2023}\natexlab{}.
\newblock \bibinfo{title}{Demystifying What Code Summarization Models Learned}.
\newblock
\newblock
\showeprint[arxiv]{2303.02333}~[cs.PL]


\bibitem[Wang et~al\mbox{.}(2023d)]%
        {WheaCha}
\bibfield{author}{\bibinfo{person}{Yu Wang}, \bibinfo{person}{Ke Wang}, {and} \bibinfo{person}{Linzhang Wang}.} \bibinfo{year}{2023}\natexlab{d}.
\newblock \showarticletitle{An Explanation Method for Models of Code}.
\newblock \bibinfo{journal}{\emph{Proc. ACM Program. Lang.}} \bibinfo{volume}{7}, \bibinfo{number}{OOPSLA2}, Article \bibinfo{articleno}{250} (\bibinfo{date}{oct} \bibinfo{year}{2023}), \bibinfo{numpages}{27}~pages.
\newblock
\urldef\tempurl%
\url{https://doi.org/10.1145/3622826}
\showDOI{\tempurl}


\bibitem[Wang et~al\mbox{.}(2021b)]%
        {wang2021codet5}
\bibfield{author}{\bibinfo{person}{Yue Wang}, \bibinfo{person}{Weishi Wang}, \bibinfo{person}{Shafiq Joty}, {and} \bibinfo{person}{Steven~C.H. Hoi}.} \bibinfo{year}{2021}\natexlab{b}.
\newblock \showarticletitle{CodeT5: Identifier-aware Unified Pre-trained Encoder-Decoder Models for Code Understanding and Generation}. In \bibinfo{booktitle}{\emph{Proceedings of the 2021 Conference on Empirical Methods in Natural Language Processing, EMNLP 2021}}.
\newblock


\bibitem[Wei et~al\mbox{.}(2022)]%
        {9916170}
\bibfield{author}{\bibinfo{person}{Moshi Wei}, \bibinfo{person}{Yuchao Huang}, \bibinfo{person}{Jinqiu Yang}, \bibinfo{person}{Junjie Wang}, {and} \bibinfo{person}{Song Wang}.} \bibinfo{year}{2022}\natexlab{}.
\newblock \showarticletitle{CoCoFuzzing: Testing Neural Code Models With Coverage-Guided Fuzzing}.
\newblock \bibinfo{journal}{\emph{IEEE Transactions on Reliability}} (\bibinfo{year}{2022}), \bibinfo{pages}{1--14}.
\newblock
\urldef\tempurl%
\url{https://doi.org/10.1109/TR.2022.3208239}
\showDOI{\tempurl}


\bibitem[Wei et~al\mbox{.}(2023)]%
        {10.1145/3611643.3616302}
\bibfield{author}{\bibinfo{person}{Xiaokai Wei}, \bibinfo{person}{Sujan~Kumar Gonugondla}, \bibinfo{person}{Shiqi Wang}, \bibinfo{person}{Wasi Ahmad}, \bibinfo{person}{Baishakhi Ray}, \bibinfo{person}{Haifeng Qian}, \bibinfo{person}{Xiaopeng Li}, \bibinfo{person}{Varun Kumar}, \bibinfo{person}{Zijian Wang}, \bibinfo{person}{Yuchen Tian}, \bibinfo{person}{Qing Sun}, \bibinfo{person}{Ben Athiwaratkun}, \bibinfo{person}{Mingyue Shang}, \bibinfo{person}{Murali~Krishna Ramanathan}, \bibinfo{person}{Parminder Bhatia}, {and} \bibinfo{person}{Bing Xiang}.} \bibinfo{year}{2023}\natexlab{}.
\newblock \showarticletitle{Towards Greener Yet Powerful Code Generation via Quantization: An Empirical Study} \emph{(\bibinfo{series}{ESEC/FSE 2023})}. \bibinfo{pages}{224–236}.
\newblock
\showISBNx{9798400703270}
\urldef\tempurl%
\url{https://doi.org/10.1145/3611643.3616302}
\showDOI{\tempurl}


\bibitem[Weisz et~al\mbox{.}(2022)]%
        {10.1145/3490099.3511157}
\bibfield{author}{\bibinfo{person}{Justin~D. Weisz}, \bibinfo{person}{Michael Muller}, \bibinfo{person}{Steven~I. Ross}, \bibinfo{person}{Fernando Martinez}, \bibinfo{person}{Stephanie Houde}, \bibinfo{person}{Mayank Agarwal}, \bibinfo{person}{Kartik Talamadupula}, {and} \bibinfo{person}{John~T. Richards}.} \bibinfo{year}{2022}\natexlab{}.
\newblock \showarticletitle{Better Together? An Evaluation of AI-Supported Code Translation}. In \bibinfo{booktitle}{\emph{27th International Conference on Intelligent User Interfaces}} (Helsinki, Finland) \emph{(\bibinfo{series}{IUI '22})}. \bibinfo{publisher}{Association for Computing Machinery}, \bibinfo{address}{New York, NY, USA}, \bibinfo{pages}{369–391}.
\newblock
\showISBNx{9781450391443}
\urldef\tempurl%
\url{https://doi.org/10.1145/3490099.3511157}
\showDOI{\tempurl}


\bibitem[Weyssow et~al\mbox{.}(2024)]%
        {weyssow2024exploring}
\bibfield{author}{\bibinfo{person}{Martin Weyssow}, \bibinfo{person}{Xin Zhou}, \bibinfo{person}{Kisub Kim}, \bibinfo{person}{David Lo}, {and} \bibinfo{person}{Houari Sahraoui}.} \bibinfo{year}{2024}\natexlab{}.
\newblock \bibinfo{title}{Exploring Parameter-Efficient Fine-Tuning Techniques for Code Generation with Large Language Models}.
\newblock
\newblock
\showeprint[arxiv]{2308.10462}~[cs.SE]


\bibitem[Wohlin(2014)]%
        {snowball2}
\bibfield{author}{\bibinfo{person}{Claes Wohlin}.} \bibinfo{year}{2014}\natexlab{}.
\newblock \showarticletitle{Guidelines for Snowballing in Systematic Literature Studies and a Replication in Software Engineering}. In \bibinfo{booktitle}{\emph{Proceedings of the 18th International Conference on Evaluation and Assessment in Software Engineering}} (London, England, United Kingdom) \emph{(\bibinfo{series}{EASE '14})}. \bibinfo{publisher}{Association for Computing Machinery}, \bibinfo{address}{New York, NY, USA}, Article \bibinfo{articleno}{38}, \bibinfo{numpages}{10}~pages.
\newblock
\showISBNx{9781450324762}
\urldef\tempurl%
\url{https://doi.org/10.1145/2601248.2601268}
\showDOI{\tempurl}


\bibitem[Wu et~al\mbox{.}(2023)]%
        {wu2023deceptprompt}
\bibfield{author}{\bibinfo{person}{Fangzhou Wu}, \bibinfo{person}{Xiaogeng Liu}, {and} \bibinfo{person}{Chaowei Xiao}.} \bibinfo{year}{2023}\natexlab{}.
\newblock \bibinfo{title}{DeceptPrompt: Exploiting LLM-driven Code Generation via Adversarial Natural Language Instructions}.
\newblock
\newblock
\showeprint[arxiv]{2312.04730}~[cs.CR]


\bibitem[Xiao et~al\mbox{.}(2023)]%
        {xiao2023devgpt}
\bibfield{author}{\bibinfo{person}{Tao Xiao}, \bibinfo{person}{Christoph Treude}, \bibinfo{person}{Hideaki Hata}, {and} \bibinfo{person}{Kenichi Matsumoto}.} \bibinfo{year}{2023}\natexlab{}.
\newblock \showarticletitle{DevGPT: Studying Developer-ChatGPT Conversations}.
\newblock \bibinfo{journal}{\emph{arXiv preprint arXiv:2309.03914}} (\bibinfo{year}{2023}).
\newblock


\bibitem[Yamamoto et~al\mbox{.}(2023)]%
        {10123655}
\bibfield{author}{\bibinfo{person}{Hiroki Yamamoto}, \bibinfo{person}{Dong Wang}, \bibinfo{person}{Gopi~Krishnan Rajbahadur}, \bibinfo{person}{Masanari Kondo}, \bibinfo{person}{Yasutaka Kamei}, {and} \bibinfo{person}{Naoyasu Ubayashi}.} \bibinfo{year}{2023}\natexlab{}.
\newblock \showarticletitle{Towards Privacy Preserving Cross Project Defect Prediction with Federated Learning}. In \bibinfo{booktitle}{\emph{2023 IEEE International Conference on Software Analysis, Evolution and Reengineering (SANER)}}. \bibinfo{pages}{485--496}.
\newblock
\urldef\tempurl%
\url{https://doi.org/10.1109/SANER56733.2023.00052}
\showDOI{\tempurl}


\bibitem[Yan et~al\mbox{.}(2023)]%
        {yan2023coco}
\bibfield{author}{\bibinfo{person}{Ming Yan}, \bibinfo{person}{Junjie Chen}, \bibinfo{person}{Jie~M. Zhang}, \bibinfo{person}{Xuejie Cao}, \bibinfo{person}{Chen Yang}, {and} \bibinfo{person}{Mark Harman}.} \bibinfo{year}{2023}\natexlab{}.
\newblock \bibinfo{title}{COCO: Testing Code Generation Systems via Concretized Instructions}.
\newblock
\newblock
\showeprint[arxiv]{2308.13319}~[cs.SE]


\bibitem[Yang et~al\mbox{.}(2022b)]%
        {goodname}
\bibfield{author}{\bibinfo{person}{Guang Yang}, \bibinfo{person}{Yu Zhou}, \bibinfo{person}{Wenhua Yang}, \bibinfo{person}{Tao Yue}, \bibinfo{person}{Xiang Chen}, {and} \bibinfo{person}{Taolue Chen}.} \bibinfo{year}{2022}\natexlab{b}.
\newblock \bibinfo{title}{How Important are Good Method Names in Neural Code Generation? A Model Robustness Perspective}.
\newblock
\newblock
\urldef\tempurl%
\url{https://doi.org/10.48550/ARXIV.2211.15844}
\showDOI{\tempurl}


\bibitem[Yang et~al\mbox{.}(2023c)]%
        {10.1145/3630010}
\bibfield{author}{\bibinfo{person}{Guang Yang}, \bibinfo{person}{Yu Zhou}, \bibinfo{person}{Wenhua Yang}, \bibinfo{person}{Tao Yue}, \bibinfo{person}{Xiang Chen}, {and} \bibinfo{person}{Taolue Chen}.} \bibinfo{year}{2023}\natexlab{c}.
\newblock \showarticletitle{How Important Are Good Method Names in Neural Code Generation? A Model Robustness Perspective}.
\newblock \bibinfo{journal}{\emph{ACM Trans. Softw. Eng. Methodol.}} (\bibinfo{date}{oct} \bibinfo{year}{2023}).
\newblock
\showISSN{1049-331X}
\urldef\tempurl%
\url{https://doi.org/10.1145/3630010}
\showDOI{\tempurl}
\newblock
\shownote{Just Accepted}.


\bibitem[Yang et~al\mbox{.}(2023d)]%
        {yang2023assessing}
\bibfield{author}{\bibinfo{person}{Guang Yang}, \bibinfo{person}{Yu Zhou}, \bibinfo{person}{Xiangyu Zhang}, \bibinfo{person}{Xiang Chen}, \bibinfo{person}{Tingting Han}, {and} \bibinfo{person}{Taolue Chen}.} \bibinfo{year}{2023}\natexlab{d}.
\newblock \bibinfo{title}{Assessing and Improving Syntactic Adversarial Robustness of Pre-trained Models for Code Translation}.
\newblock
\newblock
\showeprint[arxiv]{2310.18587}~[cs.SE]


\bibitem[Yang et~al\mbox{.}(2017)]%
        {yang2017authorship}
\bibfield{author}{\bibinfo{person}{Xinyu Yang}, \bibinfo{person}{Guoai Xu}, \bibinfo{person}{Qi Li}, \bibinfo{person}{Yanhui Guo}, {and} \bibinfo{person}{Miao Zhang}.} \bibinfo{year}{2017}\natexlab{}.
\newblock \showarticletitle{Authorship attribution of source code by using back propagation neural network based on particle swarm optimization}.
\newblock \bibinfo{journal}{\emph{PloS one}} \bibinfo{volume}{12}, \bibinfo{number}{11} (\bibinfo{year}{2017}), \bibinfo{pages}{e0187204}.
\newblock


\bibitem[Yang et~al\mbox{.}(2021)]%
        {10.1007/978-3-030-92635-9_25}
\bibfield{author}{\bibinfo{person}{Xingguang Yang}, \bibinfo{person}{Huiqun Yu}, \bibinfo{person}{Guisheng Fan}, \bibinfo{person}{Zijie Huang}, \bibinfo{person}{Kang Yang}, {and} \bibinfo{person}{Ziyi Zhou}.} \bibinfo{year}{2021}\natexlab{}.
\newblock \showarticletitle{An Empirical Study of Model-Agnostic Interpretation Technique for Just-in-Time Software Defect Prediction}. In \bibinfo{booktitle}{\emph{Collaborative Computing: Networking, Applications and Worksharing}}, \bibfield{editor}{\bibinfo{person}{Honghao Gao} {and} \bibinfo{person}{Xinheng Wang}} (Eds.). \bibinfo{publisher}{Springer International Publishing}, \bibinfo{address}{Cham}, \bibinfo{pages}{420--438}.
\newblock
\showISBNx{978-3-030-92635-9}


\bibitem[Yang et~al\mbox{.}(2022a)]%
        {alert}
\bibfield{author}{\bibinfo{person}{Zhou Yang}, \bibinfo{person}{Jieke Shi}, \bibinfo{person}{Junda He}, {and} \bibinfo{person}{David Lo}.} \bibinfo{year}{2022}\natexlab{a}.
\newblock \showarticletitle{Natural Attack for Pre-Trained Models of Code}. In \bibinfo{booktitle}{\emph{Proceedings of the 44th International Conference on Software Engineering}} (Pittsburgh, Pennsylvania) \emph{(\bibinfo{series}{ICSE '22})}. \bibinfo{publisher}{Association for Computing Machinery}, \bibinfo{address}{New York, NY, USA}, \bibinfo{pages}{1482–1493}.
\newblock
\showISBNx{9781450392211}
\urldef\tempurl%
\url{https://doi.org/10.1145/3510003.3510146}
\showDOI{\tempurl}


\bibitem[Yang et~al\mbox{.}({[n.\,d.]})]%
        {advdoor}
\bibfield{author}{\bibinfo{person}{Zhou Yang}, \bibinfo{person}{Bowen Xu}, \bibinfo{person}{Jie~M. Zhang}, \bibinfo{person}{Hong~Jin Kang}, \bibinfo{person}{Jieke Shi}, \bibinfo{person}{Junda He}, {and} \bibinfo{person}{David Lo}.} \bibinfo{year}{[n.\,d.]}\natexlab{}.
\newblock \showarticletitle{Stealthy Backdoor Attack for Code Models}.
\newblock \bibinfo{journal}{\emph{IEEE Transactions on Software Engineering}} \bibinfo{number}{01} (\bibinfo{date}{feb} \bibinfo{year}{[n.\,d.]}), \bibinfo{pages}{1--21}.
\newblock
\showISSN{1939-3520}
\urldef\tempurl%
\url{https://doi.org/10.1109/TSE.2024.3361661}
\showDOI{\tempurl}


\bibitem[Yang et~al\mbox{.}(2023a)]%
        {yang2023gotcha}
\bibfield{author}{\bibinfo{person}{Zhou Yang}, \bibinfo{person}{Zhipeng Zhao}, \bibinfo{person}{Chenyu Wang}, \bibinfo{person}{Jieke Shi}, \bibinfo{person}{Dongsum Kim}, \bibinfo{person}{Donggyun Han}, {and} \bibinfo{person}{David Lo}.} \bibinfo{year}{2023}\natexlab{a}.
\newblock \bibinfo{title}{Gotcha! This Model Uses My Code! Evaluating Membership Leakage Risks in Code Models}.
\newblock
\newblock
\showeprint[arxiv]{2310.01166}~[cs.SE]


\bibitem[Yang et~al\mbox{.}(2023b)]%
        {yang2023memorzation}
\bibfield{author}{\bibinfo{person}{Zhou Yang}, \bibinfo{person}{Zhipeng Zhao}, \bibinfo{person}{Chenyu Wang}, \bibinfo{person}{Jieke Shi}, \bibinfo{person}{Dongsun Kim}, \bibinfo{person}{DongGyun Han}, {and} \bibinfo{person}{David Lo}.} \bibinfo{year}{2023}\natexlab{b}.
\newblock \bibinfo{title}{Unveiling Memorization in Code Models}.
\newblock
\newblock


\bibitem[Yefet et~al\mbox{.}(2020)]%
        {Yefet2020}
\bibfield{author}{\bibinfo{person}{Noam Yefet}, \bibinfo{person}{Uri Alon}, {and} \bibinfo{person}{Eran Yahav}.} \bibinfo{year}{2020}\natexlab{}.
\newblock \bibinfo{journal}{\emph{Proceedings of the ACM on Programming Languages}} \bibinfo{volume}{4}, \bibinfo{number}{OOPSLA} (\bibinfo{year}{2020}).
\newblock
\showISSN{24751421}


\bibitem[Yu et~al\mbox{.}(2023)]%
        {electronics12040936}
\bibfield{author}{\bibinfo{person}{Xueqi Yu}, \bibinfo{person}{Zhen Li}, \bibinfo{person}{Xiang Huang}, {and} \bibinfo{person}{Shasha Zhao}.} \bibinfo{year}{2023}\natexlab{}.
\newblock \showarticletitle{AdVulCode: Generating Adversarial Vulnerable Code against Deep Learning-Based Vulnerability Detectors}.
\newblock \bibinfo{journal}{\emph{Electronics}} \bibinfo{volume}{12}, \bibinfo{number}{4} (\bibinfo{year}{2023}).
\newblock
\showISSN{2079-9292}
\urldef\tempurl%
\url{https://doi.org/10.3390/electronics12040936}
\showDOI{\tempurl}


\bibitem[Zeng et~al\mbox{.}(2022)]%
        {10.1145/3533767.3534390}
\bibfield{author}{\bibinfo{person}{Zhengran Zeng}, \bibinfo{person}{Hanzhuo Tan}, \bibinfo{person}{Haotian Zhang}, \bibinfo{person}{Jing Li}, \bibinfo{person}{Yuqun Zhang}, {and} \bibinfo{person}{Lingming Zhang}.} \bibinfo{year}{2022}\natexlab{}.
\newblock \showarticletitle{An Extensive Study on Pre-Trained Models for Program Understanding and Generation}. In \bibinfo{booktitle}{\emph{Proceedings of the 31st ACM SIGSOFT International Symposium on Software Testing and Analysis}} (Virtual, South Korea) \emph{(\bibinfo{series}{ISSTA 2022})}. \bibinfo{publisher}{Association for Computing Machinery}, \bibinfo{address}{New York, NY, USA}, \bibinfo{pages}{39–51}.
\newblock
\showISBNx{9781450393799}
\urldef\tempurl%
\url{https://doi.org/10.1145/3533767.3534390}
\showDOI{\tempurl}


\bibitem[Zhang et~al\mbox{.}(2023b)]%
        {Zhang_2023}
\bibfield{author}{\bibinfo{person}{Beiqi Zhang}, \bibinfo{person}{Peng Liang}, \bibinfo{person}{Xiyu Zhou}, \bibinfo{person}{Aakash Ahmad}, {and} \bibinfo{person}{Muhammad Waseem}.} \bibinfo{year}{2023}\natexlab{b}.
\newblock \showarticletitle{Practices and Challenges of Using {GitHub} Copilot: An Empirical Study}. In \bibinfo{booktitle}{\emph{International Conferences on Software Engineering and Knowledge Engineering}}. \bibinfo{publisher}{{KSI} Research Inc.}
\newblock
\urldef\tempurl%
\url{https://doi.org/10.18293/seke2023-077}
\showDOI{\tempurl}


\bibitem[Zhang et~al\mbox{.}(2023d)]%
        {zhang2023transfer}
\bibfield{author}{\bibinfo{person}{Chi Zhang}, \bibinfo{person}{Zifan Wang}, \bibinfo{person}{Ravi Mangal}, \bibinfo{person}{Matt Fredrikson}, \bibinfo{person}{Limin Jia}, {and} \bibinfo{person}{Corina Pasareanu}.} \bibinfo{year}{2023}\natexlab{d}.
\newblock \bibinfo{title}{Transfer Attacks and Defenses for Large Language Models on Coding Tasks}.
\newblock
\newblock
\showeprint[arxiv]{2311.13445}~[cs.LG]


\bibitem[Zhang et~al\mbox{.}(2022a)]%
        {zhang2022towards}
\bibfield{author}{\bibinfo{person}{Huangzhao Zhang}, \bibinfo{person}{Zhiyi Fu}, \bibinfo{person}{Ge Li}, \bibinfo{person}{Lei Ma}, \bibinfo{person}{Zhehao Zhao}, \bibinfo{person}{Hua’an Yang}, \bibinfo{person}{Yizhe Sun}, \bibinfo{person}{Yang Liu}, {and} \bibinfo{person}{Zhi Jin}.} \bibinfo{year}{2022}\natexlab{a}.
\newblock \showarticletitle{Towards robustness of deep program processing models—detection, estimation, and enhancement}.
\newblock \bibinfo{journal}{\emph{TOSEM}} \bibinfo{volume}{31}, \bibinfo{number}{3} (\bibinfo{year}{2022}), \bibinfo{pages}{1--40}.
\newblock


\bibitem[Zhang et~al\mbox{.}(2020a)]%
        {MHM}
\bibfield{author}{\bibinfo{person}{Huangzhao Zhang}, \bibinfo{person}{Zhuo Li}, \bibinfo{person}{Ge Li}, \bibinfo{person}{Lei Ma}, \bibinfo{person}{Yang Liu}, {and} \bibinfo{person}{Zhi Jin}.} \bibinfo{year}{2020}\natexlab{a}.
\newblock \showarticletitle{Generating Adversarial Examples for Holding Robustness of Source Code Processing Models}.
\newblock \bibinfo{journal}{\emph{Proceedings of the AAAI Conference on Artificial Intelligence}} \bibinfo{volume}{34}, \bibinfo{number}{01} (\bibinfo{date}{Apr.} \bibinfo{year}{2020}), \bibinfo{pages}{1169--1176}.
\newblock


\bibitem[Zhang et~al\mbox{.}({[n.\,d.]})]%
        {zhangcodebert}
\bibfield{author}{\bibinfo{person}{Huangzhao Zhang}, \bibinfo{person}{Shuai Lu}, \bibinfo{person}{Zhuo Li}, \bibinfo{person}{Zhi Jin}, \bibinfo{person}{Lei Ma}, \bibinfo{person}{Yang Liu}, {and} \bibinfo{person}{Ge Li}.} \bibinfo{year}{[n.\,d.]}\natexlab{}.
\newblock \showarticletitle{CodeBERT-Attack: Adversarial attack against source code deep learning models via pre-trained model}.
\newblock \bibinfo{journal}{\emph{Journal of Software: Evolution and Process}} (\bibinfo{year}{[n.\,d.]}), \bibinfo{pages}{e2571}.
\newblock


\bibitem[Zhang et~al\mbox{.}(2023c)]%
        {zhang2023rnns}
\bibfield{author}{\bibinfo{person}{Jie Zhang}, \bibinfo{person}{Wei Ma}, \bibinfo{person}{Qiang Hu}, \bibinfo{person}{Xiaofei Xie}, \bibinfo{person}{Yves~Le Traon}, {and} \bibinfo{person}{Yang Liu}.} \bibinfo{year}{2023}\natexlab{c}.
\newblock \bibinfo{title}{RNNS: Representation Nearest Neighbor Search Black-Box Attack on Code Models}.
\newblock
\newblock
\showeprint[arxiv]{2305.05896}~[cs.CR]


\bibitem[Zhang et~al\mbox{.}(2024)]%
        {zhang2024android}
\bibfield{author}{\bibinfo{person}{Jiwen Zhang}, \bibinfo{person}{Jihao Wu}, \bibinfo{person}{Yihua Teng}, \bibinfo{person}{Minghui Liao}, \bibinfo{person}{Nuo Xu}, \bibinfo{person}{Xiao Xiao}, \bibinfo{person}{Zhongyu Wei}, {and} \bibinfo{person}{Duyu Tang}.} \bibinfo{year}{2024}\natexlab{}.
\newblock \bibinfo{title}{Android in the Zoo: Chain-of-Action-Thought for GUI Agents}.
\newblock
\newblock
\showeprint[arxiv]{2403.02713}~[cs.CL]


\bibitem[Zhang et~al\mbox{.}(2020b)]%
        {zhang2020pegasus}
\bibfield{author}{\bibinfo{person}{Jingqing Zhang}, \bibinfo{person}{Yao Zhao}, \bibinfo{person}{Mohammad Saleh}, {and} \bibinfo{person}{Peter Liu}.} \bibinfo{year}{2020}\natexlab{b}.
\newblock \showarticletitle{Pegasus: Pre-training with extracted gap-sentences for abstractive summarization}. In \bibinfo{booktitle}{\emph{International Conference on Machine Learning}}. PMLR, \bibinfo{pages}{11328--11339}.
\newblock


\bibitem[Zhang et~al\mbox{.}(2022b)]%
        {ml-testing-survey}
\bibfield{author}{\bibinfo{person}{Jie~M. Zhang}, \bibinfo{person}{Mark Harman}, \bibinfo{person}{Lei Ma}, {and} \bibinfo{person}{Yang Liu}.} \bibinfo{year}{2022}\natexlab{b}.
\newblock \showarticletitle{Machine Learning Testing: Survey, Landscapes and Horizons}.
\newblock \bibinfo{journal}{\emph{IEEE Transactions on Software Engineering}} \bibinfo{volume}{48}, \bibinfo{number}{1} (\bibinfo{year}{2022}), \bibinfo{pages}{1--36}.
\newblock
\urldef\tempurl%
\url{https://doi.org/10.1109/TSE.2019.2962027}
\showDOI{\tempurl}


\bibitem[Zhang et~al\mbox{.}(2022c)]%
        {zhang2022does}
\bibfield{author}{\bibinfo{person}{Kechi Zhang}, \bibinfo{person}{Ge Li}, {and} \bibinfo{person}{Zhi Jin}.} \bibinfo{year}{2022}\natexlab{c}.
\newblock \bibinfo{title}{What does Transformer learn about source code?}
\newblock
\newblock
\showeprint[arxiv]{2207.08466}~[cs.SE]


\bibitem[Zhang and Li(2023)]%
        {zhang2023code}
\bibfield{author}{\bibinfo{person}{Sheng Zhang} {and} \bibinfo{person}{Hui Li}.} \bibinfo{year}{2023}\natexlab{}.
\newblock \bibinfo{title}{Code Membership Inference for Detecting Unauthorized Data Use in Code Pre-trained Language Models}.
\newblock
\newblock


\bibitem[Zhang et~al\mbox{.}(2021)]%
        {10.1145/3411764.3445646}
\bibfield{author}{\bibinfo{person}{Tianyi Zhang}, \bibinfo{person}{Zhiyang Chen}, \bibinfo{person}{Yuanli Zhu}, \bibinfo{person}{Priyan Vaithilingam}, \bibinfo{person}{Xinyu Wang}, {and} \bibinfo{person}{Elena~L. Glassman}.} \bibinfo{year}{2021}\natexlab{}.
\newblock \showarticletitle{Interpretable Program Synthesis}. In \bibinfo{booktitle}{\emph{Proceedings of the 2021 CHI Conference on Human Factors in Computing Systems}} (Yokohama, Japan) \emph{(\bibinfo{series}{CHI '21})}. \bibinfo{publisher}{Association for Computing Machinery}, \bibinfo{address}{New York, NY, USA}, Article \bibinfo{articleno}{105}, \bibinfo{numpages}{16}~pages.
\newblock
\showISBNx{9781450380966}
\urldef\tempurl%
\url{https://doi.org/10.1145/3411764.3445646}
\showDOI{\tempurl}


\bibitem[Zhang et~al\mbox{.}(2023a)]%
        {10028657}
\bibfield{author}{\bibinfo{person}{Weiwei Zhang}, \bibinfo{person}{Shengjian Guo}, \bibinfo{person}{Hongyu Zhang}, \bibinfo{person}{Yulei Sui}, \bibinfo{person}{Yinxing Xue}, {and} \bibinfo{person}{Yun Xu}.} \bibinfo{year}{2023}\natexlab{a}.
\newblock \showarticletitle{Challenging Machine Learning-Based Clone Detectors via Semantic-Preserving Code Transformations}.
\newblock \bibinfo{journal}{\emph{TSE}} \bibinfo{volume}{49}, \bibinfo{number}{5} (\bibinfo{year}{2023}), \bibinfo{pages}{3052--3070}.
\newblock
\urldef\tempurl%
\url{https://doi.org/10.1109/TSE.2023.3240118}
\showDOI{\tempurl}


\bibitem[Zhang et~al\mbox{.}(2022d)]%
        {Zhang2022diet}
\bibfield{author}{\bibinfo{person}{Zhaowei Zhang}, \bibinfo{person}{Hongyu Zhang}, \bibinfo{person}{Beijun Shen}, {and} \bibinfo{person}{Xiaodong Gu}.} \bibinfo{year}{2022}\natexlab{d}.
\newblock \showarticletitle{Diet Code is Healthy: Simplifying Programs for Pre-Trained Models of Code}. In \bibinfo{booktitle}{\emph{Proceedings of the 30th ACM Joint European Software Engineering Conference and Symposium on the Foundations of Software Engineering}} (Singapore, Singapore) \emph{(\bibinfo{series}{ESEC/FSE 2022})}. \bibinfo{publisher}{Association for Computing Machinery}, \bibinfo{address}{New York, NY, USA}, \bibinfo{pages}{1073–1084}.
\newblock
\showISBNx{9781450394130}
\urldef\tempurl%
\url{https://doi.org/10.1145/3540250.3549094}
\showDOI{\tempurl}


\bibitem[Zheng et~al\mbox{.}(2022)]%
        {ZHENG2022111245}
\bibfield{author}{\bibinfo{person}{Wei Zheng}, \bibinfo{person}{Tianren Shen}, \bibinfo{person}{Xiang Chen}, {and} \bibinfo{person}{Peiran Deng}.} \bibinfo{year}{2022}\natexlab{}.
\newblock \showarticletitle{Interpretability application of the Just-in-Time software defect prediction model}.
\newblock \bibinfo{journal}{\emph{Journal of Systems and Software}}  \bibinfo{volume}{188} (\bibinfo{year}{2022}), \bibinfo{pages}{111245}.
\newblock
\showISSN{0164-1212}
\urldef\tempurl%
\url{https://doi.org/10.1016/j.jss.2022.111245}
\showDOI{\tempurl}


\bibitem[Zheng et~al\mbox{.}(2023)]%
        {zheng2023survey}
\bibfield{author}{\bibinfo{person}{Zibin Zheng}, \bibinfo{person}{Kaiwen Ning}, \bibinfo{person}{Yanlin Wang}, \bibinfo{person}{Jingwen Zhang}, \bibinfo{person}{Dewu Zheng}, \bibinfo{person}{Mingxi Ye}, {and} \bibinfo{person}{Jiachi Chen}.} \bibinfo{year}{2023}\natexlab{}.
\newblock \bibinfo{title}{A Survey of Large Language Models for Code: Evolution, Benchmarking, and Future Trends}.
\newblock
\newblock
\showeprint[arxiv]{2311.10372}~[cs.SE]


\bibitem[Zhou et~al\mbox{.}(2023)]%
        {zhou2023concerns}
\bibfield{author}{\bibinfo{person}{Xiyu Zhou}, \bibinfo{person}{Peng Liang}, \bibinfo{person}{Beiqi Zhang}, \bibinfo{person}{Zengyang Li}, \bibinfo{person}{Aakash Ahmad}, \bibinfo{person}{Mojtaba Shahin}, {and} \bibinfo{person}{Muhammad Waseem}.} \bibinfo{year}{2023}\natexlab{}.
\newblock \bibinfo{title}{On the Concerns of Developers When Using GitHub Copilot}.
\newblock
\newblock
\showeprint[arxiv]{2311.01020}~[cs.SE]


\bibitem[Zhou et~al\mbox{.}(2019)]%
        {Devign}
\bibfield{author}{\bibinfo{person}{Yaqin Zhou}, \bibinfo{person}{Shangqing Liu}, \bibinfo{person}{Jingkai Siow}, \bibinfo{person}{Xiaoning Du}, {and} \bibinfo{person}{Yang Liu}.} \bibinfo{year}{2019}\natexlab{}.
\newblock \bibinfo{booktitle}{\emph{Devign: Effective Vulnerability Identification by Learning Comprehensive Program Semantics via Graph Neural Networks}}.
\newblock \bibinfo{publisher}{Curran Associates Inc.}, \bibinfo{address}{Red Hook, NY, USA}.
\newblock


\bibitem[Zhou et~al\mbox{.}(2022)]%
        {10.1145/3501256}
\bibfield{author}{\bibinfo{person}{Yu Zhou}, \bibinfo{person}{Xiaoqing Zhang}, \bibinfo{person}{Juanjuan Shen}, \bibinfo{person}{Tingting Han}, \bibinfo{person}{Taolue Chen}, {and} \bibinfo{person}{Harald Gall}.} \bibinfo{year}{2022}\natexlab{}.
\newblock \showarticletitle{Adversarial Robustness of Deep Code Comment Generation}.
\newblock \bibinfo{journal}{\emph{ACM Trans. Softw. Eng. Methodol.}} \bibinfo{volume}{31}, \bibinfo{number}{4}, Article \bibinfo{articleno}{60} (\bibinfo{date}{jul} \bibinfo{year}{2022}), \bibinfo{numpages}{30}~pages.
\newblock
\showISSN{1049-331X}
\urldef\tempurl%
\url{https://doi.org/10.1145/3501256}
\showDOI{\tempurl}


\bibitem[Zhu et~al\mbox{.}(2020)]%
        {10.1007/978-3-030-60029-7-32}
\bibfield{author}{\bibinfo{person}{Mingdong Zhu}, \bibinfo{person}{Xianfang Wang}, {and} \bibinfo{person}{Yang Zhang}.} \bibinfo{year}{2020}\natexlab{}.
\newblock \showarticletitle{Interpretable Text-to-SQL Generation with Joint Optimization}. In \bibinfo{booktitle}{\emph{Web Information Systems and Applications: 17th International Conference, WISA 2020, Guangzhou, China, September 23–25, 2020, Proceedings}} (Guangzhou, China). \bibinfo{publisher}{Springer-Verlag}, \bibinfo{address}{Berlin, Heidelberg}, \bibinfo{pages}{341–351}.
\newblock
\showISBNx{978-3-030-60028-0}
\urldef\tempurl%
\url{https://doi.org/10.1007/978-3-030-60029-7_32}
\showDOI{\tempurl}


\bibitem[Zhu and Zhang(2023)]%
        {10123534}
\bibfield{author}{\bibinfo{person}{Rui Zhu} {and} \bibinfo{person}{Cunming Zhang}.} \bibinfo{year}{2023}\natexlab{}.
\newblock \showarticletitle{How Robust Is a Large Pre-trained Language Model for Code Generationƒ A Case on Attacking GPT2}. In \bibinfo{booktitle}{\emph{2023 IEEE International Conference on Software Analysis, Evolution and Reengineering (SANER)}}. \bibinfo{pages}{708--712}.
\newblock
\urldef\tempurl%
\url{https://doi.org/10.1109/SANER56733.2023.00076}
\showDOI{\tempurl}


\bibitem[Zhuang et~al\mbox{.}(2022)]%
        {JIT2708}
\bibfield{author}{\bibinfo{person}{Guoqiang Zhuang}, \bibinfo{person}{Yubin Qu}, \bibinfo{person}{Long Li}, \bibinfo{person}{Xianzhen Dou}, {and} \bibinfo{person}{Mengao Li}.} \bibinfo{year}{2022}\natexlab{}.
\newblock \showarticletitle{An Empirical Study of Gradient-based Explainability Techniques for Self-admitted Technical Debt Detection}.
\newblock \bibinfo{journal}{\emph{Journal of Internet Technology}} \bibinfo{volume}{23}, \bibinfo{number}{3} (\bibinfo{year}{2022}), \bibinfo{pages}{631--641}.
\newblock
\showISSN{2079-4029}


\bibitem[Zhuo et~al\mbox{.}(2023a)]%
        {zhuo2023-robustness}
\bibfield{author}{\bibinfo{person}{Terry~Yue Zhuo}, \bibinfo{person}{Zhuang Li}, \bibinfo{person}{Yujin Huang}, \bibinfo{person}{Fatemeh Shiri}, \bibinfo{person}{Weiqing Wang}, \bibinfo{person}{Gholamreza Haffari}, {and} \bibinfo{person}{Yuan-Fang Li}.} \bibinfo{year}{2023}\natexlab{a}.
\newblock \showarticletitle{On Robustness of Prompt-based Semantic Parsing with Large Pre-trained Language Model: An Empirical Study on Codex}. In \bibinfo{booktitle}{\emph{Proceedings of the 17th Conference of the European Chapter of the Association for Computational Linguistics}}. \bibinfo{publisher}{Association for Computational Linguistics}, \bibinfo{address}{Dubrovnik, Croatia}, \bibinfo{pages}{1090--1102}.
\newblock


\bibitem[Zhuo et~al\mbox{.}(2023b)]%
        {zhuo2023source}
\bibfield{author}{\bibinfo{person}{Terry~Yue Zhuo}, \bibinfo{person}{Zhou Yang}, \bibinfo{person}{Zhensu Sun}, \bibinfo{person}{Yufei Wang}, \bibinfo{person}{Li Li}, \bibinfo{person}{Xiaoning Du}, \bibinfo{person}{Zhenchang Xing}, {and} \bibinfo{person}{David Lo}.} \bibinfo{year}{2023}\natexlab{b}.
\newblock \bibinfo{title}{Source Code Data Augmentation for Deep Learning: A Survey}.
\newblock
\newblock
\showeprint[arxiv]{2305.19915}~[cs.CL]


\bibitem[Zhuo et~al\mbox{.}(2024)]%
        {zhuo2024astraios}
\bibfield{author}{\bibinfo{person}{Terry~Yue Zhuo}, \bibinfo{person}{Armel Zebaze}, \bibinfo{person}{Nitchakarn Suppattarachai}, \bibinfo{person}{Leandro von Werra}, \bibinfo{person}{Harm de Vries}, \bibinfo{person}{Qian Liu}, {and} \bibinfo{person}{Niklas Muennighoff}.} \bibinfo{year}{2024}\natexlab{}.
\newblock \bibinfo{title}{Astraios: Parameter-Efficient Instruction Tuning Code Large Language Models}.
\newblock
\newblock
\showeprint[arxiv]{2401.00788}~[cs.CL]


\bibitem[Ziegler et~al\mbox{.}(2022)]%
        {10.1145/3520312.3534864}
\bibfield{author}{\bibinfo{person}{Albert Ziegler}, \bibinfo{person}{Eirini Kalliamvakou}, \bibinfo{person}{X.~Alice Li}, \bibinfo{person}{Andrew Rice}, \bibinfo{person}{Devon Rifkin}, \bibinfo{person}{Shawn Simister}, \bibinfo{person}{Ganesh Sittampalam}, {and} \bibinfo{person}{Edward Aftandilian}.} \bibinfo{year}{2022}\natexlab{}.
\newblock \showarticletitle{Productivity Assessment of Neural Code Completion}. In \bibinfo{booktitle}{\emph{Proceedings of the 6th ACM SIGPLAN International Symposium on Machine Programming}} (San Diego, CA, USA) \emph{(\bibinfo{series}{MAPS 2022})}. \bibinfo{publisher}{Association for Computing Machinery}, \bibinfo{address}{New York, NY, USA}, \bibinfo{pages}{21–29}.
\newblock
\showISBNx{9781450392730}
\urldef\tempurl%
\url{https://doi.org/10.1145/3520312.3534864}
\showDOI{\tempurl}


\bibitem[Zlotchevski et~al\mbox{.}(2022)]%
        {10.1145/3540250.3558959}
\bibfield{author}{\bibinfo{person}{Andrei Zlotchevski}, \bibinfo{person}{Dawn Drain}, \bibinfo{person}{Alexey Svyatkovskiy}, \bibinfo{person}{Colin~B. Clement}, \bibinfo{person}{Neel Sundaresan}, {and} \bibinfo{person}{Michele Tufano}.} \bibinfo{year}{2022}\natexlab{}.
\newblock \showarticletitle{Exploring and Evaluating Personalized Models for Code Generation} \emph{(\bibinfo{series}{ESEC/FSE 2022})}. \bibinfo{pages}{1500–1508}.
\newblock
\showISBNx{9781450394130}
\urldef\tempurl%
\url{https://doi.org/10.1145/3540250.3558959}
\showDOI{\tempurl}


\bibitem[Zou et~al\mbox{.}(2021)]%
        {10.1145/3429444}
\bibfield{author}{\bibinfo{person}{Deqing Zou}, \bibinfo{person}{Yawei Zhu}, \bibinfo{person}{Shouhuai Xu}, \bibinfo{person}{Zhen Li}, \bibinfo{person}{Hai Jin}, {and} \bibinfo{person}{Hengkai Ye}.} \bibinfo{year}{2021}\natexlab{}.
\newblock \showarticletitle{Interpreting Deep Learning-Based Vulnerability Detector Predictions Based on Heuristic Searching}.
\newblock \bibinfo{journal}{\emph{ACM Trans. Softw. Eng. Methodol.}} \bibinfo{volume}{30}, \bibinfo{number}{2}, Article \bibinfo{articleno}{23} (\bibinfo{date}{mar} \bibinfo{year}{2021}), \bibinfo{numpages}{31}~pages.
\newblock
\showISSN{1049-331X}
\urldef\tempurl%
\url{https://doi.org/10.1145/3429444}
\showDOI{\tempurl}


\end{thebibliography}

\end{document}